\documentclass[longauth]{aa} 
\usepackage{graphicx}
\usepackage{natbib}
\usepackage{amssymb}
\usepackage{longtable}
\usepackage{lscape}
\usepackage{color}
\usepackage{txfonts}

\begin{document}
   \title{International observational campaigns of the last two eclipses in
EE Cep: 2003 and 2008/9 \thanks{Tables \ref{Phot.8.dat} --
\ref{Spect.list.8910.dat} are available at the CDS via anonymous ftp to
cdsarc.u-strasbg.fr (130.79.128.5) or via
http://cdsweb.u-strasbg.fr/cgi-bin/qcat?J/A+A/544/A53}}

\author{C.\,Ga{\l}an\inst{1,2}
\and M.\,Miko{\l}ajewski\inst{1}
\and T.\,Tomov\inst{1}
\and D.\,Graczyk\inst{3}
\and G.\,Apostolovska\inst{4}		
\and I.\,Barzova\inst{5}
\and I.\,Bellas-Velidis\inst{6}		
\and B.\,Bilkina\inst{5}
\and R.M.\,Blake\inst{7}
\and C.T.\,Bolton\inst{8}
\and A.\,Bondar\inst{9}
\and L.\,Br\'at\inst{10,11}
\and T.\,Bro\.zek\inst{1}
\and B.\,Budzisz\inst{1}
\and M.\,Cika{\l}a\inst{1,12}
\and B.\,Cs\'ak\inst{13}
\and A.\,Dapergolas\inst{6}
\and D.\,Dimitrov\inst{5}
\and P.\,Dobierski\inst{1}
\and M.\,Drahus\inst{14}
\and M.\,Dr\'o\.zd\.z\inst{15}
\and S.\,Dvorak\inst{16}		
\and L.\,Elder\inst{17}
\and S.\,Fr\c{a}ckowiak\inst{1}
\and G.\,Galazutdinov\inst{18}		
\and K.\,Gazeas\inst{19}		
\and L.\,Georgiev\inst{20}
\and B.\,Gere\inst{21}
\and K.\,Go\'zdziewski\inst{1}		
\and V.P.\,Grinin\inst{22}
\and M.\,Gromadzki\inst{1,23}
\and M.\,Hajduk\inst{1,24}
\and T.A.\,Heras\inst{25}		
\and J.\,Hopkins\inst{26}		
\and I.\,Iliev\inst{5}
\and J.\,Janowski\inst{1}
\and R.\,Koci\'an\inst{27}
\and Z.\,Ko{\l}aczkowski\inst{3,28}
\and D.\,Kolev\inst{5}
\and G.\,Kopacki\inst{28}
\and J.\,Krzesi\'nski\inst{15}
\and H.\,Ku\v{c}\'akov\'a\inst{27}	
\and E.\,Kuligowska\inst{29}
\and T.\,Kundera\inst{29}
\and M.\,Kurpi\'nska-Winiarska\inst{29}
\and A.\,Ku\'zmicz\inst{29}
\and A.\,Liakos\inst{19}		
\and T.A.\,Lister\inst{30}
\and G.\,Maciejewski\inst{1}
\and A.\,Majcher\inst{1,31}
\and A.\,Majewska\inst{28}
\and P.M.\,Marrese\inst{32}
\and G.\,Michalska\inst{3,28}
\and C.\,Migaszewski\inst{1}	
\and I.\,Miller\inst{33,34}
\and U.\,Munari\inst{35}
\and F.\,Musaev\inst{36}
\and G.\,Myers\inst{37}
\and A.\,Narwid\inst{28}
\and P.\,N\'emeth\inst{38}
\and P.\,Niarchos\inst{19}		
\and E.\,Niemczura\inst{28}
\and W.\,Og{\l}oza\inst{15}
\and Y.\,\"O\v{g}men\inst{39}
\and A.\,Oksanen\inst{40}
\and J.\,Osiwa{\l}a\inst{1}
\and S.\,Peneva\inst{5}
\and A.\,Pigulski\inst{28}		
\and V.\,Popov\inst{5}
\and W.\,Pych\inst{41}
\and J.\,Pye\inst{17}			
\and E.\,Ragan\inst{1}
\and B.F.\,Roukema\inst{1}
\and P.T.\,R\'o\.za\'nski\inst{1}
\and E.\,Semkov\inst{5}			
\and M.\,Siwak\inst{15,29}		
\and B.\,Staels\inst{42}
\and I.\,Stateva\inst{5}
\and H.C.\,Stempels\inst{43}		
\and M.\,St\c{e}\'slicki\inst{28}
\and E.\,\'Swierczy\'nski\inst{1}
\and T.\,Szyma\'nski\inst{29}
\and N.\,Tomov\inst{5}			
\and W.\,Waniak\inst{29}
\and M.\,Wi\c{e}cek\inst{1,44}
\and M.\,Winiarski\inst{15,29}
\and P.\,Wychudzki\inst{1,2}
\and A.\,Zajczyk\inst{1,24}
\and S.\,Zo{\l}a\inst{15,29}
\and T.\,Zwitter\inst{45}
	}


\institute{		
Toru\'n Centre for Astronomy, Nicolaus Copernicus University, ul. Gagarina 11, 87-100 Toru\'n, Poland\\
\email{[cgalan; mamiko; tomtom]@astri.uni.torun.pl}
\and 
Olsztyn Planetarium and Astronomical Observatory, Al. Marsza{\l}ka J. Pi{\l}sudskiego 38, 10-450 Olsztyn, Poland
\and 
Universidad de Concepci\'on, Departamento de Astronomia, Casilla 160-C, Concepci\'on, Chile
\and 
Institute of Physics, Faculty of Science, Ss. Cyril and Methodius University, PO Box 162, 1000 Skopje, FYROM, Macedonia
\and 
Institute of Astronomy and National Astronomical Observatory, Bulgarian Academy of Sciences, 72 Tsarigradsko Shose Blvd., BG-1784 Sofia, Bulgaria
\and 
Institute for Astronomy, Astrophysics, Space Applications and Remote Sensing, NOA, PO Box 20048, 11810 Athens, Greece
\and 
Dept. of Physics and Earth Science, University of North Alabama, Florence, 35632 AL, USA
\and 
David Dunlap Observatory, Department of Astronomy and Astrophysics, University of Toronto, 50 St. George St., Toronto, ON M5S 3H4, Canada
\and 
International Centre for Astronomical and Medico-Ecological Research, Terskol, Russia
\and 
Variable Star and Exoplanet Section of Czech Astronomical Society, Czech Republic
\and 
Altan Observatory, Velka Upa 193, Pec pod Snezkou, Czech Republic
\and 
Tadeusz Banachiewicz Astronomical Observatory, W\c{e}gl\'owka, PL-32-412 Wi\'sniowa, Poland
\and 
Max Planck Institute for Astronomy, K\"onigstuhl 17, D-69117 Heidelberg, Germany
\and 
Department of Earth and Space Sciences, University of California at Los Angeles, 595 Charles E.  Young Dr.  East, CA 90095, USA
\and 
Mt. Suhora Observatory, Pedagogical Univ., ul. {Podchor\c{a}\.zych} 2, 30-084 Krak\'ow, Poland
\and 
Rolling Hills Observatory Clermont, FL, USA
\and 
University of Hawaii Maui College, Kahului, Hawaii
\and 
Instituto de Astronomia, Universidad Catolica del Norte, Av. Angamos 0610, Antofagasta, Chile
\and 
Department of Astrophysics, Astronomy and Mechanics, National and Kapodistrian University of Athens, GR 157 84 Zografos, Athens, Greece
\and 
Instituto de Astronom\'ıa, Universidad Nacional Aut\'onoma de M\'exico Apdo. postal 70–264, Ciudad Universitaria, M\'exico D.F. 04510, M\'exico
\and 
Department of Experimental Physics and Astronomical Observatory, University of Szeged, Dom ter 9, H-6720 Szeged, Hungary
\and 
Pulkovo Astronomical Observatory, Russian Academy of Sciences, Pulkovskoe sh. 65, St. Petersburg, 196140, Russia
\and 
Space Research Centre, Polish Academy of Sciences, Bartycka 18A, Pl-00-716 Warsaw, Poland
\and 
Nicolaus Copernicus Astronomical Center, Rabia\'nska 8, 87-100 Toru\'n, Poland
\and 
Observatorio Astron\'omico "Las Pegueras", NAVAS DE ORO (Segovia), Spain
\and 
Hopkins Phoenix Observatory, 7812 West Clayton Drive, Phoenix, Arizona 85033-2439, USA
\and 
Observatory and Planetarium of Johann Palisa, V{\v{S}}B - Technical University of Ostrava, 17. listopadu 15, 708 33 Ostrava-Poruba, Czech Republic
\and 
Instytut Astronomiczny, Uniwersytet Wroc{\l}awski, Kopernika 11, 51-622 Wroc{\l}aw, Poland
\and 
Astronomical Observatory, Jagiellonian Univ., ul. Orla 171, 30-244 Krak\'ow, Poland
\and 
Las Cumbres Observatory, 6740 Cortona Drive Suite 102, Goleta, CA 93117, USA
\and 
National Centre for Nuclear Research, Warsaw, Poland
\and 
Leiden Observatory, P.O. Box 9513, 2300 RA Leiden, The Netherlands
\and 
Furzehill House, Ilston, Swansea. SA2 7LE, UK
\and 
Variable Star Section of the British Astronomical Association
\and 
INAF, Osservatorio Astronomico di Padova, via dell’ Osservatorio 8, 36012 Asiago (VI), Italy
\and 
Special Astrophysical Observatory of the Russian AS, Nizhnij Arkhyz 369167, Russia
\and 
GRAS Observatory, Mayhill, New Mexico, USA
\and 
Department of Physics and Space Sciences, 150 W. University Blvd, Florida Institute of Technology, Melbourne, FL 32901, USA
\and 
Green Island Observatory (B34), North Cyprus
\and 
Hankasalmi Observatory, Jyvaskylan Sirius ry, Vertaalantie 419, FI-40270 Palokka, Finland
\and 
Nicolaus Copernicus Astronomical Center, Bartycka 18, 00-716 Warsaw, Poland
\and 
Sonoita Research Observatory/AAVSO, USA
\and 
Department of Physics and Astronomy, Box 516, SE-751 20 Uppsala, Sweden
\and 
Centrum Hewelianum, PKFM "Twierdza Gda\'nsk", ul. 3 Maja 9a, 80-802 Gda\'nsk, Poland
\and 
University of Ljubljana, Faculty of Mathematics and Physics, Jadranska 19, 1000 Ljubljana, Slovenia
}

   \date{Received xxxxx xx, xxxx; accepted xxxxx xx, xxxx}

  \abstract
   {EE\,Cep is~an~unusual long-period (5.6~yr) eclipsing binary discovered
during the~mid-twentieth century.  It~undergoes almost-grey eclipses that
vary in terms of both depth and duration at different epochs.  The system
consists of a~Be type star and a~dark dusty disk around an~invisible
companion.  EE\,Cep together with the~widely studied $\varepsilon$\,Aur are
the~only two~known cases of~long-period eclipsing binaries with a~dark,
dusty disk component responsible for~periodic obscurations.}
   {Two observational campaigns were~carried out~during the~eclipses
of~EE\,Cep in~2003 and~2008/9 to~verify whether the~eclipsing body in~the
system is~indeed a~dark disk and to~understand the~observed changes in~the
depths and~durations of~the~eclipses.}
   {Multicolour photometric data and~spectroscopic observations performed
at~both low and~high resolutions were~collected with~several dozen
instruments located in~Europe and~North America.  We~numerically modelled
the~variations in~brightness and~colour during the~eclipses.  We~tested
models with different disk structure, taking into consideration
the~inhomogeneous surface brightness of~the~Be~star.  We~considered the
possibility of~disk precession.}
   {The~complete set~of~observational data collected during the~last three
eclipses are~made available to~the~astronomical community.  The 2003
and~2008/9 eclipses of~EE\,Cep were~very shallow.  The~latter
is~the~shallowest among all~observed.  The~very high quality photometric
data illustrate in~detail the~colour evolution during the~eclipses
for~the~first time.  Two blue maxima in~the~colour indices were~detected
during these two eclipses, one before and one after the~photometric minimum. 
The~first (stronger) blue maximum is~simultaneous with a~``bump''
that~is~very clear in~all the~$UBV$($RI$)$_{\mathrm C}$ light curves. 
A~temporary increase in~the~$I$-band brightness at~the~orbital phase $\sim
0.2$ was~observed after each of~the~last three eclipses.  Variations
in~the~spectral line profiles seem to~be~recurrent during each cycle. 
The~\ion{Na}{i} lines always show at~least three absorption components
during the~eclipse minimum and~strong absorption is~superimposed
on~the~H$\alpha$ emission.}
   {These observations confirm that the~eclipsing object in~EE\,Cep system
is~indeed a~dark, dusty disk around a~low luminosity object.  The~primary
appears to~be~a~rapidly rotating Be star that is~strongly darkened
at~the~equator and brightened at~the~poles.  Some of~the~conclusions
of~this~work require verification in~future studies: (i) a~complex, possibly
multi-ring structure of~the~disk in~EE\,Cep; (ii) our~explanation of~the
``bump'' observed during the~last two eclipses in~terms of~the~different
times of~obscuration of~the~hot polar regions of~the~Be star by~the~disk;
and (iii) our suggested period of~the~disk precession ($\sim$ 11--12
$P_{\mathrm {orb}}$) and predicted depth of~about 2\fm~for the~forthcoming
eclipse in~2014.}

   \authorrunning {C. Ga{\l}an et al.}

   \titlerunning {International observational campaigns of the last two eclipses in EE\,Cep}


   \maketitle
%
\section{Introduction}

\begin{figure}
  \resizebox{\hsize}{!}{\includegraphics{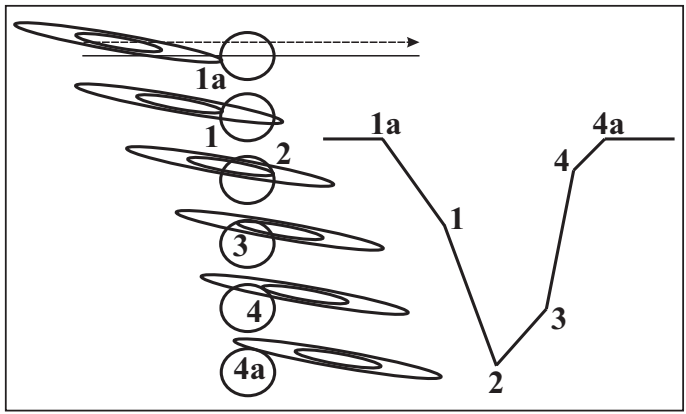}}
  \caption{Schematic representation of the eclipse geometry in the
EE\,Cep system.  The inner opaque and outer semi-transparent regions of the
disk are separated.  The characteristic positions of the disk and the star
configuration during the eclipses ({\it left}) correspond to the contact
times (1a, 1, 2, 3, 4, 4a) distinguished in the light curve ({\it right}). 
The figure shows a~highly simplified case that ignores a~number of
issues such as e.g.  possible inhomogeneities in the distribution of
brightness on the star's surface or of the actual size of the disk.}
  \label{fig.schem}
\end{figure}

The 11th magnitude star \object{EE\,Cep} is a~unique object among the about
40 well-known eclipsing systems with orbital periods longer than one year. 
The primary B5 III star is obscured by an invisible, dark secondary
component of very low luminosity every 5.6~yr.  The variability of the star
was discovered in 1952 (epoch $E$\,=\,0) by \citet{Rom1956} and soon
confirmed by \citet{Web1956}, who reported observations obtained during
a~previous eclipse in 1947 ($E$\,=\,-1).  Since then, ten consecutive
primary eclipses have been observed, while a~secondary eclipse has never
been detected.  The depths of the eclipses vary across a~wide range of
magnitudes from about $0\fm5$ to $2\fm0$ \citep[see][]{Gra2003}.  However,
all of them seem to have the same features: they are almost grey and have
a~similar asymmetric shape (the descending branch of every eclipse has
a~longer duration than the ascending one).  In the light curves of all the
eclipses, it is possible to distinguish five characteristic phases (shown in
Fig.\,\ref{fig.schem}): ingress (1-2) and egress (3-4) are preceded and
followed, respectively, by extended atmospheric eclipse parts (1a-1 and
4-4a), and in the middle of the eclipses a~bottom phase of variable slope
\mbox{(2-3)} occurs.

The most plausible hypothesis to explain the observed shape of the light
curve, as well as the changes in the eclipse depth during successive
conjunctions and their weak dependence on the photometric band was proposed
by \citet{MiGr1999}.  Their model considers the eclipses of a~hot B5-type
primary by an invisible, dark companion, which is most probably a~dusty disk
around a~low-luminosity central object.  The disk is slightly inclined to
the orbital plane.  The obscurations of the star by the opaque interior of
the disk can explain the deep central parts of the eclipses, while the
semi-transparent exterior areas are responsible for the observed external
wings, which are similar to wings caused by atmospheric eclipses in
$\zeta$\,Aur type variable stars.  The projection of the inclined disk onto
the sky plane produces oblong shape of an obscuring body, which is tilted
with respect to the direction of motion during most of the occultations. 
Since the eclipses are not central (the impact parameter is non-zero), the
light curves observed during the eclipses have an asymmetric shape
(Fig.~\ref{fig.schem}).  A~possible precession of the disk can change both
the inclination of the disk to the line of sight and the tilt of its
cross-section to the transit direction.  This leads to changes in the depth
and the duration of the eclipses.  The model briefly described above can
explain the shallow ($0\fm6$), flat-bottomed eclipse observed in 1969, if we
assume a~nearly edge-on and non-tilted projection of the disk
\citep[][]{Gra2003}.  This very specific configuration in 1969 is very
similar to the geometry of the eclipses in the \object{$\varepsilon$\,Aur}
system \citep[see][]{MiGr1999}.  Wide eclipsing binaries of this kind, i.e. 
those containing a~nearly edge-on dusty disk as an eclipsing object, are
very rare and apart from the two above-mentioned cases we know of only about
one additional system -- \object{M2-29} -- that may show some similarities
\citep[][]{Haj2008}.

In this paper, we present the results of two observational campaigns
organized for the eclipses that occured in 2003 and 2008/9, mainly to test
the hypothesis of a~precessing disk.  The results of the second campaign
were systematically presented during the eclipse at a~web
page\footnote{http://www.astri.uni.torun.pl/$\sim$cgalan/EECep/}.

%
\section{Observations}

      \subsection{The 1997 eclipse}

\begin{figure}
  \resizebox{\hsize}{!}{\includegraphics{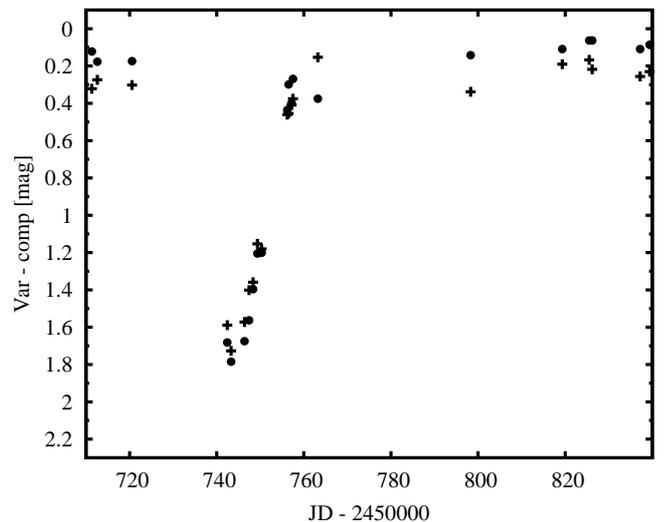}}
\caption{$U$ (dots) and $i$ (crosses) light curves obtained during the 1997
eclipse of EE\,Cep.}
  \label{fig.LC1997}
\end{figure}

During the 1997 eclipse (epoch $E$\,=\,8), the first multicolour
$UBVR_{\mathrm C}i$ ($\bar\lambda_{\mathrm i} \approx 7420$~\AA) photometric
observations were made using a~60\,cm Cassegrain telescope at the Piwnice
Observatory near Toru\'n (Poland) equipped with a~one-channel photometer
\citep{MiGr1999}.  This eclipse was one of the deepest eclipses of all those
observed in EE\,Cep.  However, the amplitude of the minimum changes quite
weakly with wavelength from about $1\fm75$ in the $U$ passband to about
$1\fm45$ in $i$ (Fig.\,\ref{fig.LC1997}).  These observations thus provided
the first evidence that the eclipsing body cannot be an ordinary evolved
cool star, motivating us to organize a~special international observational
campaign for the next two minima.  A~complete set of $UBVR_{\mathrm C}i$
photometry obtained in 1997 is shown in our Appendix (available online) in
Table\,\ref{Phot.8.dat}.

\begin{table*}
\caption{Overview of instruments and their involvement in photometric
observations of the EE\,Cep eclipses since 1997.}
\label{tab.phot.sum}
\centering			
\begin{tabular}{llllllrl}
\hline\hline
Observatory		& Country		& Telescope type	& Diameter [m]	& Bands     & Epoch & N & Table\\
\hline
Altan, Mt Giant		& Czech Republic	& Reflector		    & 0.2m  	&$B,V,R,I  $&10    &60	& \ref{Phot.Altan.10.dat}\\
Athens			& Greece		& Cassegrain		    & 0.4m 	&$B,V,R,I  $&9, 10 &176	& \ref{Phot.Athens.9.dat}, \ref{Phot.Athens.10.dat}\\
Bia\l{}k\'ow		& Poland		& Cassegrain		    & 0.6m 	&$B,V,R,I,H\alpha^{{\mathrm W} \star},H\alpha^{{\mathrm N} \star}$&9, 10	&109 & \ref{Phot.Bialkow.9.dat}, \ref{Phot.Bialkow.10.dat}\\
Green Island		& North Cyprus		& Ritchey-Chr\'etien 	    & 0.35m 	&$B,V,R,I  $&10    &35	& \ref{Phot.GreenIsl.10.dat}\\
Hankasalmi		& Finland		& RCOS			    & 0.4m 	&$B,V,R,I  $&10    &28	& \ref{Phot.Hankasalmi.10.dat}\\
Furzehill, Swansea	& United Kingdom	& Schmidt-Cassegrain	    & 0.35m 	&$B,V,R,I  $&10    &68	& \ref{Phot.Ilston.10.dat}\\
Krak\'ow		& Poland		& Cassegrain		    & 0.5m 	&$U,B,V,R,I$&9, 10 &336	& \ref{Phot.Krakow.9.dat}, \ref{Phot.Krakow.10.dat}\\
Kryoneri		& Greece		& Cassegrain		    & 1.2m 	&$U,B,V,R,I$&9, 10 &42	& \ref{Phot.Kryoneri.9.dat}, \ref{Phot.Kryoneri.10.dat}\\
GRAS, Mayhill		& USA (NM)		& Reflector		    & 0.3m 	&$B,V,I    $&10    &127	& \ref{Phot.GRAS.10.dat}\\
Navas de Oro, Segovia	& Spain			& Reflector		    & 0.35m 	&$V        $&10    &16	& \ref{Phot.NavasDeOro.10.dat}\\
Ostrava			& Czech Republic	& Newton		    & 0.2m 	&$B,V,R,I  $&10    &24	& \ref{Phot.Kucakova.10.dat}\\
Ostrava			& Czech Republic	& Schmidt-Cassegrain	    & 0.3m 	&$B,V,R,I  $&10    &4	& \ref{Phot.Kocian.10.dat}\\
Piszk\'estet\"o		& Hungary		& Schmidt		    & 0.6/0.9m	&$B,V,R,I  $&9     &12	& \ref{Phot.Piszkesteto.9.dat}\\
Piwnice$^{\star\star}$	& Poland		& Cassegrain & 0.6m 	    &$U,B,V,R,I,c^{\star},H\beta^{\star}$&8, 9 &612 & \ref{Phot.8.dat},  \ref{Phot.Piwnice.9.dat}\\
Piwnice			& Poland		& Cassegrain		    & 0.6m 	&$U,B,V,R,I$&10    &470	& \ref{Phot.Piwnice.9.dat}\\
Rolling Hills, Clermont	& USA (FL)		& Reflector		    & 0.25m 	&$B,V      $&10    &80	& \ref{Phot.Dvorak.10.dat}\\
Rozhen			& Bulgaria		& Ritchey-Chr\'etien	    & 2m	&$U,B,V,R,I$&9, 10 &20	& \ref{Phot.Rozhen.2m.9.dat}, \ref{Phot.Rozhen2m.10.dat}\\
Rozhen			& Bulgaria		& Schmidt		    & 0.5/0.7m	&$U,B,V,R,I$&9, 10 &33	& \ref{Phot.Rozhen.Schm.9.dat}, \ref{Phot.Rozhen0507m.10.dat}\\
Rozhen$^{\star\star}$	& Bulgaria		& Cassegrain 		    & 0.6m 	&$U,B,V    $&9     &18	& \ref{Phot.Rozhen0.6m.9.dat}\\
Rozhen			& Bulgaria		& Cassegrain		    & 0.6m 	&$U,B,V,R,I$&10    &34	& \ref{Phot.Rozhen06m.10.dat}\\
Skinakas		& Greece		& Ritchey-Chr\'etien	    & 1.3m 	&$U,B,V,R,I$&9     &44	& \ref{Phot.Skinakas.9.dat}\\
Sonoita			& USA (AZ)		& Reflector		    & 0.5m 	&$B,V,R,I  $&10    &349	& \ref{Phot.SRO.Staels.10.dat}, \ref{Phot.SRO.HPO.10.dat}\\
Suhora			& Poland		& Cassegrain		    & 0.6m 	&$U,B,V,R,I$&10    &196	& \ref{Phot.Suhora.10.dat}\\
Tenagra-II		& USA (AZ)		& Ritchey-Chr\'etien	    & 0.81m 	&$U,B,V,R,I$&10    &20	& \ref{Phot.Lister.10.dat}\\
\hline
\end{tabular}
\begin{list}{}{}
\item[{\bf Notes.}] $N$ is the number of individual brightness
determinations summed over all the photometric bands.  The last column
specifies the number of the table with the original data.  ($^{\star}$)
$H\alpha^{\mathrm W}$ and $H\alpha^{\mathrm N}$ are intermediate width (FWHM
$\approx 200$\,\AA) and narrow (FWHM $\approx 30$\,\AA) photometric bands,
both centred at the H$\alpha$ spectroscopic line.  $c$ and $H\beta$ are
narrow (FWHM $\approx 100$\,\AA) photometric bands centred on $\lambda =
4804$\,\AA and at the H$\beta$ spectroscopic line, respectively. 
($^{\star\star}$) In these two cases, the photomultipliers were used as the
light receiver instead of CCD.
\end{list}
\end{table*}

      \subsection{International photometric campaigns in 2003 and 2008/9}

Observers from four European countries responded to the appeal of
\citet{Mik2003} to perform a~precise monitoring of the subsequent eclipse of
EE\,Cep anticipated in 2003 (epoch $E$\,=\,9).  During the organized
campaign, ten telescopes were used to acquire the photometric observations
(Table\,\ref{tab.phot.sum}).  Very high quality photometric
$UBV(RI)_{\mathrm C}$ data were obtained with very fine sampling.  The
eclipse turned out to be quite shallow and in accordance with the
expectations, almost grey.  The eclipse reached depths from about $0\fm7$ in
$U$ to $0\fm5$ in $I_{\mathrm C}$.  The preliminary photometric results of
the 2003 campaign were described by \citet{Mik2005a}.  The results of this
fruitful campaign in 2003 did not however significantly constrain the
precessing disk model, and the nature of the central part of the disk and
its contribution to the total flux remained uncertain.  The next opportunity
for resolving these uncertainties came with the most recent eclipse, which
took place at the turn of 2008 (epoch $E$\,=\,10), with a~minimum on January
10, 2009.  An invitation to participate in an observational campaign
\citep{Gal2008} attracted strong interest.  Twenty telescopes located in
Europe and North America were involved in the photometric observations
(Table\,\ref{tab.phot.sum}), which provided a more comprehensive multicolour
and temporal photometric coverage than for any previous eclipse.  The first
results and the $UBV(RI)_{\mathrm C}$ light curves in graphical form were
published by \citet{Gal2009}.  Surprisingly, the last eclipse turned out to
be the shallowest in the observing history of EE\,Cep, reaching a~depth of
only $\sim0\fm5$ in $U$ and nearly $\sim0\fm4$ in $I_{\mathrm C}$.

\begin{figure*}
  \centering
    \includegraphics[width=17cm]{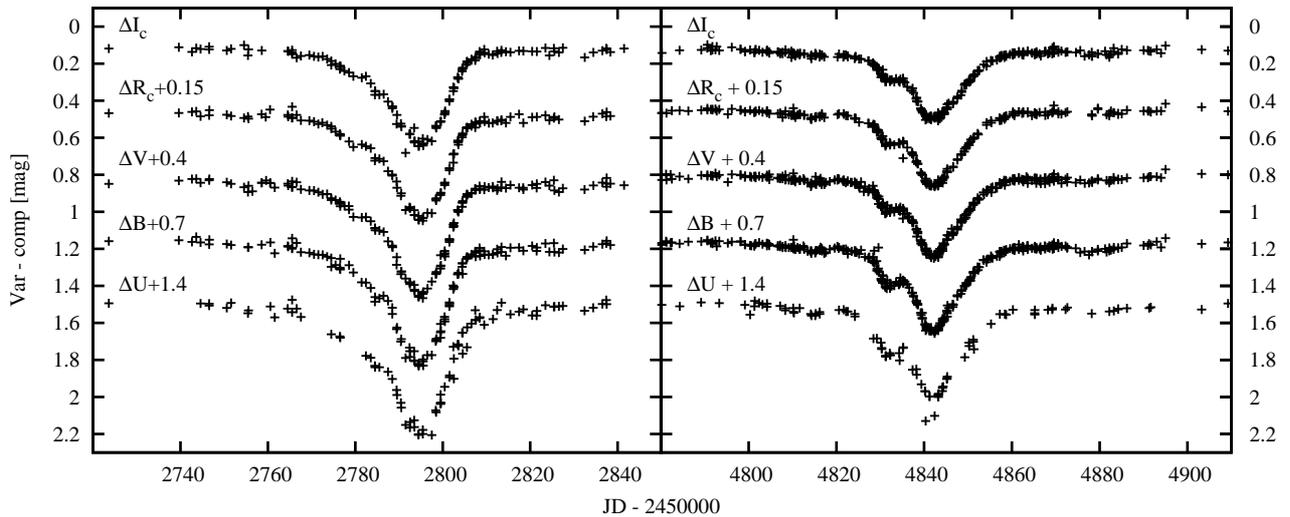}
\caption{All of the approximately 800 individual photometric
$UBV(RI)_{\mathrm C}$ measurements obtained during the 2003 eclipse ({\it
left}; Tables\,\ref{Phot.Athens.9.dat}-\ref{Phot.Piszkesteto.9.dat}) and all
of the more than 1600 observations obtained during the 2008/9 eclipse ({\it
right}; Tables\,\ref{Phot.Altan.10.dat}-\ref{Phot.Dvorak.10.dat}).}
  \label{fig.LC.2003i9}
\end{figure*}

The strong interest inspired by \citet{Mik2003} and \citet{Gal2008} resulted
in many observations.  The original data and the observatories that sent
them are presented in
Tables\,\ref{Phot.Athens.9.dat}-\ref{Phot.Piszkesteto.9.dat} for the 2003
eclipse ($E$\,=\,9) and
Tables\,\ref{Phot.Altan.10.dat}-\ref{Phot.Dvorak.10.dat} for the 2008/9
($E$\,=\,10) eclipse, respectively.  The three standard stars ``$a$'' =
BD\,+55$\degr$2690, ``$b$'' = GSC-3973:2150, and ``$c$'' =
BD\,+55$\degr$2691 have been recommended by \citet{Mik2003}.  Most
magnitudes were evaluated with respect to either the standard star "$a$"
(Tables\,\ref{Phot.Athens.9.dat}-\ref{Phot.Rozhen06m.10.dat}) or all three
standard stars ``$a$'', ``$b$'', and ``$c$'' independently
(Tables\,\ref{Phot.Athens.10.dat}-\ref{Phot.Dvorak.10.dat}).  One set of
data (Table\,\ref{Phot.Kryoneri.10.dat}) were obtained only with respect to
standard star ``$c$'' because of the small field of view of the instrument
used.  The data in
Tables\,\ref{Phot.Athens.9.dat}-\ref{Phot.Kryoneri.10.dat} and
\ref{Phot.Athens.10.dat}-\ref{Phot.Dvorak.10.dat} are shown in differential
form.  Two sets of data, from Sonoita Research Observatory
(Tables\,\ref{Phot.SRO.Staels.10.dat}-\ref{Phot.SRO.HPO.10.dat}), were
obtained partly with respect to other standard stars (see
Table\,\ref{Phot.SRO.Staels.10.dat}), and we show them as apparent
magnitudes.  They were transformed to a differential form using the known
brightness of star ``$a$'' from \citet{Mik2003}.  The original differential
magnitudes obtained with the three standard stars were calculated with
respect to star ``$a$'', using the average differences between the standard
stars $(\overline{a-b})$ and $(\overline{a-c})$.  All of the mean values in
Tables\, \ref{Phot.Athens.10.dat}--\ref{Phot.Dvorak.10.dat} for variable
star ``$v$'' were calculated according to the expression $\left[(v-a) +
(v-b)-(\overline{a-b}) + (v-c)-(\overline{a-c})\right]/3$ for each filter,
excluding $\Delta V$ and $\Delta R_{\mathrm C}$ data in
Table\,\ref{Phot.Hankasalmi.10.dat}.  For this last set of data, the
differences $(v-a)$ were not recorded by the observers, hence the reduced
average values were calculated using the expression
$\left[(v-b)-(\overline{a-b}) + (v-c)-(\overline{a-c})\right]/2$, where the
$(\overline{a-b})$ and $(\overline{a-c})$ $VR$ magnitudes were adopted from
\citet{Mik2003}.  All data were corrected for the differences between the
particular photometric systems.  The CCD data from Krak\'ow were adopted as
a~zero-point, owing to their high quality and good coverage during both the
eclipses.  Original individual data points obtained close to the eclipses
are presented in Fig.\,\ref{fig.LC.2003i9}, which is composed of about 2500
measurements.  The phases were calculated with ephemeris \citep[][]{MiGr1999}

\begin{center}
\begin{equation}
JD(Min) = 2434344.1 + 2049\fd94 \times E.   \label{eq.efem}
\end{equation}
\end{center}

\begin{table*}
\caption{Overview of the instruments involved in the spectroscopic
observations during the three last eclipses at epochs $E$\,=\,8, $E$\,=9,
and $E$\,=10.  $N$ is the number of spectra.}
\label{tab.spec.sum}
\centering			
\begin{tabular}{llllllllr}
\hline\hline
Observatory	& Country	& Telescope type	& Diameter [m]	& Spectrograph	& Res. power	& Spectral reg.		&Epoch	   &  N \\
\hline
NOT, La Palma	& Canary Isl.	& Ritchey-Chr\'etien	& 2.56	& FIES-Echelle	& 48000		& 3680-7280 \AA			& 10	   &  4 \\
Rozhen		& Bulgaria	& Ritchey-Chr\'etien	& 2.0	& Coude		& 15000, 30000	& H$\alpha$, H$\beta$, \ion{Na}{i}& 8, 9, 10 & 38 \\
Asiago		& Italy		& Cassegrain		& 1.82	& Echelle	& 20000		& 4600-9200 \AA			& 9        &  5 \\
SPM		& Mexico	& Ritchey-Chr\'etien	& 2.12	& Echelle	& 18000		& 3700-6800 \AA			& 9        &  1 \\
DDO		& Canada	& Cassegrain		& 1.88	& Cassegrain	& 16000		& H$\alpha$, \ion{Na}{i}	& 9        & 16 \\
Terskol		& Russia	& Ritchey-Chr\'etien	& 2.0	& Echelle	& 13500		& 4200-6700 \AA			& 9        &  7 \\
Asiago		& Italy		& Cassegrain		& 1.82	& AFOSC/echelle	& 3600		& 3600-8800 \AA			& 9        &  3 \\
Piwnice		& Poland	& Schmidt-Cassegrain	& 0.9	& CCS		& 2000 -- 4000  & 3500-10500 \AA		& 9, 10    & 26 \\
\hline
\end{tabular}
\end{table*}

\noindent The photometric observational data were further processed by
averaging the groups of neighbouring points.  In the case of the previous
eclipse at $E$\,=\,9, for which the photometric measurements were obtained
only in Europe, each point in the light curves represents the average of all
measurements obtained in a~given passband during a~single night.  The $V$
light curve constructed in this way was complemented by the data obtained
independently for this eclipse by \citet{Sam2004}, which we shifted by
$+0\fm02$ onto the reference system.  The last eclipse at $E$\,=\,10 was
observed from two continents, Europe and North America.  The measurements
obtained during each day form groups of points separated by about one-third
of a~day and should not be averaged together.  In the light curves of this
eclipse, each point represents the average of all measurements obtained in
a~given filter during the first or second part of a~particular Julian day. 
The accuracy of the photometry obtained in this way is excellent, reaching
a~few mmag.  The resulting mean points of the average light curves, together
with the formal standard deviations for particular observations, are shown
in Table\,\ref{Mean.E.9.10.dat}.

      \subsection{Spectroscopic data collected in 2003 and 2008/9}

For several decades until the eclipse at epoch $E$\,=\,9 in 2003, changes in
EE\,Cep's spectrum outside and during eclipses had been poorly studied,
whereas the photometric behaviour during the eclipses had been relatively
well characterized.  The situation improved significantly after the
observational campaign in 2003.  Seven observatories located in Europe and
North America took part in the observations (using the instruments listed in
Table\,\ref{tab.spec.sum}), collecting spectra at low and high resolution. 
The spectral observations covered various phases of the eclipse, revealing
changes in the line profiles (mainly H$\alpha$, \ion{Na}{i}, H$\beta$, and
\ion{Fe}{ii}) not only during the photometric eclipse but even more than two
months before and after the minimum \citep{Mik2005b}.  Unfortunately, during
the last campaign at the turn of 2008 (epoch $E$\,=\,10) only a~small number
of spectra were obtained, with only three instruments.  The new spectra
complement those obtained during the previous epoch, because a~significant
number of these spectra were acquired during orbital phases that had not
been previously covered.

With this paper, we make available a~large number (100) of spectra, most of
which were, however, obtained with moderate resolution and/or cover a~narrow
spectral range, containing mainly H$\alpha$ or \ion{Na}{i} spectral lines. 
In addition, they were clustered near the eclipse -- the spectroscopic
observations have insufficient temporal coverage throughout the orbital
phase to use them in studying changes in the radial velocities.  The list of
spectra and instruments together with some additional information are given
in Table\,\ref{Spect.list.8910.dat}.  All these spectra were heliocentric
corrected and some of them (obtained at Rozhen Observatory and DDO), which
cover a~narrow spectral range ($\sim100$-$200$\,\AA), were normalized to the
continuum.  The low resolution spectra obtained at Piwnice Observatory were
flux calibrated.  All spectra are available as FITS files at the
CDS.\footnote{Centre de Donn\'ees astronomiques de Strasbourg}

\begin{figure*}
  \centering
    \includegraphics[width=17cm]{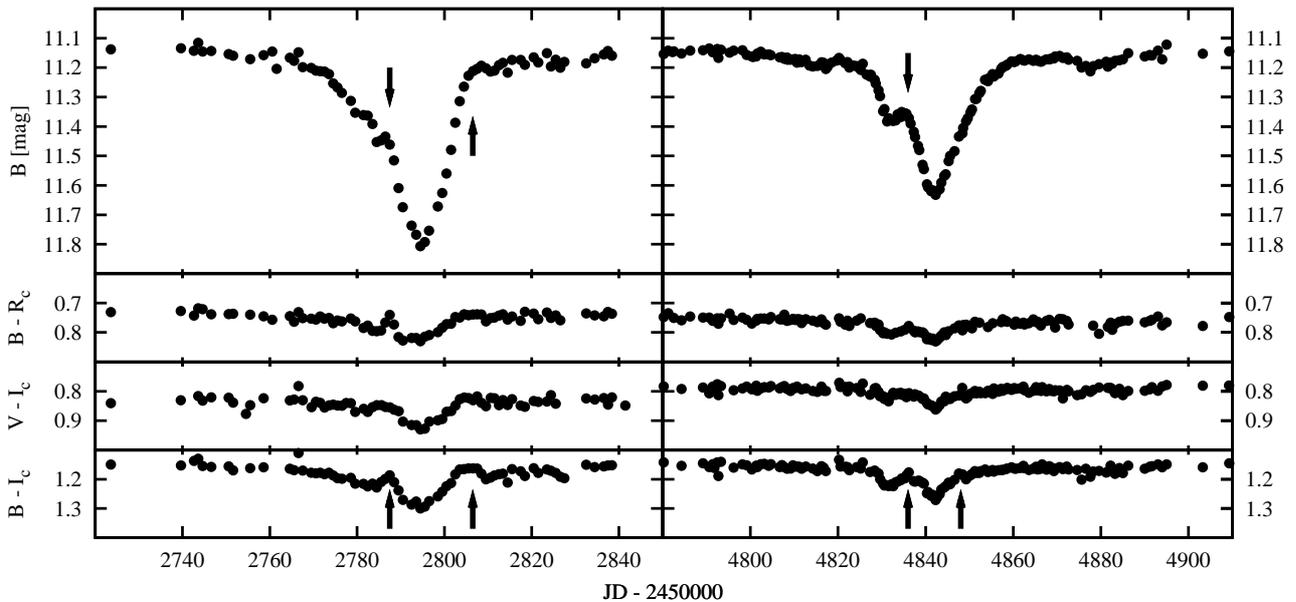}
\caption{Average points from Table\,\ref{Mean.E.9.10.dat} of the 2003
eclipse ({\it left}) and the 2008/9 eclipse ({\it right}).  The $B$ light
curves ({\it top}) and three colour indices ({\it bottom}) are presented. 
Arrows denote times of blue maxima.}
  \label{fig.Bcol2003i9}
\end{figure*}

%
\section{Results}

	\subsection{Light and colour changes during the last two eclipses}

\begin{figure}
  \resizebox{\hsize}{!}{\includegraphics{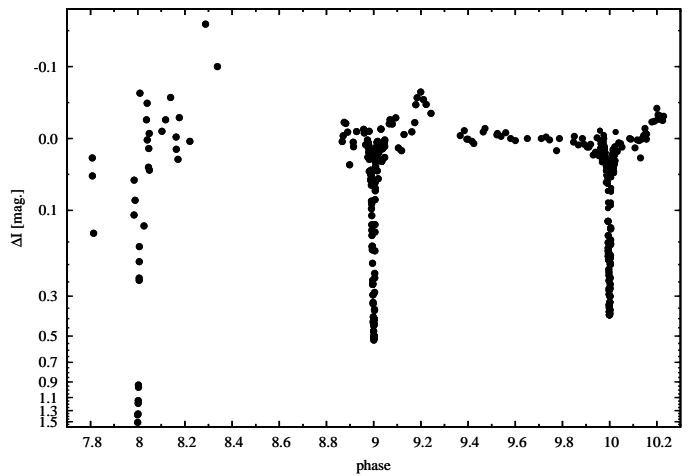}}
     \caption{Differential $I$ magnitudes of EE\,Cep obtained at Piwnice
Observatory during the 13 years from 1996 to 2010.  The zero value
corresponds to the level of the average brightness outside (phases 0.4--0.8)
eclipses.  Below +0\fm1 (i.e.  for magnitude changes smaller than +0\fm1) an
artificial, strongly non-linear scale is used to reduce the contrast in the
amplitude of the changes during and outside eclipses (thus, the relatively
small variations outside the eclipses can be seen and compared with the
depth of the eclipses).}
  \label{fig.I8_9_10}
\end{figure}

Thanks to the aforementioned observational campaigns, it has been possible
for the first time to analyse colour evolution during the eclipses.  In
Fig.\,\ref{fig.Bcol2003i9}, the mean $B$ light curves and the colour indices
for both of the last eclipses are presented.  The 2003 and 2008/9 eclipses
reached their minima on Julian days $JD = 2452795$ and $JD = 2454842$,
respectively.  The small timing residuals ($O - C$, observations minus
calculations) $+1\fd44$ and $-1\fd5$ with respect to the ephemeris
[Eq.\,(\ref{eq.efem})] did not change this significantly.  The
\citet[][]{MiGr1999} ephemeris was used (unchanged) for orbital phasing to
produce all the observational data in this paper.  The colour indices for
the 2003 eclipse show two blue maxima, about nine days before and after the
mid-eclipse.  Two weak maxima in the $B$ light curve are also clearly
visible.  Similar features also occur in the 2008/9 eclipse but the ``bump''
(at JD\,2454836) preceding the minimum
(Fig.\,\ref{fig.LC.2003i9}\,and\,\ref{fig.Bcol2003i9}) is much more
pronounced than previously.  The differences in the phase and strength of
these features can be caused, such as the depth of eclipses, by changes in
the spatial orientation of the disk.\\ 
\indent The observed variations in the $I$ passband after the eclipses could
give additional support to this idea.  In Fig.\,\ref{fig.I8_9_10}, the
$I$-band light curve obtained over 13 years, from 1996 to 2010, is shown. 
About one year after each eclipse, near the orbital phase $\sim 0.2$, an
increase in $I$-band brightness appears.  The recurrence and rapid variation
during these events allow us to speculate that this increase may be caused
by proximity effects when the components are close to periastron.  If this
is true, then the orbit in the EE\,Cep system must be significantly
eccentric.  An interesting correlation -- the brightening events appear to
be stronger when the eclipses are deeper -- may indicate that there are
changes in the disk projection and this may be an additional observational
argument for precession of the disk.  The quite large amplitude of
variability outside the eclipse in the $I$ passband (which is not clearly
visible at shorter wavelengths) also indicates that the contribution of
a~dark component (disk or/and central object) to this band has to be
significant.  The cool component becomes readily noticeable at the red edge
of the visible spectrum, and in the near infrared (the $JHK$ bands) it might
dominate the observed fluxes.

	\subsection{Variations in the spectrum}

\begin{figure}
  \resizebox{\hsize}{!}{\includegraphics{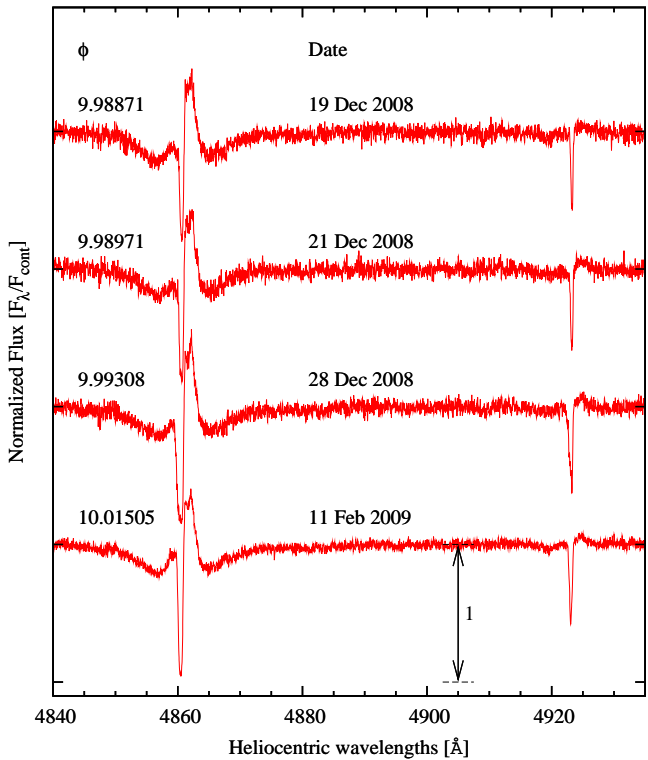}}
     \caption{H$\beta$ and \ion{Fe}{ii} line profiles in the spectra
obtained during the 2008/9 eclipse with the Nordic Optical Telescope (NOT,
La\,Palma).  The spectra are vertically offset for clarity.}
  \label{fig.hb.FeII}
\end{figure}

\begin{figure}
  \resizebox{\hsize}{!}{\includegraphics{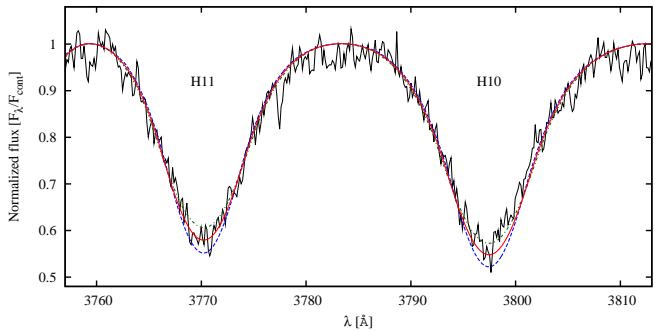}}
     \caption{The Balmer H10 and H11 line profiles in the spectrum of
EE\,Cep taken on 11\,Aug\,2003 with an Echelle spectrograph on the 2.12\,m
telescope in SPM Observatory in Mexico.  The best fit was obtained for
a~synthetic spectrum with: $T_{\mathrm {eff}} = 15000$K, $\log{g} = 3.5$,
$[Fe/H] = 0$, and $v \sin{i} = 350$\,km\,s$^{-1}$ (solid line).  Two poorer
fits calculated with different values of the rotational velocity, $v \sin{i}
= 300$\,km\,s$^{-1}$ (dashed line) and $v \sin{i} = 400$\,km\,s$^{-1}$
(dash-dotted line), are shown for comparison purposes.}
  \label{fig.H10H11}
\end{figure}

The most important results of the spectroscopic observations obtained during
the 2003 campaign seem to be the conclusions regarding the nature of the hot
component.  The emission and absorption components of the Balmer and
\ion{Fe}{ii} line profiles in the spectra obtained around the 2003 eclipse
imply that the hot component is a~rapidly rotating Be star surrounded by
a~highly inclined emitting gaseous ring \citep{Mik2005b}.  These line
profiles show the same pattern during the 2008/9 eclipse \citep[compare
Fig.\,\ref{fig.hb.FeII} with Fig.\,2 of][]{Mik2005b}.  A~comparison of the
Balmer H8-H11 absorption lines in the spectrum of EE\,Cep with theoretical
profiles (Fig.\,\ref{fig.H10H11}) gives $v \sin{i} \approx
350$\,km\,s$^{-1}$ \citep[][]{Gal2008}, which implies that there has been
a\,strong equatorial gravitational darkening.  The rotational velocity of
the Be star in the EE\,Cep system is very close to the critical value.  It
must lead to a~continuously strong radial outflow of the gas stream from the
equator, which is confirmed by the existence of the gaseous ring --
a~characteristic feature of Be-type stars \citep{Mik2005b}.

\begin{figure}
  \resizebox{\hsize}{!}{\includegraphics{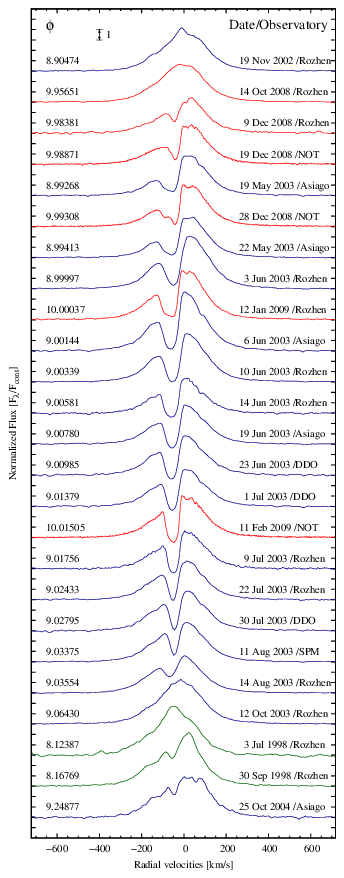}}
     \caption{Representative examples of H$\alpha$ line profiles in the
spectra obtained near the eclipses at epochs $E$\,=\,10, $E$\,=\,9, and
$E$\,=\,8 (in the electronic version of this paper, the profiles have
different colours: red, blue, and green, respectively).  The spectra are
vertically offset for clarity.}
  \label{fig.ha}
\end{figure}

\begin{figure}
  \resizebox{\hsize}{!}{\includegraphics{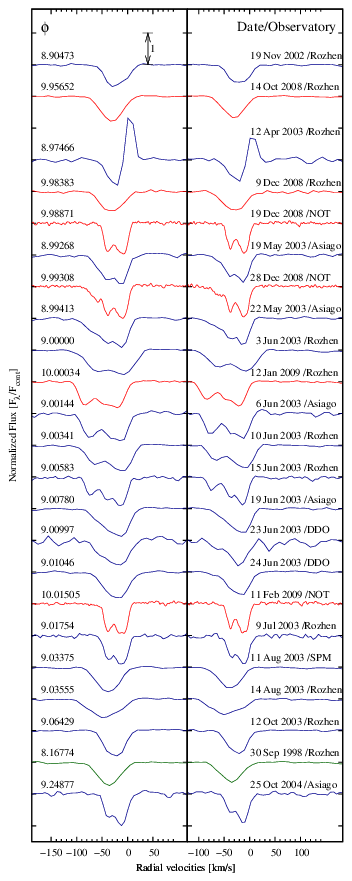}}
     \caption{Representative examples of \ion{Na}{i} doublet line profiles
in the spectra obtained near the eclipses at epochs $E$\,=\,10, $E$\,=\,9,
and $E$\,=\,8 (in the electronic version of this paper, the profiles have
different colours: red, blue, and green, respectively).  The spectra are
vertically offset for clarity.}
  \label{fig.na}
\end{figure}

\begin{figure}
  \resizebox{\hsize}{!}{\includegraphics{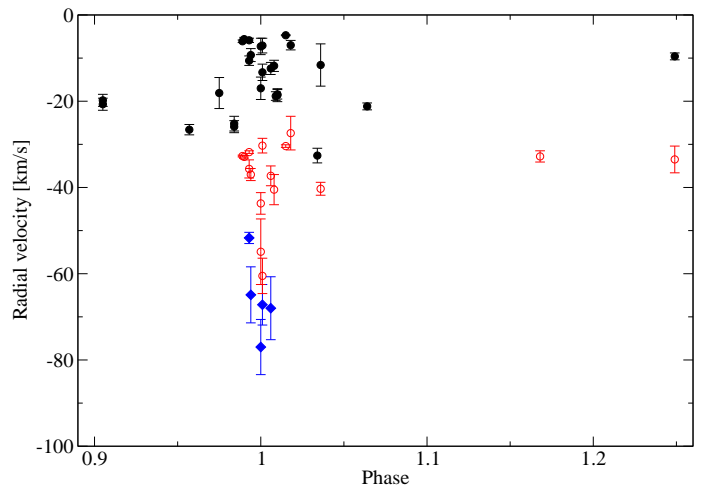}}
     \caption{Radial velocities of the three components of the sodium
doublet \ion{Na}{i} lines obtained from the spectra of five observatories
(NOT, SPM, DDO, Asiago, and Rozhen).  The main stellar component is shown
with filled circles, and two components from the disk blue-shifted by about
-40 and -70\,km\,s$^{-1}$ are shown with open circles and filled diamonds,
respectively.}
  \label{fig.na.rv}
\end{figure}

Figures \ref{fig.ha}--\ref{fig.PhotSpec} show the evolution of the H$\alpha$
and \ion{Na}{i} line profiles in which additional absorption components
appeared during both of the last two eclipses.  Towards the mid-eclipse, an
absorption component in the H$\alpha$ profile grows, and during the minimum
it is very deep and broad.  The sodium doublet line profiles in the minimum
show a~multi-component structure and we can discern at least two additional
absorption components shifted towards blue wavelengths, first at a~velocity
of about -40\,km\,s$^{-1}$ and then at about -70\,km\,s$^{-1}$
(Fig.\,\ref{fig.na.rv}).

In a~few high-resolution spectra from the Rozhen, NOT, Asiago, and Terskol
observatories, we can see H$\alpha$ and H$\beta$ lines.  The spectra from
NOT and Terskol contain higher order Balmer lines, which are, however,
underexposed, permitting us to see only that the emission is weakening and
the broad and strong absorption features of the Be star begin to dominate. 
The only spectrum displaying absorption from H$\alpha$ to H13-H14 is the SPM
spectrum (see Table\,\ref{tab.spec.sum}) obtained during the 2003 eclipse. 
In this spectrum, strong absorption in the Be star dominates and the higher
order emission Balmer lines are absent (see Fig.\,\ref{fig.H10H11}).  In the
case of other lines of the Be star, in a~few spectra, the \ion{He}{i}
5876\,\AA, 4471\,\AA, and \ion{Mg}{ii} 4481\,\AA\ lines appear to be present
but are barely visible.  In the spectra from CCS and Asiago, the emission
triplet \ion{Ca}{ii} (8498\,\AA\, 8542\,\AA\, 8662\,\AA\ ) and the
\ion{O}{i} 8446\,\AA\ line are visible.  Because of the small number of
lines in the Be star spectrum and their weakness and large width, it was
impossible to extract reliable information about changes in the radial
velocities of the hot component.

The spectra obtained during the two most recent eclipses suggest that the
behaviour of the spectral line profiles might not change between eclipses. 
A~unique spectrum was obtained at phase $\sim -0.025$ before the 2003
eclipse when both lines of the \ion{Na}{i} doublet showed a\,P\,Cyg profile. 
If this is a~sign of outflow from the Be star, then this implies that the
eclipses occur relatively close to periastron.  At orbital phases far from
the eclipses, absorption structures are indeed sometimes appear imposed on
the emission lines indicating that there are large amounts of loose gaseous
clouds in the system, which could support such a~scheme (see e.g. 
Fig.\,\ref{fig.ha} -- H$\alpha$ line profiles at phases $\sim 0.17$ and
$\sim 0.25$).  On the other hand, these structures are observed during the
brightening event observed in the $I$ band at an orbital phase $\sim 0.2$.

The changes of the spectral line profiles during phases close to the
photometric eclipses allowed us to estimate the sizes of the eclipsing dusty
disk and gaseous ring around the Be star \citep[][]{Mik2005b}.  The
disappearance of the bluest component of the \ion{Na}{i} doublet at a\,phase
of about 0.011 suggests that the radius of the eclipsing cloud producing the
\ion{Na}{i} lines is at least $6R_{\mathrm {Be}}$.  The shell absorption in
H$\alpha$ rapidly decays about 2.5 months after minimum (at phase $\sim
0.036$), which suggests that the gaseous ring around the Be star producing
the H$\alpha$ emission is almost twice the size of the eclipsing cloud, i.e. 
$\ga 10R_{\mathrm {Be}}$.

%
\section{Modelling of the eclipse light curves, precession of the disk,
and discussion}

	\subsection{Numerical code and basic assumptions} \label{basass}

Although similar to $\varepsilon$\,Aur and maybe even M2-29, EE Cep is
nevertheless quite a~unique system, and existing tools do not seem to be
suitable for analysing this system.  To model the brightness and colour
variations during the eclipses and changes from epoch to epoch, possibly
depending on precession, it was necessary to develop our own, original
numerical code.

\begin{figure*}
\centering
  \includegraphics[width=17cm]{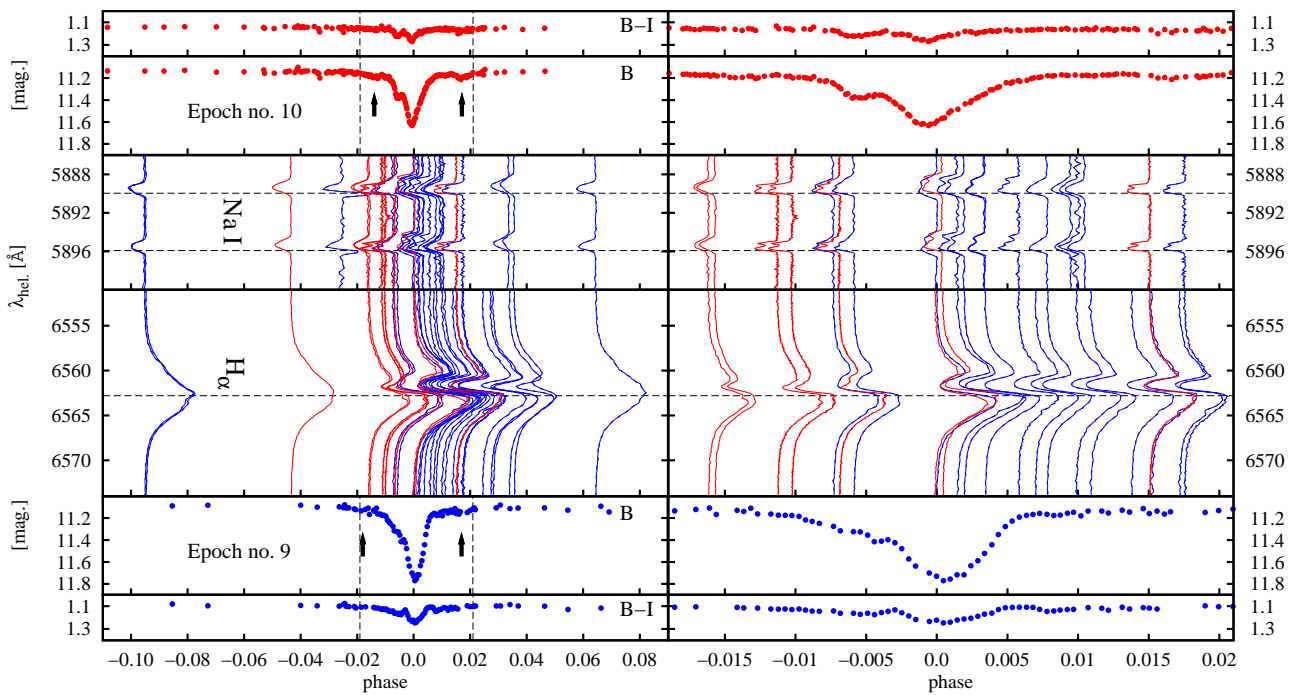}
     \caption{The $B$ and $B-V$ light curves of the 2003 ({\it bottom}) and
2008/9 ({\it top}) eclipses.  In the middle panel, we present all available
line profiles of the \ion{Na}{i} doublet and H$\alpha$ for epoch\,9
(superposition of two lines: thick dashed with thin solid) and\,10 (solid
line).  The positions of the continuum levels refer to the orbital phases. 
The right-hand panels show a~zoom of the central part of left-hand panels as
indicated by vertical dashed lines.  The arrows indicate the shallow minima
in the external parts of the eclipses observed about 35 days before and
after mid-eclipse at both last epochs (both last eclipses seem to be longer
than expected and lasted about 90 days).}
     \label{fig.PhotSpec}
\end{figure*}

The models require the adoption of some quite simplistic assumptions.  An
axially symmetric, circular, flat disk with an $r^{-n}$ density profile was
assumed in all cases.  The disk was considered to be geometrically thin,
although we highlight that two different approaches are possible with our
code.  One is to assume a~disk thickness $H$ and integrate the density of
the matter in the disk.  The second approach, which is more efficient for
the calculation, is to assume that the disk has a\,negligible thickness (in
reality zero thickness in the model).  Changes in the optical depth
depending on the disk inclination could also be taken into account ($\tau
\sim |\cos{i_{\mathrm d}}^{-1}|$).  The outer disk radius was assumed to be
six times the equatorial radius of the Be star ($R_{\mathrm {d0}} = 63.4
R_{\sun}$), i.e.  approximately the minimal possible radius that can be
estimated from the contact times in the eclipse light curves
\citep{Mik2005a}.  A~possible additional contribution of radiation from the
eclipsing body (disk and/or its central object) to the total flux (commonly
called ``the third light'') and scattering of Be star radiation off the disk
particles have been neglected.  We assumed that the matter of the disk
absorbs radiation selectively in accordance with interstellar extinction. 
The passband-dependent absorption coefficients were estimated based on the
total value of the reddening, which increased by $\Delta E_{\mathrm {B-V}}
\approx 0.05$ during mid-eclipse in 2003.

The Be star parameters that could not be reliably calculated during the
modelling process had to be entered as inputs into the model.  Our spectra
show that the hot component has an effective temperature $T_{\mathrm{eff}} =
15000$\,K and $\log{g} = 3.5$, implying that it is a~B5III or B4II type
star.  With the spectral type and luminosity class, the stellar effective
temperature and luminosity $L = 3500 L_{\sun}$ can be determined using the
statistical relations of \citet{deJN1987}.  Comparing the resulting values
with the theoretical evolutionary models for stars with moderate and high
masses \citep{Cla2004}, via interpolation, the mass range $M_{Be} = 8.0 \pm
2.2 M_{\sun}$ was estimated for the mass of the Be star in EE\,Cep.  A~mean
stellar radius was estimated with the Stefan-Boltzmann law.  For
a~description of the star's surface, its shape and radiation, we used the
model described by \citet{CrOw1995} and \citet{OwCr1994} in our program. 
The model takes into account both the oblateness of the star and gravity
darkening using a~Roche model and a~von~Zeipel ($F\sim g \rightarrow
T_{\mathrm {eff}}\sim g^{0.25}$) law.  By comparing of the critical
rotational velocities that characterize each pair of mass and radius with
the observed rotational velocity, which for the adopted inclination $i =
90\degr$ is $v = 350 \pm 50$ km s$^{-1}$, we constrained the possible masses
to the range 5.9$M_{\sun}$ $\la M_{Be} \la$ 7.9 $M_{\sun}$, i.e.  the range
in which stars do not disintegrate as a~result of rapid rotation.  We
eventually decided to fix the basic parameters of the Be star in our model
to a~mass $M = 6.7 M_{\sun}$, mean radius $\bar{R}_{Be} \approx 9.0
R_{\sun}$ (with equatorial to polar radius ratio $R_{eq}R_p^{-1} \approx
1.44$, giving equatorial and polar radii, respectively, $R_{eq} \approx
10.57 R_{\sun}$ and $R_p \approx 7.34 R_{\sun}$), luminosity $L = 3500
L_{\sun}$, and rotational velocity at the equator $V_{\mathrm {eq}} =
325$\,km\,s$^{-1}$.  To perform a~$\chi^{2}$ minimization, our numerical
code was equipped with the simplex algorithm.  This procedure used the
method described in \citet{Bra1998} and the flowchart of \citet{KaMi1999}.

\begin{figure}
  \resizebox{\hsize}{!}{\includegraphics{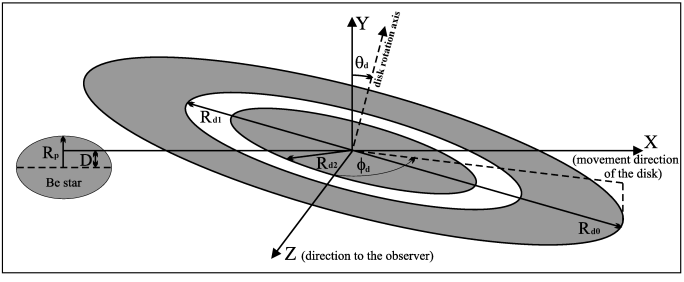}}
     \caption{Schematical explanation of the geometrical parameters in the
special case when the precession axis of the disk (the symmetry axis of the
conical surface over which the rotation axis of the disk moves cyclically
with the precession period $P_{\mathrm{prec}}$) is parallel to the $Y$ axis
of the coordinate system (i.e.  it is perpendicular to the orbital plane). 
Our code allows this axis to be inclined by angles $\theta_{\mathrm {prec}}$
and $\phi_{\mathrm {prec}}$ in a~similar way to the inclination of the
rotation axis.}
  \label{fig.SchemPar}
\end{figure}

The solutions were carried out using the $UBV(RI)_{\mathrm C}$ light curves. 
In general, several parameters were treated as free parameters: the impact
parameter $D$, the mid-eclipse moment $T_0$ (in the sense of the minimum
proximity of the star and the disk centres in the projection on the sky
plane), the relative tangential velocity of the star and the disk
$V_{\mathrm t}$, the inclination of the disk ($90\degr - \theta_{\mathrm
d}$), an angle related to the disk precession phase $\phi_{\mathrm d}$, the
absorption coefficient $\kappa_{\mathrm s}$ representing the contribution by
the grey extinction and the central disk density $\rho_{\mathrm c}$
expressed in arbitrary units.  The geometrical parameters ($D$,
$\theta_{\mathrm d}$, $\phi_{\mathrm d}$, $R_{\mathrm {d0}}$, $R_{\mathrm
{d1}}$, and $R_{\mathrm {d2}}$) describing the disk sizes and orientation in
the models are presented in Fig.\,\ref{fig.SchemPar}.  The code allows
additional disk radii $R_{\mathrm {di}}$ for $i \le 5$ to be defined, making
it possible to take into account the presence of one or two gaps in the disk
and the central opening.

	\subsection{A~solid or a~gapped disk model?}

	We used our code to model the last two eclipses with a~solid disk
causing the eclipses.  This model is consistent with the global changes in
the light curves and colours (it fits the depth and the shape of the
eclipses), especially for the 2003 eclipse (see Appendix \ref{AppendixM},
Fig.\,\ref{fig.Model2003i9.LC}).  However, this model has trouble in
explaining the two blue maxima in the colour evolution that appeared during
the last two eclipses, roughly symmetrically a~few days before and after the
photometric minimum.

	We tried to explain these blue maxima based on a~hot star being
rotationally darkened at the equator and brightened at the poles, and
assuming that the eclipsing disk is divided into two parts by a~gap.  For
a~hot Be star such as EE\,Cep, convection is impossible in the envelope and
we can expect the darkening effect to occur in pure \citet{Zei1924} form. 
Since the star rotates with a~velocity very close to the critical value, the
gravity darkening effect can result in a difference between the polar and
the equatorial temperatures of as much as 5--6 thousand Kelvins.  Thus, the
appearance of the hot polar area in the gap could be observed as the blue
maxima.  \citet{Gal2008} considered and briefly described a~model with
a~disk that has a~concentric gap for the 2003 eclipse.  In the current work,
an attempt has been made to create a~similar model for the last two eclipses
together, by taking into account the precession period $P_{\mathrm {prec}}$
as an additional free parameter.  Although this model seemed to be
appropriate when the 2003 eclipse alone is considered, it does not work well
when we consider the two eclipses together (see the online Appendix
\ref{AppendixM}: Fig.\,\ref{fig.Model.9and10}).  The model indeed generates
colour changes and blue maxima of the same order as observed during the 2003
eclipse.  However, its predictions are inconsistent with the observations,
because the shallower eclipses should be accompanied by a~less pronounced
``bump'' and associated blue maximum, in contradiction to the case of the
2003 and 2008/9 eclipses taken together.  The ``bump'' at JD\,2454836 in the
last eclipse ($E$\,=\,10) seems to be too strong to be explained entirely by
a~gap in the disk.

\begin{figure}
  \resizebox{\hsize}{!}{\includegraphics{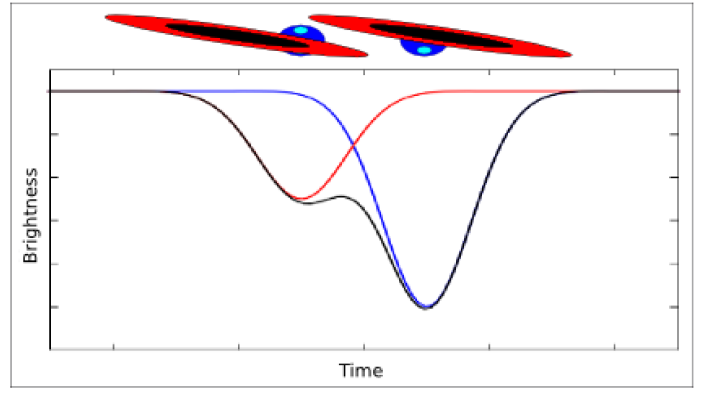}}
     \caption{Schematic explanation for the formation of the bump in the
light curves.  The configurations of the star and the disk at the first and
second minima are shown at the top.}
  \label{2min.expl_m}
\end{figure}

\begin{figure*}
\centering
  \includegraphics[width=17cm]{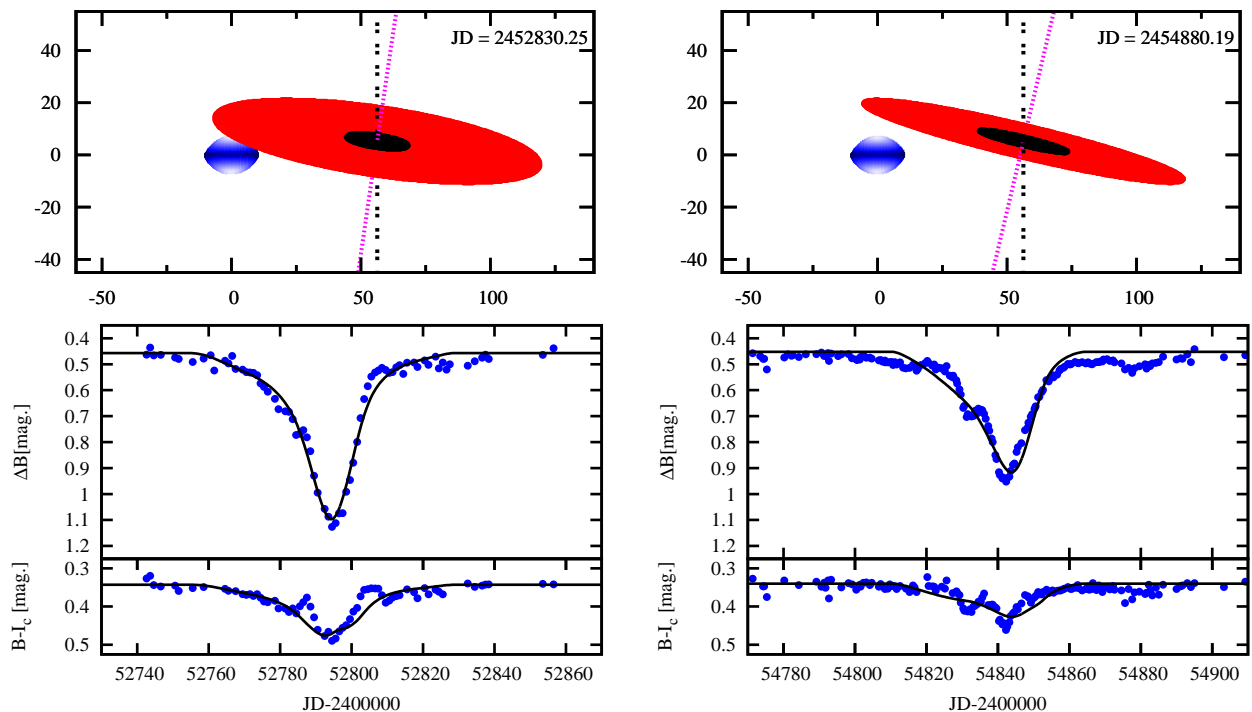}
     \caption{Modelling of the eclipses of a~rapidly rotating Be star by
a~solid disk, considering both the 2003 ({\it left}) and 2008/9 ({\it
right}) eclipses.  The precession period is a~free parameter.  The top
panels show projections of the system onto the plane of the sky.  The
polar\,(hot) and equatorial\,(cool) areas of the star are shown by changing
shades.  The inner (opaque) and outer (semi-transparent) areas of the disk
are shown by dark and light shades, respectively.  The sizes are expressed
in solar radii.  The lower panels show the $B$ light curve ({\it middle})
and the $B-I_{\mathrm C}$ colour index ({\it bottom}) together with
the\,synthetic fits (lines).  The Julian day in the upper right corner
represents the moment at which (according to the model) the spatial
configuration of the system is the same as shown in the relevant panel.}
  \label{fig.Model.prec.2003i9.LC}
\end{figure*}

After many attempts to model the light changes during the eclipses,
especially the shallow ones, we realized that another mechanism, connected
with the flattening of the Be star, could be helpful in explaining the
``bump''.  For example, we consider the case when the disk has an
inclination close to 90 degrees (when the eclipse is shallow) and the tilt
with respect to the direction of motion is high (we expect that this could
occur for the eclipses at epochs $E$\,=\,9, $E$\,=\,10).  The temporal
superposition of the two minima should be observed, corresponding to the
successive obscuration of the two hot poles separately, at significantly
different times (see Fig.~\ref{2min.expl_m}).  The first minimum is
shallower and the second deeper, because of the non-zero value of the impact
parameter, which reduces the obscuration of one of the poles, and because
the outer part of the disk is more transparent than the inner part.  The
condition for the occurrence of the ``bump'' is a~sufficiently long temporal
separation of the two minima.  Models generated using our numerical code at
various precession phases suggest that the ``bump'' would disappear in the
deep eclipses and be stronger in the shallow ones, matching what we observe. 
Although this scenario is very promising, it may not match the observed
colour changes.  The main problem is the second blue maximum, which appears
in both of the last eclipses (during 2003 and 2008/9), although it is
already very weak during the last one.  We speculate that the reason for the
occurrence of the second blue maximum may be (i) a~concentric gap (or
a~local, concentric depression of the density) or (ii) a~central opening in
the disk structure.  The former would be adequate for eclipse models similar
to those presented here, i.e.  in which the impact parameter is small enough
for both sides of the entire disk to be involved in the eclipses.  The
latter would require the impact parameter to be larger, so that the eclipses
could be caused by about half of the disk.  Which of these two cases is
true?  We are unable to establish this based on the photometric
observations.  The behaviour of the \ion{Na}{i} sodium doublet lines
suggests that case (ii) may be correct.  In this case, the disk would be
about twice as large as in case (i) and it would move on an orbit with an
appropriately higher speed.

	\subsection{Precession solutions from the 2003 and 2008/9 eclipses} \label{pr9i10}

As stated above, we propose that the precession of the disk is responsible
for the observed differences between the eclipses from epoch to epoch. 
Since the number of eclipses observed is small, the constraints on
precession are weak, so other processes, perhaps connected with changes in
the disk size and/or its internal structure, might be needed to explain the
rapid changes.  Nevertheless, we decided to study our hypothesis of
precession as the most likely hypothesis that can presently be constrained.

In one approach, we used the highest quality photometric data obtained
during the last two eclipses, when the disk was nearly edge-on, so that its
optical thickness was high, but it projected to a~small solid angle at that
time.  Using our code, we determined the best-fit solution for the solid
disk for the two last eclipses, taking precession into account.  The basic
assumptions about the nature of the Be star, and the both fixed and free
parameters were the same as those in Section~\ref{basass}.  The disk was
treated as having negligible thickness (infinitesimally small with all its
mass concentrated in the plane) with an $r^{-2}$ density profile.  The
precession period of the disk $P_{\mathrm{prec}}$ was adopted as an
additional free parameter.  For simplicity, the precession axis was assumed
to be perpendicular to the orbital plane.  The resulting model is presented
in Fig.\,\ref{fig.Model.prec.2003i9.LC} and its parameters are shown in
Table\,\ref{tab.mod.9and10.solid}.  The best-fit solution was obtained for
the precession period $P_{\mathrm {prec}} \approx 61.94$~yr (about
11\,$P_{\mathrm {orb}}$) for which the angle related to the disk precession
phase $\phi_{\mathrm d}$ changes from $34.88${\degr} at epoch $E$\,=\,9 to
$67.5${\degr} at epoch $E$\,=\,10.  Such a~fast precession seems to be
necessary to explain the observed rapid changes in the eclipse depths at
consecutive epochs.  For example, the very shallow eclipse at $E$\,=\,3
occurred very close in time to two very deep eclipses at $E$\,=\,2 and
$E$\,=\,4, and the deep minimum at $E$\,=\,8 was followed by two shallow
ones at $E$\,=\,9 and 10.  An alternative solution exists according to which
the disk would achieve, at epoch $E$\,=\,10, the precession phase
$\phi_{\mathrm d} = 112.5${\degr} with about half the precession period of
the former solution, $P_{\mathrm {prec}} \approx 26.03$~yr
($\sim\,5\,P_{\mathrm {orb}}$).  Since we are only able to observe
a~projection of the disk, we are unable to distinguish between these two
cases using only the photometric data of these two eclipses.

\begin{table}
\caption{Parameters obtained from the solution of the solid disk model when
applied to the last two eclipses together, taking into account disk
precession.}
\label{tab.mod.9and10.solid}
\centering			
\begin{tabular}{lrll}
\hline\hline
Parameter                  & Value    & $\pm   $ & Unit          \\
\hline
$D$                        & 4.91     & 0.15     & $R_{\sun}$    \\
$T_{0 \mathrm{(E=9)}}$     & 52795.25 & 0.27     & day           \\
$T_{0 \mathrm{(E=10)}}$    & 54845.19 & $-_{''}-$ & day           \\
$V_{\mathrm t}$            & 1.78     & 0.12     & $R_{\sun}$/day\\
$\theta_{\mathrm d}$       & 14.39    & 2.59     & degree        \\
$\phi_{\mathrm {d (E=9)}}$ & 34.88    & 2.63     & degree        \\
$\phi_{\mathrm {d (E=10)}}$& 67.50    & $-_{''}-$ & degree        \\
$P_{\mathrm {prec}}$       & 61.94    & 1.66     & yr            \\
$\kappa_{\mathrm s}$       & 0.175    & 0.019    & 1             \\
$\rho_{\mathrm c}$         & 92.8     & 11.2     & 1\,$R_{\sun}^{-3}$\\
\hline
\end{tabular}
\end{table}

	\subsection{Precession solutions using all the eclipses} \label{prall}

However, although the precession period in EE\,Cep should indeed be rather
short, its lower limit of about five orbital periods inferred from colour
variations during two successive eclipses, as presented in
Section~\ref{pr9i10}, seems unrealistic.  One credible argument for a~longer
period of precession is the following, which favours a~$P_{\mathrm {prec}}$
of the order of $14 P_{\mathrm {orb}}$ ($\approx 78.5$ yr).  If the shallow
minimum with a~flat bottom observed in 1969 was indeed caused by an edge-on,
non-tilted disk \citep{MiGr1999}, then the precession axis is not
perpendicular to the orbital plane.  For perpendicular orientations, two
edge-on positions with opposite tilt angles should be observed.  More
generally, two edge-on positions can occur when the precession axis lies
nearly in the sky plane and is inclined with respect to the orbital plane. 
One of these positions may be non-tilted with respect to the orbital motion,
but both should produce similar (shallow) eclipse depths, despite the very
different tilt angles, because an edge-on disk obscures at most only a~small
part of the Be star.  This situation might have arisen in 1969 ($E$\,=\,3)
and 2008/9 ($E$\,=\,10), if the time interval between these minima was about
half a~precession period.  This hypothesis provides a~solution what may
satisfy the data of all the eclipses.  Thus, we propose this as a~possible
way of explaining the seemingly chaotic changes that occur in successive
eclipses.  The next key step in understanding this system was to realize
that the deepest eclipses are not necessarily those that occur when the
projected disk size is greatest (as we assumed at first); the deepest
eclipses must instead be those where the column density in the disk and the
projected disk size are together high enough to obscure most of the Be star
surface.  This situation must occur close in time to the most shallow
eclipses in order to be consistent with the rapid changes in the eclipse
depths.  According to this line of reasoning, during the shallowest
eclipses, the projection of the disk becomes small so that there is very
little obscuration of the stellar flux, even though the column density in
the disk is greatest at that time.  Similarly, the eclipses should have
intermediate depths at those precession phases at which the projected disk
size is largest, since, despite the eclipsing of nearly the whole surface of
the star, this eclipsing is performed by the highly transparent part of the
disk.

\begin{figure}
  \resizebox{\hsize}{!}{\includegraphics{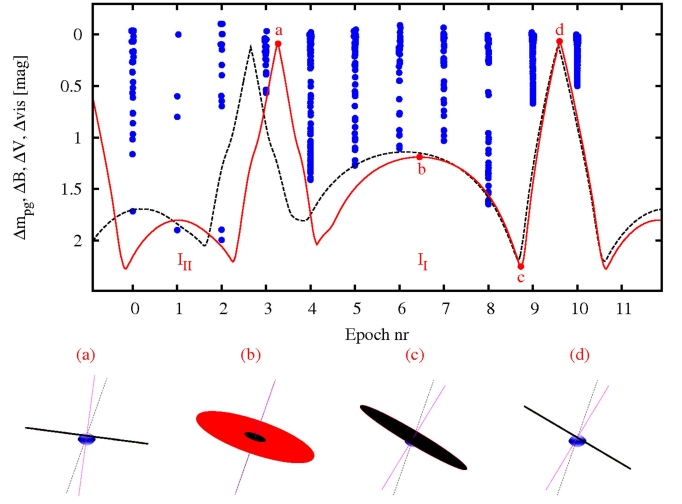}}
     \caption{Dependence of the depths of eclipses in EE\,Cep system on
precession.  Photometry obtained during epochs 0--10 is shown as circular
symbols.  The solid and dashed lines delineate two models of the changes in
eclipse depth as a~function of orbital phase, generated using our numerical
code.  In these two cases, the precession periods $P_{\mathrm {prec}}$ are
10.8\,$P_{\mathrm {orb}}$ and 11.8\,$P_{\mathrm {orb}}$, respectively.  At
the bottom, the spatial configurations of the disk and the star in four
special cases, denoted by the letters a, b, c, and d, are shown.}
  \label{Prec11Porb}
\end{figure}

We tested this hypothesis using our numerical code for the case with
a~precession axis inclined relative to the normal of the orbital plane by
several different small values of $\theta_{\mathrm{prec}}$.  We found that
as a~first approximation, $\phi_{\mathrm{prec}}$ should be around
70\degr--80{\degr} (see Fig.~\ref{fig.SchemPar} for the definition of
$\theta_{\mathrm{prec}}$ and $\phi_{\mathrm{prec}}$).  When
$\theta_{\mathrm{prec}}$ is non-zero (i.e.  the precession axis is
inclined), the value of $\phi_{\mathrm{prec}}$ determines how unequal the
time intervals between the two shallowest minima in the precession cycle
will be and how different the eclipse depths will be between these two
intervals.  We define $I_{\mathrm I}$ as the time interval from the eclipse
with the minimal tilt of the projected disk (as in 1969) to the eclipse with
a~maximal disk tilt (as in 2008/9) and vice versa, $I_{\mathrm{II}}$ as the
time interval from the eclipse with maximal disk tilt to the eclipse with
minimal disk tilt.  When $\phi_{\mathrm{prec}}$ = 90\degr, the time
intervals $I_{\mathrm I}$ and $I_{\mathrm{II}}$ are equal and the changes in
eclipse depth as a~function of precession phase proceed symmetrically with
respect to the times of the shallowest eclipses.  In general, however,
$\phi_{\mathrm{prec}}$ differs from 90\degr, in which case asymmetry
appears.  For example, when $\phi_{\mathrm{prec}} <$ 90\degr, the time
interval $I_{\mathrm{II}}$ is shorter than $I_{\mathrm I}$ and the eclipses
during $I_{\mathrm{II}}$ are deeper, especially during the central parts of
this interval.  This seems to be the case for EE\,Cep.  We know that
$I_{\mathrm I} \sim 7 P_{\mathrm {orb}}$ and that $I_{\mathrm{II}}$ seems to
be shorter.  Hence, if this scenario were correct, we would have succeeded
in constraining the full precession period $P_{\mathrm {prec}} \la 14
P_{\mathrm {orb}}$.

\begin{table}
\caption{Subjective (``eye'') and quantitative (``fit'') optimal solutions
of our precession model using all the eclipses.}
\label{tabprec}
\centering			
\begin{tabular}{lrrll}
\hline\hline
Parameter                 & eye      & fit      & $\pm$   & Unit          \\
\hline
$R_{\mathrm{d0}}$         & 75.0     & 73.8     & 2.7     & $R_{\sun}$    \\
$H$                       & 1.0      & 0.6      & ...     & $R_{\sun}$    \\
$D$                       & 4.0      & 4.0      & ...     & $R_{\sun}$    \\
$T_0$                     & 40493.92 & 40493.92 & ...     & day           \\
$V_{\mathrm t}$           & 2.0      & 2.0      & ...     & $R_{\sun}$/day\\
$\theta_{\mathrm d}$      & 12.5     & 12.1     & 0.7     & degree        \\
$\phi_{\mathrm d}^{\star}$&175.0     &173.7     & 1.9     & degree        \\
$\theta_{\mathrm{prec}}$  & 18.0     & 19.5     & 1.0     & degree        \\
$\phi_{\mathrm{prec}}$    & 79.0     & 80.0     & 2.0     & degree        \\
$P_{\mathrm{prec}}$       & 11.8     & 10.8     & 0.4     & $P_{\mathrm{orb}}$\\
$\kappa_{\mathrm s}$      & 0.328    & 0.328    & ...     & 1             \\
$\rho_{\mathrm c}$        & 1.18     & 1.13     & 1.21    & 1\,$R_{\sun}^{-3}$\\
n$^{\star\star}$          & 0.3      & 0.27     & 0.09    & 1             \\
\hline
\end{tabular}
\begin{list}{}{}
\item[{\bf{Notes.}} ($^{\star}$)] At the time of minimal disk tilt.
\item[($^{\star\star}$)] Exponent in the function describing the disk density
distribution.
\end{list}
\end{table}

We considered many combinations of sets of parameters, for different disk
sizes, starting from the large ($R_{\mathrm{d0}} \sim 200$ R\sun) and
geometrically thick ($H \sim R_{\mathrm p}$) disks.  Because the adopted
artificial density distribution does not provide a~good solutions for the
external parts of the light curves during the eclipses, we concentrated on
their central parts, i.e.  we searched for the set of parameters that could
explain the dependence of eclipse depth on precession phase.  By visual
inspection of plots comparing the synthetic curves that represent the
dependence of the eclipse depth on precession with the observational data,
we chose subjectively a~few optimal sets of parameters.  One of these fits,
perhaps the best of them, is shown in Table\,\ref{tabprec} (left) and
Fig.\,\ref{Prec11Porb} (dashed line).  Using the simplex algorithm, we
performed a~$\chi^2$ minimization over the parameter space of the system
input parameters.  The best-fit solution is that shown in
Table\,\ref{tabprec} (right) and the synthetic fit is presented as the solid
line in Fig.\,\ref{Prec11Porb}.  Our solutions were obtained for a~disk
radius $R_{\mathrm{d0}} \approx 75 R_{\sun}$ and a~geometrically very thin
disk $H = 0.6 R_{\sun}$.  We cannot exclude, however, the possibility that a
smaller or larger disk could provide more reliable results.  On the other
hand, the disk thickness in the optimal solution equals the spatial
resolution adopted in the model (which represents the grid size) and in
reality, the disk could be even thinner.  The disk must be extremely thin in
order for this model to work correctly at times close to the shallow minima. 
Some of the eclipses in the EE\,Cep system could be even shallower than
those in 2008/9 (the disk could sometimes almost completely disappear, as
happens with Saturn's rings when they have an edge-on orientation).

Thus, our model (Table~\ref{tabprec}) suggests that the value of the
precession period should be about 11--12\,$P_{\mathrm{orb}}$.  Is it
possible that the precession period is really so short?  The development of
a~mechanical model of a~precessing circumstellar disk would be a~useful
follow-up study to this paper to test our preferred solution theoretically. 
Empirically, at least one binary system that appears to have a~similarly
short precession period exists.  This system, \object{SS433}, is very
different from EE\,Cep -- it is far more compact and the disk has to be
smaller and perhaps more massive.  \citet{Mar1980} suggested that the SS433
system, with $P_{\mathrm {orb}} \approx 13$\fd1, has an accretion disk that
can precess with a~period of 164\fd, which is just 12.5 times its orbital
period.

Applying our model of precession to predict the depth of the next EE\,Cep
minimum, we find that it should be similar to the deepest of the previous
eclipses, reaching about 2\fm.  According to the ephemeris
(Eq.\,\ref{eq.efem}), this minimum should occur on 23\,Aug~2014.  To some
degree, this tests our proposed model, although this model has a~serious
problem.  The existing set of observations span a~time interval that is
almost identical to the expected precession period.  A~statistically more
significant test would require much more than one full precession cycle, and
preferably at least two cycles.  Several more decades of observations would
be required.  Nevertheless, there is some hope that old photographic surveys
might contain a~sufficient amount of extra data.  For example, the database
of the DASCH\footnote{http://hea-www.harvard.edu/DASCH/} (Digital Access to
a~Sky Century\,@\,Harvard) project \citep[see][]{Gri2009} provides data from
more than 5000 old Harvard photographic plates obtained between JD\,2411556
and JD\,2447823 (nearly exactly 100 years) that contain EE\,Cep in their
field of view.  By checking the Julian dates when these photographs were
obtained using EE\,Cep's ephemeris, it seems likely that we may be able to
extract data for the eclipses from epochs $E = -9$ to $E = -1$, with which
we can test our model.

	\subsection{On the similarity of EE\,Cep and $\varepsilon$\,Aur}

Our observations show that the disk in the EE\,Cep system may be similar to
(though smaller than) the multi-ring structure observed in
$\varepsilon$\,Aurigae.  \citet{LeSt2010} found that the equivalent width of
the \ion{K}{i} potassium line at 7699\,{\AA} increased step-wise during the
ingress of the last $\varepsilon$\,Aur eclipse.  They interpreted this
pattern as a~manifestation of the complex structure of the disk as an
alternating series of concentric rings and gaps, which had already been
suggested based on the observations of the previous eclipse during the
1980's \citep{Fer1990}.  According to what until recently has been the
dominant interpretation, based on $\varepsilon$\,Aur observations during the
eclipse of the 1980's, a~quite close binary system should exist at the disk
centre \citep{Liss1984}.  On the basis of their spectral energy distribution
(SED) analysis, \citet{Hoa2010} suggested that the $\varepsilon$\,Aur system
is composed of a~massive B5V type primary embedded in a~dusty disk with
a~radius of about $3.8 AU$ and an F\,type post-AGB secondary of about half
the mass of the primary.  In this model, the disk has to be a~byproduct of
mass transfer from the initially more massive star (currently an F\,type
post-AGB) to what was initially the secondary (and is now a~more massive B5V
star).  \citet{Klo2010}, via interferometric observations during the ingress
of the 2009 eclipse, detected and measured movement of the disk with respect
to the F star.  They confirmed the existence of an optically thick, inclined
disk in the system and provided the first direct evidence that the disk is
geometrically thin.  The mass of the disk is dynamically negligible ($\la$
15 M$_{\oplus}$), but is sufficient to cause eclipses.  \citet{Hoa2010}
pointed out that the dust content of the disk must be largely confined to
grains larger than $\sim$10\,$\mu$m to explain the grey nature of eclipses,
from the optical range up to the infrared, and the lack of broad dust
emission features in the mid-infrared spectra.  Owing to the important role
that grey extinction plays in our models, we can conclude that the disk in
the EE\,Cep system should also be dominated by particles of quite large
diameters -- a~mixture of grains and dust, and our results concerning our
precession model also suggest that this disk should be geometrically very
thin.

In both cases, $\varepsilon$\,Aur and EE\,Cep, the disks are inclined to the
orbital plane.  The presence of a~binary system at the disk centre might
help to explain the inclination of the disk and the rapid precession. 
However, the statistical likelihood of a~binary system at the centre of the
disk in the case of $\varepsilon$\,Aur is low, and for EE\,Cep even lower. 
Another possible way of explaining the inclination of the disk is that since
the main component in EE\,Cep is a~rapidly rotating Be star, it is very
probable that it is the donor star that supports a disk around its
companion.  Therefore, to explain the inclination of this disk relative to
the orbit, we have to assume that the orbital plane is not coplanar with the
equatorial plane of the Be star.  In this case, the disk around the
companion will also not be coplanar with the orbit.  In principle, to
introduce disk precession into the system in a~way in which the precession
axis is inclined to the orbital plane, it is sufficient to add a~third body
as the perturber if it satisfies two conditions: (i) its orbit should not be
coplanar with the orbital plane of the disk, and (ii) it should have a~high
enough mass (and/or the disk should have a~low enough mass).  In the light
of these two conditions, we have many possibilities as to what could
constitute a~third body.  It could be an object orbiting the Be star, either
closer to or further away from than the eclipsing object, but it might also
be an object on an orbit around the body at the disk centre, either outside
or within the disk.  The latter case would be equivalent to the presence of
a~binary system at the disk centre.  At the moment, we probably do not have
enough observations of the system to decide which of these possibilities is
most likely to be correct.

In addition to the differences in the sizes of the systems, there are
additional indications that the geometries of the eclipses could be very
different.  While the consecutive eclipses in $\varepsilon$\,Aur are nearly
identical in terms of their depth \citep[see eg.][]{Ste2009}, the eclipse
depths in EE\,Cep are highly variable.  Some variations in the durations of
the entire eclipse and the various stages of the eclipse (ingress, totality,
and egress) in $\varepsilon$\,Aur \citep[see][]{Hop2008} are observed. 
Perhaps these changes could be explained in terms of disk precession.  In
this case, the large differences between the eclipse depths of EE\,Cep and
$\varepsilon$\,Aur would be explained by variations in the direction of the
rotation axis due to the strong precession in the case of EE\,Cep and
a~small $\theta_{\mathrm {prec}}$ (of Fig.\ref{fig.SchemPar} caption) in
$\varepsilon$\,Aur.  There are also interesting differences between the
\ion{Na}{i} doublet line profiles observed during the eclipses of these
systems.  During the last eclipse of $\varepsilon$\,Aur, an additional
component appeared, which was redshifted during the first part of the
eclipse and blueshifted during the second part \citep[see][]{Tom2012}, in
contrast to EE\,Cep, where only blueshifted additional components were
observed.  This may indicate that in the case of EECep, the impact parameter
may be so large that roughly only half of the disk is involved in the
eclipses.

%
\section{Conclusions}

We have presented our observational data obtained during the last three
eclipses of EE\,Cep.  We release these data for use by the astronomical
community.  For the two latest minima, our investigations were carried out
as international campaigns that provided data of unprecedented quality for
this object, especially in the case of the photometry, where an accuracy of
a~few thousandths of a~magnitude was achieved.  These minima turned out to
be the shallowest EE\,Cep eclipses observed.  The grey character of these
eclipses, i.e.  the weak dependence of the eclipses' depth on the
photometric band, reinforces our belief that the eclipsing object is indeed
a~dark, dusty debris-disk around a~low-luminosity central body that is
visible in neither the spectra nor photometry.

The results of these campaigns shed new light on our understanding of the
EE\,Cep system.  Our spectroscopic data have demonstrated that the main
component of the system is a~rapidly rotating ($v \sin{i} \approx
350$\,km\,s$^{-1}$) Be-type star.  The oblateness of the star leads, via the
von Zeipel effect, to a~highly inhomogeneous temperature distribution across
its surface.  The spectra obtained during the last two eclipses suggest that
the absorption lines change in the same way during each eclipse.  During the
minima of both eclipses, we were able to detect at least three absorption
components in the \ion{Na}{i} lines and the same strong absorption
superimposed on the H$\alpha$ emission.

By analysing all the photometric and spectroscopic data together, we have
proposed several hypotheses that provide predictions for future eclipses.

Using high quality photometry, it was possible to detect two blue maxima in
the colour indices during the 2003 and 2008/9 eclipses, that occurred from
about six to nine days before and after the photometric minimum.  The first
(stronger) blue maximum occurred simultaneously with a~``bump'' in the light
curves, which is very clear in all the $UBV$($RI$)$_{\mathrm C}$ photometric
bands.  This ``bump'' seems to be caused by a~temporal offset between the
two minima in a~single eclipse, which can be explained by the
non-simultaneous obscuration of the hot polar regions of the Be star by the
elliptical, tilted shape of the projected disk.

The durations of the last two eclipses were longer than expected (both
lasting about 90 days).  In the external parts of these long minima, two
shallow minima were observed about 35 days before and after mid-eclipse
during both epochs (arrows\,in\,Fig.\,\ref{fig.PhotSpec}).  This could be
explained by the presence of a~gap near the outer border of the disk.  The
second blue maximum, which could not be explained by the mechanism proposed
for the ``bump'', may indicate the existence of either an inner gap or
a~central opening.  Thus, the disk in the EE\,Cep system could have
a~complex, possibly multi-ring structure.  The behaviour of the \ion{Na}{i}
line profiles gives some support to this idea.  Another hypothesis that
follows from the behaviour of these lines and the recurrent asymmetry of the
eclipses is that maybe only half the disk is involved in the eclipses.

Considering all the eclipses together, from the 1950's to the present, we
estimated the duration of the disk precession period to be about 62--67
years ($\sim$\,11--12\,$P_{\mathrm {orb}}$).  Using our new model of
precession, we predict that the depth of the forthcoming eclipse in 2014
should be one of its deepest, reaching about 2\fm.

More spectroscopic observations during the next eclipse would be needed to
more clearly understand the nature of the EE\,Cep system.  Photometry in the
infrared {\it JHK} bands during and after the eclipse would be very useful. 
This could make it possible to detect the secondary companion of EE\,Cep
(disk and/or central star/stars), as it seems to make a~significant
contribution to the total flux at the red edge of the visible spectrum
(a~brightening event by about 0\fm05 at the phase $\sim$\,0.2 was observed). 
The radial velocity variations of the hot component, which would be a~real
challenge to obtain, may be of crucial importance in constraining the
parameters of this system.


\begin{acknowledgements}
E. Semkov would like to thank the Director of Skinakas Observatory, Prof. 
I.  Papamastorakis, and Dr.  I.  Papadakis for granting telescope time.  We
thank T.  Karmo, Stefan Mochnacki, and G.  Conidis for contributing their
data.  Some of the observations used here were taken courtesy of the AAVSO
and the Sonoita Research Observatory.  This study was supported by MNiSW
grant No.  N203 018 32/2338.
\end{acknowledgements}


\bibliographystyle{aa} 

%
\Online

%
\begin{appendix}

\section{The models of eclipses with a~solid or a~gapped disk} \label{AppendixM}

\begin{figure*}
\centering
  \includegraphics[width=17cm]{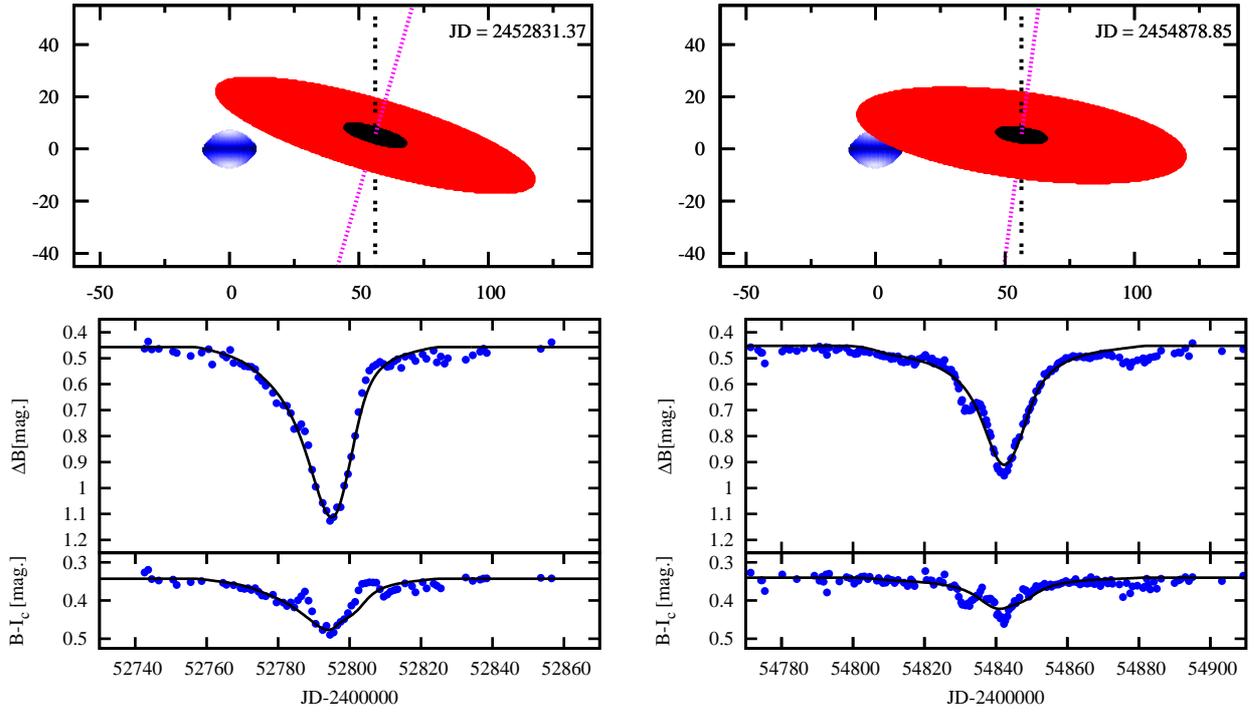}
     \caption{Modelling of the eclipse of a~rapidy rotating Be star as
a~solid disk during the 2003 eclipse ({\it left}) and the 2008/9 eclipse
({\it right}).  The top panels show the system projected onto the plane of
the sky.  The polar\,(hot) and equatorial\,(cool) areas of the star are
shown by different shades.  The inner (opaque) and outer (semi-transparent)
areas of the disk are shown by dark and light shades, respectively.  The
sizes are expressed in solar radii.  The lower panels show the $B$ light
curve ({\it middle}) and the $B-I_{\mathrm C}$ colour index ({\it bottom})
together with the~synthetic fits (lines).  The Julian day in the upper right
corner represents a~moment at which (according to the model) the spatial
configuration of the system is the same as that shown in the relevant
panel.}
  \label{fig.Model2003i9.LC}
\end{figure*}

Using our numerical code, we fitted models of the last two eclipses
separately, using a~solid disk as the eclipsing object.  We made the same
assumptions about the nature of the Be star, and we used the same fixed and
free parameters as in Section~\ref{basass}.  The disk was treated as having
negligible thickness (infinitesimally small with all the mass concentrated
in the plane) and an $r^{-2}$ density profile.  For simplicity, the
precession axis was assumed to be perpendicular to the orbital plane.  The
resulting solution is shown in Table\,\ref{tab.2mod} together with the error
estimates.  In Fig.\,\ref{fig.Model2003i9.LC}, we present models of two
eclipses using a~solid disk, for the 2003 eclipse on the left and for the
2008/9 eclipse on the right.  The models containing a~solid disk provide
quite a~good fit to the light curve and the global colour changes,
reproducing both the depth and the shape of the eclipses, especially for the
2003 eclipse.  This model, however, cannot explain the two maxima in the
colour evolution during the eclipses.

\begin{figure*}
\centering
  \includegraphics[width=17cm]{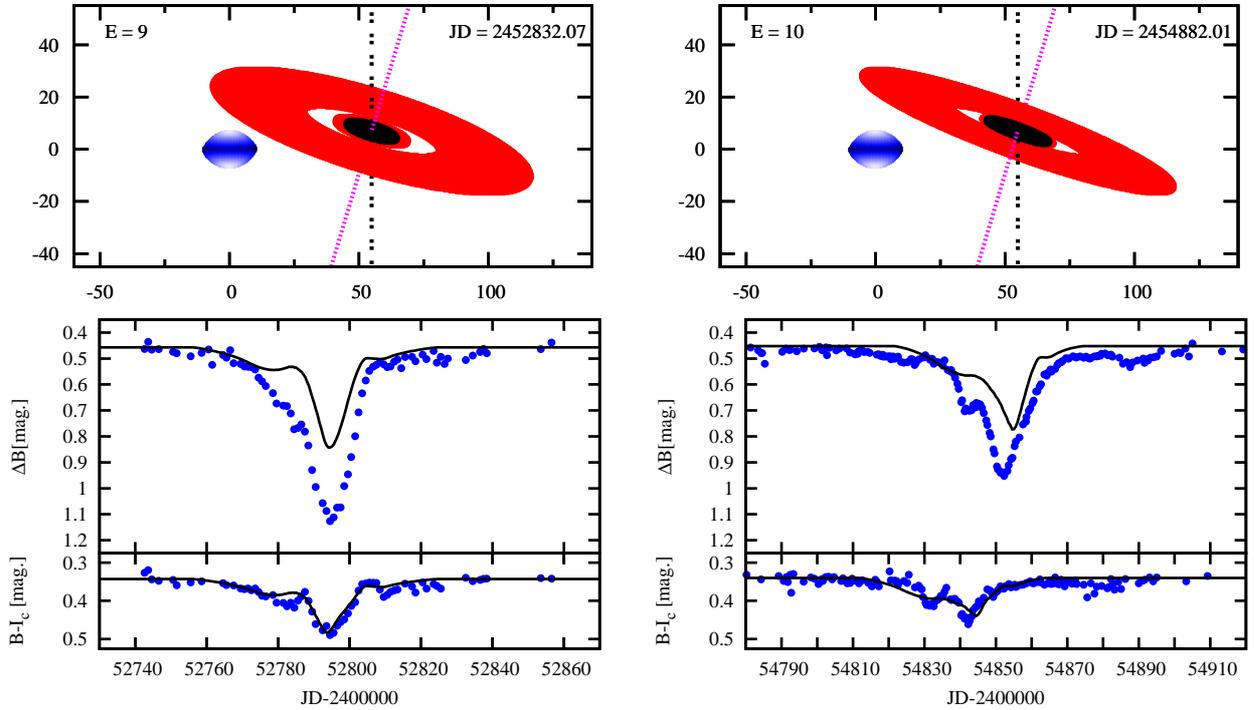}
     \caption{Modelling of the last two eclipses together with the
precession period taken as an additional free parameter, when a~gapped disk
is considered.  The sky plane projections of the system during the last two
eclipses (at $E$\,=\,9 and $E$\,=\,10) ({\it top}), the $B$ light curve
({\it middle}), and the $B-I_{\mathrm C}$ colour index ({\it bottom}),
together with the synthetic fits (lines) are shown.  The symbols and the
shades of colour have the same meaning as those in
Fig.\,\ref{fig.Model2003i9.LC}.}
  \label{fig.Model.9and10}
\end{figure*}

In the present study, we adopted a~model containing a~disk that has
a~concentric gap for the two last eclipses, taking into account the
precession of the disk.  We assumed the same disk diameter, disk density
distribution, and orthogonality of the precession axis to the orbital plane,
and the same Be star parameters as in the case of the solid disk.  This
model was based solely on the $B-I_{\mathrm C}$, $V-I_{\mathrm C}$, and
$V-R_{\mathrm C}$ colour indices.  We chose the same free parameters as in
the case of the solid disk model for the 2008/9 eclipse (the tangential
velocity was fixed at $V_{\mathrm t}=1.57 R{_{\sun}} \mathrm{day}^{-1}$) but
added three more free parameters: the precession period $P_{\mathrm {prec}}$
and two parameters specifying the outer $R_{\mathrm {d1}}$ and inner
$R_{\mathrm {d2}}$ radii of the gap.  The resulting model is presented in
Fig.\,\ref{fig.Model.9and10} and Table\,\ref{tab.mod.9and10.gapped}.  The
best results were obtained for the precession period $P_{\mathrm {prec}}
\approx 31.91$yr (about 5--6\,$P_{\mathrm {orb}}$), for which the angle
related to the disk precession phase $\phi_{\mathrm d}$ changes from
$50.00${\degr} at epoch $E$\,=\,9 to $113.32${\degr} at epoch $E$\,=\,10. 
An alternative solution was found to exist in which the precession phase
$\phi_{\mathrm d} = 66.68${\degr} at epoch $E$\,=\,10 has a~precession
period $P_{\mathrm {prec}} \approx 121.13$~yr, which is almost four times
longer (being about 22\,$P_{\mathrm {orb}}$).  In the light of the results
of Sections \ref{pr9i10} and \ref{prall}, both these periods of precession
seem to be unrealistic.  Comparison of this model with the \citet{Gal2008}
model for the 2003 eclipse alone reveals a~problem.  The gapped disk model
seemed to be very promising for explaining the colour changes that occurred
during the 2003 eclipse, but fails in the case of the 2008/9 eclipse, since
it cannot explain either the colour changes during an eclipse or the strong
``bump'' in the light curve.

\begin{table}[h!]
\caption{The parameters of the solutions obtained for the solid disk model,
derived independently for the 2003 eclipse (left) and the 2008/9 eclipse
(right).}
\label{tab.2mod}
\centering			
\begin{tabular}{lrlrll}
\hline\hline
Parameter            & 2003     & $\pm$    & 2008/9         & $\pm$    & Unit          \\
                     & eclipse  &          & eclipse        &          &               \\
\hline
$D$                  & 4.74     & 0.24     & 6.32           & 0.59     & $R_{\sun}$    \\
$T_0$                & 52795.98 & 0.29     & 54843.85       & 0.50     & day           \\
$V_{\mathrm t}$      & 1.57     & 0.06     & 1.57$^{\star}$ & ...      & $R_{\sun}$/day\\
$\theta_{\mathrm d}$ & 20.05    & 0.93     & 16.36          & 2.05     & degree        \\
$\phi_{\mathrm d}$   & 52.85    & 1.28     & 27.32          & 2.00     & degree        \\
$\kappa_{\mathrm s}$ & 0.171    & 0.022    & 0.346          & 0.034    & 1             \\
$\rho_{\mathrm c}$   & 94.8     & 8.6      & 45.5           & 3.9      & 1\,$R_{\sun}^{-3}$\\
\hline
\end{tabular}
\begin{list}{}{}
\item[$^{\star}$] {For the 2008/9 eclipse model the tangential velocity
$V_{\mathrm t}$ was adopted to be identical to that obtained for the 2003
eclipse.}
\end{list}
\end{table}

\begin{table}[h!]
\caption{Parameters of the solution of the gapped disk model when
applied to the last two eclipses together, taking into account disk
precession.}
\label{tab.mod.9and10.gapped}
\centering			
\begin{tabular}{lrll}
\hline\hline
Parameter            & Value    & $\pm   $     & Unit          \\
\hline
$R_{\mathrm {d1}}$   & 27.61    & 0.85         & $R_{\sun}$    \\
$R_{\mathrm {d2}}$   & 14.19    & 0.46         & $R_{\sun}$    \\
$D$                  & 6.97     & 0.37         & $R_{\sun}$    \\
$T_{0 \mathrm{(E=9)}}$  & 52797.07 & 0.48      & day           \\
$T_{0 \mathrm{(E=10)}}$ & 54847.01 & $-_{''}-$ & day           \\
$V_{\mathrm t}$      & 1.57$^{\star}$ & ...    & $R_{\sun}$/day\\
$\theta_{\mathrm d}$ & 21.48    & 0.47         & degree        \\
$\phi_{\mathrm {d (E=9)}}$ &  50.00 & 1.92     & degree        \\
$\phi_{\mathrm {d (E=10)}}$& 113.32 & $-_{''}-$& degree        \\
$P_{\mathrm {prec}}$ & 31.91    & 1.14         & yr            \\
$\kappa_{\mathrm s}$ & 0.056    & 0.014        & 1             \\
$\rho_{\mathrm c}$   & 113.6    & 5.9          & 1\,$R_{\sun}^{-3}$\\
\hline
\end{tabular}
\end{table}

\end{appendix}

%
\newpage
\begin{appendix} 

\section{Online photometric and spectroscopic data}\label{AppendixT}

\begin{table*} 
   \caption{Photometry in standard $UBVR_{\rm C}$ filters and a~non-standard
$i$ (${\bar{\lambda}}_{\rm i} \approx 7420$~\AA) filter obtained at Piwnice
Observatory$^\star$ (Poland) during and near the 1997 eclipse ($E$~=~8). 
A~one-channel photometer with an uncooled EMI~9558B photomultiplier on the
0.6~m Cassegrain telescope was used.  The differential magnitudes are given
with respect to our standard star BD~+55$\degr$2690, together with the
corresponding standard deviations.}
   \label{Phot.8.dat}
\centering 
\begin{tabular}{lrrrrrrrrrr}
\hline\hline
$JD$ & $\Delta U$ & $\Delta B$ & $\Delta V$ & $\Delta R_{\rm C}$ & $\Delta i$ & $\sigma_{\rm U}$ & $\sigma_{\rm B}$ & $\sigma_{\rm V}$ & $\sigma_{{\rm R}_{C}}$ & $\sigma_{\rm i}$ \\
\hline
2450348.616 & 0.050 & 0.450 & 0.415 & 0.310 & 0.243 & 0.025 & 0.025 & 0.015 & 0.009 & 0.054 \\
2450349.618 & 0.040 & 0.481 & 0.414 & 0.320 & 0.268 & 0.026 & 0.024 & 0.014 & 0.009 & 0.043 \\
2450360.540 & 0.045 & 0.426 & 0.392 & 0.298 & 0.350 & 0.032 & 0.022 & 0.011 & 0.016 & 0.055 \\
2450474.366 & 0.176 & 0.515 & 0.407 & 0.225 & 0.132 & 0.081 & 0.055 & 0.031 & 0.026 & 0.041 \\
2450512.336 &-0.034 & 0.239 & 0.205 & 0.196 & 0.132 & 0.139 & 0.104 & 0.075 & 0.077 & 0.119 \\
2450711.327 & 0.122 & 0.506 & 0.441 & 0.331 & 0.322 & 0.033 & 0.017 & 0.010 & 0.008 & 0.056 \\
2450712.606 & 0.177 & 0.461 & 0.437 & 0.345 & 0.274 & 0.028 & 0.016 & 0.007 & 0.008 & 0.022 \\
2450720.565 & 0.174 & 0.585 & 0.527 & 0.441 & 0.302 & 0.040 & 0.030 & 0.030 & 0.030 & 0.040 \\
2450742.436 & 1.682 & 2.089 & 1.948 & 1.791 & 1.589 & 0.035 & 0.025 & 0.018 & 0.017 & 0.040 \\
2450743.298 & 1.785 & 2.107 & 2.013 & 1.765 & 1.727 & 0.035 & 0.024 & 0.015 & 0.009 & 0.035 \\
2450746.393 & 1.675 & 1.988 & 1.874 & 1.651 & 1.572 & 0.035 & 0.022 & 0.012 & 0.007 & 0.020 \\
2450747.421 & 1.563 & 1.807 & 1.700 & 1.557 & 1.401 & 0.035 & 0.018 & 0.008 & 0.011 & 0.041 \\
2450748.364 & 1.396 & 1.730 & 1.650 & 1.463 & 1.359 & 0.044 & 0.020 & 0.010 & 0.012 & 0.039 \\
2450749.339 & 1.205 & 1.613 & 1.524 & 1.346 & 1.153 & 0.045 & 0.028 & 0.016 & 0.009 & 0.025 \\
2450750.329 & 1.201 & 1.476 & 1.406 & 1.271 & 1.180 & 0.031 & 0.024 & 0.020 & 0.017 & 0.042 \\
2450756.203 & 0.436 & 0.695 & 0.674 & 0.559 & 0.461 & 0.023 & 0.023 & 0.016 & 0.017 & 0.027 \\
2450756.511 & 0.299 & 0.684 & 0.624 & 0.522 & 0.454 & 0.027 & 0.013 & 0.010 & 0.008 & 0.022 \\
2450757.202 & 0.398 & 0.620 & 0.588 & 0.509 & 0.409 & 0.021 & 0.015 & 0.012 & 0.010 & 0.031 \\
2450757.519 & 0.269 & 0.631 & 0.591 & 0.485 & 0.375 & 0.022 & 0.010 & 0.007 & 0.008 & 0.015 \\
2450763.209 & 0.375 & 0.624 & 0.452 & 0.313 & 0.153 & 0.050 & 0.040 & 0.030 & 0.020 & 0.050 \\
2450798.300 & 0.142 & 0.556 & 0.498 & 0.361 & 0.338 & 0.028 & 0.028 & 0.016 & 0.012 & 0.023 \\
2450819.339 & 0.109 & 0.447 & 0.396 & 0.288 & 0.190 & 0.028 & 0.019 & 0.009 & 0.010 & 0.020 \\
2450825.516 & 0.063 & 0.462 & 0.396 & 0.307 & 0.167 & 0.033 & 0.013 & 0.010 & 0.009 & 0.026 \\
2450826.201 & 0.063 & 0.467 & 0.408 & 0.312 & 0.218 & 0.035 & 0.015 & 0.012 & 0.010 & 0.021 \\
2450837.216 & 0.109 & 0.465 & 0.410 & 0.290 & 0.256 & 0.024 & 0.015 & 0.012 & 0.015 & 0.030 \\
2450839.386 & 0.086 & 0.428 & 0.390 & 0.276 & 0.230 & 0.045 & 0.030 & 0.009 & 0.008 & 0.004 \\
2450843.266 & 0.079 & 0.469 & 0.415 & 0.310 & 0.209 & 0.014 & 0.015 & 0.008 & 0.008 & 0.022 \\
2450845.397 & 0.142 & 0.434 & 0.401 & 0.314 & 0.260 & 0.042 & 0.019 & 0.008 & 0.009 & 0.024 \\
2450954.455 & 0.085 & 0.435 & 0.398 & 0.309 & 0.206 & 0.042 & 0.018 & 0.011 & 0.005 & 0.020 \\
2450985.472 & 0.009 & 0.395 & 0.375 & 0.242 & 0.190 & 0.026 & 0.012 & 0.017 & 0.027 & 0.018 \\
2451029.502 & 0.195 & 0.452 & 0.347 & 0.306 & 0.159 & 0.040 & 0.025 & 0.016 & 0.016 & 0.067 \\
2451077.352 & 0.171 & 0.495 & 0.399 & 0.330 & 0.214 & 0.047 & 0.011 & 0.015 & 0.014 & 0.027 \\
2451078.470 & 0.075 & 0.465 & 0.402 & 0.292 & 0.231 & 0.025 & 0.012 & 0.015 & 0.010 & 0.029 \\
2451093.412 & 0.085 & 0.450 & 0.414 & 0.340 & 0.245 & 0.022 & 0.026 & 0.014 & 0.016 & 0.022 \\
2451105.418 & 0.101 & 0.447 & 0.409 & 0.293 & 0.187 & 0.010 & 0.009 & 0.010 & 0.005 & 0.011 \\
2451196.285 & 0.109 & 0.435 & 0.398 & 0.283 & 0.220 & 0.019 & 0.013 & 0.010 & 0.006 & 0.019 \\
2451332.418 & 0.090 & 0.476 & 0.389 & 0.290 & 0.057 & 0.040 & 0.019 & 0.009 & 0.011 & 0.032 \\
2451435.577 & 0.118 & 0.464 & 0.424 & 0.310 & 0.116 & 0.020 & 0.011 & 0.008 & 0.006 & 0.029 \\
\hline
\end{tabular}
\begin{list}{}{}
\item[$^\star$ Observers:] D. Graczyk, J. Janowski, M. Miko{\l}ajewski.
\end{list}
\end{table*}

\begin{table*} 
   \caption{Photometry obtained at Athens Observatory$^\star$ (Greece) with
standard $BV(RI)_{\rm C}$ (Bessell) filters during and near the 2003 eclipse
($E$~=~9).  The 0.4~m Cassegrain telescope with an SBIG~ST8~CCD camera was
used.  Differential magnitudes are given with respect to BD~+55$\degr$2690. 
Each point is the mean value obtained from several to tens of frames.  The
columns labelled $HJD+$ denote the fraction of the day.}
   \label{Phot.Athens.9.dat}
\centering 
\begin{tabular}{lrrrrrrrr}
\hline\hline
$HJD$ & $HJD+$ & $\Delta B$ & $HJD+$ & $\Delta V$ & $HJD+$ & $\Delta R_{\rm C}$ & $HJD+$ & $\Delta I_{\rm C}$\\
\hline
2452739 & .62750 & 0.488 & .62337 & 0.437 & .62475 & 0.316 & .62541 & 0.144\\
2452764 & .64212 & 0.522 & .64044 & 0.464 & .64775 & 0.336 & .64316 & 0.166\\
2452765 & .53587 & 0.529 & .54065 & 0.472 & .54330 & 0.342 & .54433 & 0.172\\
2452769 & .49517 & 0.555 & .49336 & 0.490 & .48959 & 0.357 & .49209 & 0.186\\
2452770 & .56873 & 0.564 & .57578 & 0.491 & .57405 & 0.363 & .57508 & 0.193\\
2452771 & .55519 & 0.566 & .54796 & 0.499 & .54933 & 0.376 & .55270 & 0.197\\
2452772 & .53586 & 0.567 & .54233 & 0.511 & .46700 & 0.369 & .52149 & 0.193\\
2452773 & .56504 & 0.576 & .57164 & 0.509 & .56991 & 0.381 & .57401 & 0.207\\
2452774 & .50343 & 0.584 & .50238 & 0.532 & .50405 & 0.388 & .49545 & 0.218\\
2452775 & .47683 & 0.621 & .47800 & 0.542 & .48459 & 0.417 & .47315 & 0.233\\
2452776 & .47378 & 0.636 & .47824 & 0.561 & .47607 & 0.430 & .48915 & 0.246\\
2452781 & .45747 & 0.715 & .46881 & 0.636 & .47640 & 0.486 & .47743 & 0.308\\
2452786 & .47119 & 0.788 & .49102 & 0.714 & .49241 & 0.577 & .49783 & 0.398\\
2452788 & .46608 & 0.852 & .46946 & 0.795 & .46176 & 0.654 & .46171 & 0.463\\
2452792 & .42979 & 1.068 & .44118 & 0.988 & .44256 & 0.825 & .44825 & 0.611\\
2452793 & .57410 & 1.141 & .57307 & 1.033 & .57445 & 0.868 & .57859 & 0.651\\
2452794 & .51185 & 1.160 & .51545 & 1.065 & .51156 & 0.894 & .53862 & 0.678\\
2452795 & .53744 & 1.165 & .54773 & 1.046 & .54494 & 0.882 & .54615 & 0.675\\
2452800 & .42771 & 0.901 & .42606 & 0.836 & .43365 & 0.695 & .42845 & 0.507\\
2452801 & .53422 & 0.823 & .53600 & 0.752 & .53604 & 0.614 & .54151 & 0.435\\
2452802 & .40815 & 0.732 & .41125 & 0.674 & .41528 & 0.540 & .43071 & 0.359\\
2452803 & .44077 & 0.671 & .44744 & 0.600 & .44330 & 0.465 & .44978 & 0.305\\
2452804 & .45837 & 0.612 & .46491 & 0.555 & .46056 & 0.438 & .46753 & 0.263\\
2452805 & .56707 & 0.577 & .57346 & 0.510 & .57175 & 0.393 & .57591 & 0.220\\
2452806 & .57030 & 0.567 & .57443 & 0.503 & .57952 & 0.377 & .57786 & 0.208\\
2452807 & .53823 & 0.576 & .54110 & 0.487 & .54125 & 0.375 & .54398 & 0.196\\
2452808 & .53893 & 0.572 & .53822 & 0.482 & .53939 & 0.370 & .53724 & 0.184\\
2452809 & .57845 & 0.565 & .57985 & 0.470 & .57969 & 0.362 & .57967 & 0.174\\
2452810 & .54475 & 0.567 & .54468 & 0.474 & .55963 & 0.371 & .54444 & 0.181\\
2452811 & .42696 & 0.564 & .42732 & 0.469 & .42716 & 0.370 & .42753 & 0.186\\
2452812 & .53275 & 0.556 & .53775 & 0.479 & .54214 & 0.359 & .52927 & 0.178\\
2452813 & .38111 & 0.551 & .38625 & 0.473 & .38870 & 0.358 & .38856 & 0.172\\
2452814 & .47151 & 0.571 & .47200 & 0.477 & .47231 & 0.370 & .47257 & 0.168\\
\hline
\end{tabular}
\begin{list}{}{}
\item[$^\star$ Observers:] K. Gazeas, P. Niarchos.
\end{list}
\end{table*}

\begin{table*} 
   \caption{Photometry in standard $VI_{\rm C}$ filters and wide
$H\alpha^{\rm W}$ and narrow $H\alpha^{\rm N}$ filters (FWHM $\approx$
200~\AA~and 30~\AA, respectively) obtained at Bia{\l}k\'ow
Observatory$^\star$ (Poland) during and near the 2003 eclipse ($E$~=~9). 
The 0.6~m Cassegrain telescope with a~Photometrics~Star~I CCD camera was
used.  Differential magnitudes are given with respect to BD~+55$\degr$2690. 
The columns labelled $HJD+$ denote the fraction of the day.}
   \label{Phot.Bialkow.9.dat}
\centering 
\begin{tabular}{lrrrrrrrr}
\hline\hline
$HJD$ & $HJD+$ & $\Delta V$ & $HJD+$ & $\Delta I_{\rm C}$ & $HJD+$ & $\Delta H\alpha^{{\rm W} \star}$ & $HJD+$ & $\Delta H\alpha^{{\rm N} \star}$\\
\hline
2452776 & .5778 & 0.518 & .5743 & 0.223 & .5810 & 0.208 & .5807 & -0.390\\
2452777 & .5320 & 0.531 & .5175 & 0.244 & .5503 & 0.222 & .5555 & -0.426\\
2452784 & .5616 & 0.662 & .5487 & 0.354 & .5598 & 0.350 & .5598 & -0.301\\
2452788 & .5578 & 0.754 & .5582 & 0.438 & .5592 & 0.421 & .5592 & -0.227\\
2452790 & .5551 & 0.885 & .5540 & 0.545 & .5549 & 0.524 & .5546 & -0.175\\
2452793 & .5550 & 0.983 & .5533 & 0.624 & .5562 & 0.600 & .5565 & -0.137\\
2452795 & .5448 & 0.995 & .5345 & 0.641 & .5449 & 0.601 & .5488 & -0.166\\
2452798 & .5410 & 0.887 & .5431 & 0.552 & .5420 & 0.524 & .5423 & -0.205\\
2452800 & .5436 & 0.804 & .5471 & 0.469 & .5542 & 0.481 & .5583 & -0.237\\
2452837 & .5224 & 0.427 & .5243 & 0.128 & .5312 & 0.118 & .5308 & -0.473\\
2452840 & ...   & ...   & ...   & ...   & .4921 & 0.125 & .4921 & -0.511\\
2452841 & .4810 & 0.419 & .4777 & 0.123 & .4883 & 0.111 & .4881 & -0.500\\
2452872 & .4676 & 0.425 & .4652 & 0.129 & .4732 & 0.091 & .4731 & -0.572\\
2452885 & .6077 & 0.424 & .6058 & 0.121 & .6110 & 0.084 & .6109 & -0.597\\
2452887 & .6238 & 0.422 & .6238 & 0.119 & .6238 & 0.090 & .6238 & -0.617\\
2452888 & .6174 & 0.412 & .6146 & 0.110 & .6236 & 0.098 & .6236 & -0.637\\
2452889 & .6265 & 0.416 & .6287 & 0.117 & .6390 & 0.089 & .6386 & -0.606\\
\hline
\end{tabular}
\begin{list}{}{}
\item[$^\star$ Observers:] Z. Ko{\l}aczkowski, G. Michalska, A. Pigulski
\end{list}
\end{table*}

\begin{table*} 
   \caption{$UBV(RI)_{\rm C}$ photometry obtained at Krak\'ow
Observatory$^{\star}$ (Poland) during and near the 2003 eclipse ($E$~=~9). 
The 0.5~m Cassegrain telescope with a~Photometrics CH350 CCD camera was
used.  Differential magnitudes are given with respect to BD~+55$\degr$2690. 
Each point is the mean value obtained from several to tens of frames.  The
columns labelled $HJD+$ denote the fraction of the day.}
   \label{Phot.Krakow.9.dat}
\centering 
\begin{tabular}{lrrrrrrrrrr}
\hline\hline
$HJD$ & $HJD+$ & $\Delta U$ & $HJD+$ & $\Delta B$ & $HJD+$ & $\Delta V$ & $HJD+$ & $\Delta R_{\rm C}$ & $HJD+$ & $\Delta I_{\rm C}$\\
\hline
2452742 & ...    & ...   & .61300 & 0.463 & .61483 & 0.423 & .61444 & 0.310 & .61482 & 0.136\\
2452743 & ...    & ...   & .61000 & 0.436 & .60964 & 0.423 & .60846 & 0.308 & .60989 & 0.116\\
2452744 & .61238 & 0.095 & .59863 & 0.466 & .59958 & 0.443 & .59856 & 0.335 & .60249 & 0.122\\
2452746 & .58829 & 0.107 & .59760 & 0.451 & .59757 & 0.423 & .59779 & 0.301 & .59065 & 0.109\\
2452750 & .60107 & 0.119 & .58647 & 0.474 & .59057 & 0.440 & .58976 & 0.326 & .59308 & 0.129\\
2452751 & .60010 & 0.092 & .59110 & 0.480 & .59252 & 0.449 & .59116 & 0.333 & .59189 & 0.120\\
2452755 & .58115 & 0.133 & .58724 & 0.491 & .58649 & 0.460 & .58530 & 0.347 & .58756 & 0.123\\
2452758 & ...    & ...   & .58841 & 0.478 & .58998 & 0.443 & .58900 & 0.323 & .58976 & 0.129\\
2452764 & .54462 & 0.124 & .54941 & 0.485 & .54591 & 0.448 & .54599 & 0.327 & .54659 & 0.132\\
2452765 & .55173 & 0.142 & .55733 & 0.499 & .55698 & 0.467 & .55731 & 0.343 & .55851 & 0.149\\
2452766 & .56137 & 0.123 & .56430 & 0.468 & .56418 & 0.440 & .56502 & 0.327 & .56445 & 0.168\\
2452767 & .54455 & 0.168 & .54632 & 0.518 & .54654 & 0.479 & .54529 & 0.357 & .54675 & 0.159\\
2452776 & .53959 & 0.280 & .54328 & 0.611 & .54325 & 0.570 & .54319 & 0.439 & .54210 & 0.212\\
2452778 & ...    & ...   & ...    & ...   & .52319 & 0.569 & ...    & ...   &  ...   &  ... \\
2452783 & .53779 & 0.388 & .53285 & 0.712 & .53142 & 0.651 & .53316 & 0.507 & .52791 & 0.307\\
2452784 & .53914 & 0.439 & .52331 & 0.760 & .52563 & 0.689 & .52620 & 0.559 & .52332 & 0.347\\
2452785 & .52699 & 0.439 & .51793 & 0.768 & .51579 & 0.701 & .51617 & 0.564 & .51684 & 0.368\\
2452789 & .53035 & 0.589 & .52150 & 0.925 & .51801 & 0.845 & .52549 & 0.720 & .52472 & 0.477\\
2452790 & .53799 & 0.658 & .52399 & 0.988 & .53132 & 0.922 & .53065 & 0.750 & .53227 & 0.537\\
2452792 & .52509 & 0.736 & .49760 & 1.053 & .50715 & 0.981 & .50787 & 0.828 & .51185 & 0.582\\
2452793 & .51916 & 0.763 & .51716 & 1.103 & .52349 & 1.021 & .52716 & 0.867 & .52136 & 0.626\\
2452794 & .48100 & 0.804 & .51445 & 1.133 & .51573 & 1.054 & .49771 & 0.892 & .51477 & 0.645\\
2452795 & .47197 & 0.800 & .52219 & 1.108 & .46593 & 1.043 & .46248 & 0.883 & .52356 & 0.630\\
2452796 & ...    & ...   & .49899 & 1.074 & .50321 & 1.014 & .51733 & 0.855 & .52067 & 0.609\\
2452798 & .48089 & 0.683 & .48471 & 0.990 & .48100 & 0.936 & .48271 & 0.790 & .48434 & 0.546\\
2452799 & .49353 & 0.637 & .49693 & 0.951 & .49907 & 0.894 & .50182 & 0.749 & .50152 & 0.515\\
2452800 & ...    & ...   & .51947 & 0.892 & .52276 & 0.825 & .52514 & 0.692 & .52758 & 0.456\\
2452801 & ...    & ...   & .51660 & 0.804 & .51731 & 0.755 & .51561 & 0.626 & .51059 & 0.395\\
2452803 & .43021 & 0.317 & .44238 & 0.637 & .44233 & 0.590 & .44656 & 0.474 & .44356 & 0.277\\
2452805 & ...    & ...   & .40439 & 0.554 & .40822 & 0.520 & .40338 & 0.397 & .40517 & 0.206\\
2452807 & .48001 & 0.185 & .46494 & 0.514 & .46306 & 0.478 & .46424 & 0.365 & .46265 & 0.170\\
2452808 & .46784 & 0.169 & .46858 & 0.508 & .46863 & 0.479 & .46801 & 0.361 & .46931 & 0.144\\
2452815 & .51562 & 0.154 & .50313 & 0.494 & .49738 & 0.455 & .50087 & 0.338 & .50353 & 0.139\\
2452818 & ...    & ...   & ...    & ...   & .39964 & 0.471 & ...    & ...   &  ...   &  ... \\
2452820 & .45161 & 0.160 & .45133 & 0.485 & .45258 & 0.456 & .44952 & 0.339 & .45251 & 0.133\\
2452821 & .54204 & 0.136 & .54376 & 0.493 & .54900 & 0.449 & .54549 & 0.337 & .54868 & 0.134\\
2452824 & ...    & ...   & .40417 & 0.516 & .40485 & 0.459 & .40368 & 0.355 & .40434 & 0.155\\
2452832 & .47416 & 0.135 & .49666 & 0.506 & .47412 & 0.480 & .49257 & 0.360 & .50195 & 0.166\\
2452834 & .44905 & 0.118 & .45297 & 0.488 & .45749 & 0.458 & .44761 & 0.336 & .45349 & 0.140\\
2452836 & .54793 & 0.099 & .53299 & 0.475 & .53872 & 0.443 & .53637 & 0.320 & .54058 & 0.130\\
2452837 & .51369 & 0.099 & .51689 & 0.457 & .51298 & 0.417 & .51053 & 0.311 & .51141 & 0.122\\
2452838 & ...    & ...   & .39151 & 0.480 & .38864 & 0.448 & .38661 & 0.333 & .38459 & 0.138\\
\hline
\end{tabular}
\begin{list}{}{}
\item[$^\star$Observers:] M. Drahus, M. Kurpi\'nska-Winiarska, A. Majewska, M. Siwak, W. Waniak, M. Winiarski, S. Zo{\l}a.
\end{list}
\end{table*}

\begin{table*}
   \caption{$UBV(RI)_{\rm C}$ photometry obtained at Kryoneri
Observatory$^{\star}$ (Greece) during the 2003 eclipse ($E$~=~9).  The 1.2~m
Cassegrain telescope with a~CCD camera was used.  The differential
magnitudes are given with respect to BD~+55$\degr$2690.  Each point is the
mean value obtained from several to tens of frames.  The columns labelled
$JD+$ denote the fraction of the day.}
   \label{Phot.Kryoneri.9.dat}
\centering 
\begin{tabular}{lrrrrrrrrrr}
\hline\hline
$JD$ & $JD+$ & $\Delta U$ & $JD+$ & $\Delta B$ & $JD+$ & $\Delta V$ & $JD+$ & $\Delta R_{\rm C}$ & $JD+$ & $\Delta I_{\rm C}$\\
\hline
2452802 & .5545 & 0.510 & .5534 & 0.680 & .5537 & 0.640 & .5552 & 0.530 & .5554 & 0.340\\
2452803 & .5742 & 0.325 & .5737 & 0.600 & .5739 & 0.560 & .5742 & 0.465 & .5737 & 0.290\\
2452804 & .5712 & 0.375 & .5715 & 0.550 & .5717 & 0.505 & .5719 & 0.425 & .5721 & 0.235\\
2452805 & .5726 & 0.340 & .5729 & 0.520 & .5738 & 0.470 & .5726 & 0.390 & .5728 & 0.200\\
2452825 & .5731 & 0.120 & .5731 & 0.489 & .5731 & 0.448 & .5731 & 0.331 & .5731 & 0.143\\
2452826 & .5707 & 0.119 & .5707 & 0.492 & .5707 & 0.456 & .5707 & 0.345 & .5707 & 0.149\\
2452827 & .4723 & 0.117 & .4723 & 0.472 & .4723 & 0.436 & ...   & ...   & .4723 & 0.124\\
\hline
\end{tabular}
\begin{list}{}{}
\item[$^{\star}$ Observers:] I. Bellas-Velidis, A. Dapergolas.
\end{list}
\end{table*}

\begin{table*} 
   \caption{Photometry in standard $UBV(RI)_{\rm C}$ and narrow (FWHM
$\approx 100$~\AA) $c$ (continuum at ${\bar{\lambda}} = 4804$~\AA) and
H$\beta$ filters obtained at Piwnice Observatory$^{\star}$ (Poland) during
and near the 2003 eclipse ($E$~=~9).  A~one-channel photometer with a~cooled
Burle~C31034 photomultiplier on the 0.6~m Cassegrain telescope was used. 
Differential magnitudes are given with respect to BD~+55$\degr$2690,
together with the corresponding standard deviations.}
   \label{Phot.Piwnice.9.dat}
\centering 
{\tiny
\begin{tabular}{lrrrrrrrrrrrrrr}
\hline\hline
$JD$ & $\Delta U$ & $\Delta B$ & $\Delta V$ & $\Delta R_{\rm C}$ & $\Delta I_{\rm C}$ & $\Delta$c$^{\star}$ & $\Delta H\beta^{\star}$ & $\sigma_{\rm U}$ & $\sigma_{\rm B}$ & $\sigma_{\rm V}$ & $\sigma_{{\rm R}_{C}}$ & $\sigma_{{\rm I}_{C}}$ & $\sigma_{\rm c}$ & $\sigma_{\rm H\beta}$\\
\hline
2452520.4438 & 0.056 & 0.479 & 0.423 & 0.253 & 0.098 & 0.495 & 0.326 & 0.013 & 0.010 & 0.008 & 0.009 & 0.010 & 0.019 & 0.031\\
2452528.4847 & 0.036 & 0.449 & 0.389 & 0.240 & 0.090 & 0.439 & 0.358 & 0.012 & 0.011 & 0.010 & 0.009 & 0.007 & 0.018 & 0.026\\
2452537.3925 & 0.015 & 0.455 & 0.386 & 0.232 & 0.072 & 0.405 & 0.371 & 0.014 & 0.015 & 0.012 & 0.008 & 0.013 & 0.023 & 0.033\\
2452550.4774 & 0.041 & 0.477 & 0.404 & 0.247 & 0.073 & 0.435 & 0.358 & 0.011 & 0.008 & 0.008 & 0.006 & 0.008 & 0.013 & 0.024\\
2452567.4421 & 0.008 & 0.431 & 0.404 & 0.239 & 0.085 & 0.426 & 0.426 & 0.013 & 0.026 & 0.007 & 0.008 & 0.007 & 0.034 & 0.027\\
2452584.5783 & 0.125 & 0.501 & 0.497 & 0.309 & 0.131 & 0.427 & 0.365 & 0.043 & 0.031 & 0.013 & 0.012 & 0.017 & 0.020 & 0.024\\
2452615.4933 &-0.007 & 0.496 & 0.461 & 0.252 & 0.099 & 0.420 & 0.355 & 0.010 & 0.024 & 0.018 & 0.016 & 0.019 & 0.005 & 0.008\\
2452618.4436 & 0.049 & 0.464 & 0.399 & 0.257 & 0.106 & 0.412 & 0.433 & 0.011 & 0.011 & 0.008 & 0.008 & 0.010 & 0.009 & 0.013\\
2452644.2981 & 0.043 & 0.458 & 0.399 & 0.233 & 0.084 & 0.426 & 0.346 & 0.021 & 0.012 & 0.013 & 0.010 & 0.011 & 0.008 & 0.006\\
2452706.3175 & ...   & ...   & ...   & 0.227 & 0.081 & 0.280 & ...   & ...   & ...   & ...   & 0.029 & 0.024 & 0.029 & ...  \\
2452711.6256 & 0.022 & 0.458 & 0.406 & 0.245 & 0.086 & 0.417 & 0.395 & 0.014 & 0.014 & 0.013 & 0.011 & 0.012 & 0.007 & 0.015\\
2452723.5955 & 0.085 & 0.476 & 0.428 & 0.264 & 0.103 & 0.465 & 0.321 & 0.012 & 0.011 & 0.013 & 0.010 & 0.011 & 0.008 & 0.016\\
2452746.5724 & 0.087 & 0.494 & 0.409 & 0.276 & 0.107 & 0.433 & 0.386 & 0.011 & 0.011 & 0.006 & 0.007 & 0.009 & 0.006 & 0.020\\
2452755.4434 & 0.139 & 0.509 & 0.472 & 0.283 & 0.140 & 0.543 & 0.432 & 0.029 & 0.025 & 0.027 & 0.032 & 0.027 & 0.027 & 0.020\\
2452765.5459 & 0.067 & 0.461 & 0.415 & 0.230 & 0.111 & 0.515 & 0.427 & 0.026 & 0.025 & 0.025 & 0.020 & 0.029 & 0.016 & 0.009\\
2452774.5113 & 0.253 & 0.615 & 0.516 & 0.348 & 0.192 & 0.636 & 0.406 & 0.010 & 0.023 & 0.017 & 0.023 & 0.016 & 0.006 & 0.010\\
2452776.4954 & 0.264 & 0.623 & 0.525 & 0.376 & 0.214 & 0.576 & 0.589 & 0.013 & 0.009 & 0.015 & 0.005 & 0.014 & 0.020 & 0.003\\
2452781.5073 & 0.333 & 0.765 & 0.698 & 0.499 & 0.255 & 0.669 & 0.653 & 0.009 & 0.026 & 0.029 & 0.029 & 0.031 & 0.021 & 0.040\\
2452782.4471 & 0.369 & 0.701 & 0.608 & 0.443 & 0.253 & 0.616 & 0.654 & 0.016 & 0.019 & 0.006 & 0.018 & 0.011 & 0.018 & 0.058\\
2452784.4702 & 0.423 & 0.804 & 0.667 & 0.521 & 0.352 & 0.749 & 0.703 & 0.014 & 0.001 & 0.012 & 0.012 & 0.001 & 0.022 & 0.014\\
2452787.4748 & 0.455 & 0.800 & 0.739 & 0.578 & 0.389 & 0.785 & 0.693 & 0.002 & 0.010 & 0.002 & 0.009 & 0.022 & 0.013 & 0.010\\
2452788.4596 & 0.494 & 0.870 & 0.762 & 0.596 & 0.425 & 0.888 & 0.760 & 0.004 & 0.009 & 0.006 & 0.018 & 0.007 & 0.022 & 0.006\\
2452789.4708 & 0.554 & 0.952 & 0.866 & 0.634 & 0.510 & 0.879 & 0.897 & 0.017 & 0.001 & 0.001 & 0.019 & 0.001 & 0.008 & 0.023\\
2452790.4560 & 0.623 & 1.020 & 0.916 & 0.711 & 0.508 & 0.933 & 0.914 & 0.028 & 0.027 & 0.033 & 0.038 & 0.001 & 0.001 & 0.018\\
2452792.4784 & 0.757 & 1.102 & 0.970 & 0.778 & 0.564 & 0.969 & 1.038 & 0.003 & 0.002 & 0.018 & 0.011 & 0.011 & 0.026 & 0.053\\
2452793.4881 & 0.717 & 1.071 & 0.974 & 0.788 & 0.531 & 1.026 & 0.884 & 0.006 & 0.021 & 0.023 & 0.019 & 0.030 & 0.017 & 0.040\\
2452794.4838 & 0.792 & 1.140 & 1.037 & 0.819 & 0.604 & 1.087 & 0.978 & 0.016 & 0.008 & 0.008 & 0.001 & 0.001 & 0.063 & 0.001\\
2452795.4733 & 0.769 & 1.116 & 1.045 & 0.845 & 0.589 & 1.056 & 1.015 & 0.004 & 0.014 & 0.001 & 0.015 & 0.006 & 0.012 & 0.014\\
2452797.5101 & 0.797 & 1.092 & 0.956 & 0.805 & 0.602 & 1.009 & 1.149 & 0.034 & 0.023 & 0.028 & 0.017 & 0.018 & 0.016 & 0.025\\
2452798.4117 & 0.664 & 1.018 & 0.915 & 0.728 & 0.520 & 0.969 & 0.848 & 0.015 & 0.015 & 0.014 & 0.010 & 0.007 & 0.019 & 0.048\\
2452799.5068 & 0.621 & 0.999 & 0.886 & 0.706 & 0.497 & 0.899 & 0.860 & 0.014 & 0.017 & 0.014 & 0.008 & 0.008 & 0.014 & 0.020\\
2452800.4813 & 0.536 & 0.898 & 0.791 & 0.652 & 0.454 & 0.868 & 0.799 & 0.002 & 0.049 & 0.012 & 0.017 & 0.026 & 0.017 & 0.048\\
2452802.4563 & 0.320 & 0.736 & 0.662 & 0.532 & 0.332 & 0.733 & 0.685 & 0.026 & 0.025 & 0.027 & 0.003 & 0.008 & 0.016 & 0.004\\
2452804.4661 & 0.244 & 0.615 & 0.553 & 0.395 & 0.226 & 0.559 & 0.518 & 0.001 & 0.011 & 0.013 & 0.008 & 0.018 & 0.006 & 0.018\\
2452807.4662 & 0.123 & 0.538 & 0.454 & 0.342 & 0.167 & 0.515 & 0.484 & 0.024 & 0.001 & 0.019 & 0.011 & 0.017 & 0.026 & 0.022\\
2452808.4152 & 0.165 & 0.514 & 0.441 & 0.309 & 0.119 & 0.485 & 0.431 & 0.013 & 0.016 & 0.010 & 0.012 & 0.008 & 0.028 & 0.013\\
2452809.5060 & 0.202 & 0.532 & 0.461 & 0.284 & 0.107 & 0.565 & 0.447 & 0.030 & 0.028 & 0.008 & 0.010 & 0.007 & 0.038 & 0.044\\
2452812.4715 & 0.121 & 0.520 & 0.464 & 0.306 & 0.117 & 0.457 & 0.397 & 0.008 & 0.011 & 0.010 & 0.005 & 0.005 & 0.016 & 0.029\\
2452813.4397 & 0.082 & 0.509 & 0.418 & 0.300 & 0.112 & 0.454 & 0.418 & 0.020 & 0.011 & 0.005 & 0.007 & 0.006 & 0.021 & 0.032\\
2452817.4821 & 0.130 & 0.512 & 0.449 & 0.270 & 0.115 & 0.474 & 0.500 & 0.026 & 0.020 & 0.016 & 0.016 & 0.020 & 0.023 & 0.019\\
2452818.4616 & 0.118 & 0.529 & 0.459 & 0.318 & 0.116 & 0.464 & 0.516 & 0.010 & 0.009 & 0.010 & 0.007 & 0.004 & 0.011 & 0.032\\
2452823.5078 & 0.096 & 0.489 & 0.419 & 0.276 & 0.099 & 0.464 & 0.390 & 0.013 & 0.010 & 0.011 & 0.013 & 0.007 & 0.022 & 0.030\\
2452825.4532 & 0.104 & 0.487 & 0.409 & 0.291 & 0.101 & 0.456 & 0.358 & 0.010 & 0.015 & 0.008 & 0.012 & 0.005 & 0.015 & 0.040\\
2452837.4346 & 0.068 & 0.489 & 0.427 & 0.282 & 0.101 & 0.511 & 0.470 & 0.008 & 0.008 & 0.010 & 0.008 & 0.006 & 0.025 & 0.042\\
2452853.5289 & 0.072 & 0.482 & 0.417 & 0.255 & 0.107 & 0.436 & 0.414 & 0.010 & 0.012 & 0.006 & 0.006 & 0.004 & 0.014 & 0.030\\
2452856.5146 & 0.047 & 0.457 & 0.416 & 0.258 & 0.081 & 0.456 & 0.415 & 0.010 & 0.009 & 0.006 & 0.007 & 0.006 & 0.005 & 0.007\\
2452863.4907 & 0.045 & 0.489 & 0.412 & 0.282 & 0.126 & 0.511 & 0.433 & 0.016 & 0.024 & 0.014 & 0.023 & 0.015 & 0.027 & 0.031\\
2452869.4528 & 0.096 & 0.486 & 0.446 & 0.283 & 0.108 & 0.464 & 0.470 & 0.005 & 0.013 & 0.006 & 0.006 & 0.008 & 0.026 & 0.028\\
2452888.5588 & 0.054 & 0.482 & 0.418 & 0.248 & 0.120 & 0.490 & 0.417 & 0.007 & 0.011 & 0.007 & 0.009 & 0.010 & 0.013 & 0.028\\
2452929.4190 & 0.071 & 0.468 & 0.409 & 0.275 & 0.074 & 0.457 & 0.327 & 0.015 & 0.011 & 0.011 & 0.017 & 0.025 & 0.012 & 0.021\\
2452935.5038 & 0.097 & 0.520 & 0.421 & 0.212 & 0.068 & 0.432 & 0.350 & 0.040 & 0.038 & 0.032 & 0.023 & 0.020 & 0.018 & 0.001\\
2452954.4270 &-0.060 & 0.394 & 0.324 & 0.240 & 0.074 & 0.296 & 0.268 & 0.019 & 0.014 & 0.012 & 0.007 & 0.010 & 0.023 & 0.041\\
2452985.2883 & 0.050 & 0.427 & 0.379 & 0.263 & 0.065 & ...   & ...   & 0.010 & 0.008 & 0.008 & 0.007 & 0.007 & ...   &  ... \\
2453008.3696 & 0.072 & 0.445 & 0.413 & 0.264 & 0.107 & ...   & ...   & 0.017 & 0.017 & 0.009 & 0.009 & 0.009 & ...   &  ... \\
2453035.2914 & 0.113 & 0.468 & 0.427 & 0.279 & 0.111 & 0.447 & 0.403 & 0.014 & 0.011 & 0.015 & 0.010 & 0.008 & 0.020 & 0.022\\
2453057.2583 & 0.045 & 0.461 & 0.400 & 0.254 & 0.088 & ...   & ...   & 0.009 & 0.010 & 0.009 & 0.004 & 0.006 & ...   &  ... \\
2453124.5278 & 0.114 & 0.464 & 0.378 & 0.243 & 0.085 & ...   & ...   & 0.012 & 0.009 & 0.009 & 0.006 & 0.010 & ...   &  ... \\
2453150.4178 & 0.095 & 0.485 & 0.415 & 0.257 & 0.072 & ...   & ...   & 0.021 & 0.010 & 0.014 & 0.008 & 0.009 & ...   &  ... \\
2453159.4468 & 0.027 & 0.460 & 0.379 & 0.231 & 0.047 & ...   & ...   & 0.011 & 0.008 & 0.011 & 0.010 & 0.006 & ...   &  ... \\
2453170.4703 & 0.055 & 0.461 & 0.409 & 0.219 & 0.037 & ...   & ...   & 0.013 & 0.014 & 0.009 & 0.008 & 0.011 & ...   &  ... \\
2453202.4900 & 0.018 & 0.465 & 0.390 & 0.234 & 0.029 & ...   & ...   & 0.018 & 0.009 & 0.008 & 0.008 & 0.007 & ...   &  ... \\
2453226.4641 & 0.050 & 0.482 & 0.386 & 0.224 & 0.040 & ...   & ...   & 0.016 & 0.009 & 0.008 & 0.008 & 0.007 & ...   &  ... \\
2453249.5761 & 0.025 & 0.470 & 0.365 & 0.213 & 0.047 & ...   & ...   & 0.010 & 0.011 & 0.013 & 0.009 & 0.008 & ...   &  ... \\
2453291.4884 & 0.008 & 0.422 & 0.379 & 0.224 & 0.059 & ...   & ...   & 0.018 & 0.009 & 0.010 & 0.008 & 0.013 & ...   &  ... \\
\hline
\end{tabular}
}
\begin{list}{}{}
\item[$^{\star}$ Observers:] C. Ga{\l}an, A. Majcher, M. Miko{\l}ajewski.
\end{list}
\end{table*}

\begin{table*} 
   \caption{$UBV(RI)_{\rm C}$ photometry obtained at Rozhen
Observatory$^{\star}$ (Bulgaria) during the 2003 eclipse ($E$~=~9).  The 2~m
Ritchey-Chr\'etien telescope with a~CCD camera was used.  The differential
magnitudes are given with respect to BD~+55$\degr$2690.  The columns
labelled $JD+$ denote the fraction of the day.}
   \label{Phot.Rozhen.2m.9.dat}
\centering 
\begin{tabular}{lrrrrrrrrrr}
\hline\hline
$JD$ & $JD+$ & $\Delta U$ & $JD+$ & $\Delta B$ & $JD+$ & $\Delta V$ & $JD+$ & $\Delta R_{\rm C}$ & $JD+$ & $\Delta I_{\rm C}$\\
\hline
2452754 & ...   & ...   & ...   & ...   & .5367 & 0.399 & .5364 & 0.343 & .5368 & 0.100\\
2452759 & ...   & ...   & ...   & ...   & ...   & ...   & .5023 & 0.338 & ...   & ...  \\
2452760 & ...   & ...   & ...   & ...   & .5024 & 0.376 & .5023 & 0.298 & ...   & ...  \\
2452791 & .5259 & 0.751 & .5222 & 1.089 & .5231 & 0.892 & .5276 & 0.840 & .5298 & 0.682\\
2452792 & .5465 & 0.680 & .5529 & 0.959 & .5410 & 0.797 & .5392 & 0.534 & .5356 & 0.451\\
\hline
\end{tabular}
\begin{list}{}{}
\item[$^{\star}$ Observers:] M. Gromadzki, D. Kolev.
\end{list}
\end{table*}

\begin{table*} 
   \caption{Photometry in the $V$ filter obtained at Rozhen
Observatory$^{\star}$ (Bulgaria) during the 2003 eclipse ($E$~=~9).  The
0.5/0.7~m Schmidt telescope with a~CCD camera was used.  The differential
magnitudes are given with respect to the BD~+55$\degr$2690 comparison star.}
   \label{Phot.Rozhen.Schm.9.dat}
\centering 
\begin{tabular}{lr}
\hline\hline
$JD$ & $\Delta V$ \\
\hline
2452756.4912  &  0.427 \\
2452758.5752  &  0.381 \\
2452759.5894  &  0.369 \\
\hline
\end{tabular}
\begin{list}{}{}
\item[$^{\star}$ Observers:] G. Apostolovska, B. Bilkina, M. Gromadzki, D. Kolev.
\end{list}
\end{table*}

\begin{table*}
   \caption{$UBV$ photometry obtained at Rozhen Observatory$^{\star}$
(Bulgaria) during the 2003 eclipse ($E$~=~9).  The 0.6~m Cassegrain
telescope with a~one-channel photomultiplier was used.  The differential
magnitudes are given with respect to BD~+55$\degr$2690.}
   \label{Phot.Rozhen0.6m.9.dat}
\centering 
\begin{tabular}{lrrr}
\hline\hline
$JD$ & $\Delta U$ & $\Delta B$ & $\Delta V$\\
2452760.575 & 0.021 & 0.404 & 0.412\\
2452761.571 & 0.079 & 0.463 & 0.437\\
2452799.505 & 0.497 & 0.845 & 0.802\\
2452801.516 & 0.402 & 0.749 & 0.731\\
2452811.510 & 0.089 & 0.469 & 0.438\\
2452821.503 & 0.070 & 0.450 & 0.439\\
\hline
\hline
\end{tabular}
\begin{list}{}{}
\item[$^{\star}$ Observer:] D. Dimitrov.
\end{list}
\end{table*}

\begin{table*} 
   \caption{$UBV(RI)_{\rm C}$ photometry obtained at Skinakas
Observatory$^{\star}$ (Crete, Greece) during the 2003 eclipse ($E$~=~9). 
The 1.3~m Ritchey-Chr\'etien telescope with a~CCD camera was used.  The
differential magnitudes are given with respect to BD~+55$\degr$2690.}
   \label{Phot.Skinakas.9.dat}
\centering 
\begin{tabular}{llllll}
\hline\hline
$JD$ & $\Delta U$ & $\Delta B$ & $\Delta V$ & $\Delta R_{\rm C}$ & $\Delta I_{\rm C}$\\
\hline
2452798.578 & ...  & 0.98  & 0.91 & 0.77 & 0.55\\
2452799.585 & 0.67 & 0.94  & 0.87 & 0.74 & 0.52\\
2452801.577 & 0.52 & 0.79  & 0.73 & 0.61 & 0.40\\
2452802.579 & 0.43 & 0.70  & 0.64 & 0.53 & 0.34\\
2452803.599 & 0.36 & 0.63  & 0.58 & 0.48 & 0.29\\
2452804.587 & 0.29 & 0.63  & 0.53 & 0.42 & 0.23\\
2452805.587 & 0.26 & 0.54  & 0.49 & 0.39 & 0.205\\
2452806.579 & 0.24 & 0.525 & 0.48 & 0.38 & 0.19\\
2452807.579 & 0.22 & 0.515 & 0.46 & 0.36 & 0.175\\
\hline
\end{tabular}
\begin{list}{}{}
\item[$^{\star}$ Observer:] E. Semkov.
\end{list}
\end{table*}

\begin{table*} 
   \caption{$BV(RI)_{\rm C}$ photometry obtained at Piszk\'estet\"o
Observatory$^{\star}$ (Hungary) during the 2003 eclipse ($E$~=~9).  The
0.6/0.9~m Schmidt telescope with an AT200 CCD camera was used.  The
differential magnitudes are given with respect to BD~+55$\degr$2690.}
   \label{Phot.Piszkesteto.9.dat}
\centering 
\begin{tabular}{lrrrr}
\hline\hline
$HJD$ & $\Delta B$ & $\Delta V$ & $\Delta R_{\rm C}$ & $\Delta I_{\rm C}$\\
\hline
2452776.563     & 0.592      & 0.516      & 0.393              & 0.215\\
2452778.565     & 0.620      & 0.545      & 0.430              & 0.245\\
2452779.568     & 0.660      & 0.590      & 0.460              & 0.270\\
\hline
\end{tabular}
\begin{list}{}{}
\item[$^{\star}$ Observers:] B. Cs\'ak, B. Gere, P. N\'emeth.
\end{list}
\end{table*}

\begin{table*} 
   \caption{$BV(RI)_{\rm C}$ photometric data obtained at Altan
Observatory$^{\star}$ (Czech Republic) during the 2008/9 eclipse ($E$~=~10). 
The 0.2 meter RL~Vixen~VMC200L telescope with a~G2-0402 CCD camera was used. 
Differential magnitudes are given with respect to BD~+55$\degr$2690
together, with the corresponding standard deviations.}
   \label{Phot.Altan.10.dat}
\centering 
\begin{tabular}{lrrrrrrrrrrrrrr}
\hline\hline
$HJD$& $HJD+$ & $\Delta$B & $\sigma_{\rm B}$ & $HJD+$ & $\Delta$V & $\sigma_{\rm V}$ & $HJD+$ & $\Delta$R & $\sigma_{{\rm R}_C}$ & $HJD+$ & $\Delta$I & $\sigma_{{\rm I}_C}$\\
\hline
2454810 & .1966 & 0.476 & 0.029 & .1989 & 0.422 & 0.004 & .1966 & 0.298 & 0.010 & .1895 & 0.161 & 0.013\\
2454824 & .3541 & 0.536 & 0.012 & .3596 & 0.477 & 0.006 & .3533 & 0.352 & 0.005 & .3533 & 0.183 & 0.005\\
2454828 & .3614 & 0.569 & 0.014 & .3529 & 0.490 & 0.013 & .3595 & 0.393 & 0.006 & .3485 & 0.215 & 0.009\\
2454829 & .1900 & 0.591 & 0.014 & .1927 & 0.551 & 0.002 & .1917 & 0.424 & 0.002 & .1896 & 0.245 & 0.004\\
2454830 & .1965 & 0.693 & 0.008 & .1975 & 0.602 & 0.003 & .1952 & 0.461 & 0.004 & .1945 & 0.281 & 0.002\\
2454831 & .1936 & 0.730 & 0.011 & .1926 & 0.640 & 0.005 & .1917 & 0.503 & 0.005 & .1946 & 0.317 & 0.004\\
2454832 & .2061 & 0.738 & 0.005 & .2070 & 0.627 & 0.005 & .2061 & 0.490 & 0.005 & .2045 & 0.310 & 0.004\\
2454835 & .1856 & 0.711 & 0.068 & .1792 & 0.667 & 0.009 & .1879 & 0.569 & 0.049 & .1820 & 0.299 & 0.012\\
2454840 & .3251 & 0.937 & 0.014 & .3222 & 0.836 & 0.008 & .3212 & 0.699 & 0.013 & .3203 & 0.524 & 0.022\\
2454843 & .2002 & 0.961 & 0.009 & .1967 & 0.842 & 0.007 & .1958 & 0.686 & 0.009 & .1986 & 0.484 & 0.007\\
2454844 & .2440 & 0.915 & 0.010 & .2431 & 0.842 & 0.005 & .2441 & 0.702 & 0.007 & .2450 & 0.490 & 0.003\\
2454845 & .2532 & 0.891 & 0.021 & .2541 & 0.779 & 0.004 & .2532 & 0.645 & 0.005 & .2542 & 0.467 & 0.015\\
2454854 & .2473 & 0.597 & 0.002 & .2483 & 0.525 & 0.004 & .2474 & 0.389 & 0.004 & .2464 & 0.208 & 0.004\\
2454857 & .2190 & 0.565 & 0.013 & .2161 & 0.474 & 0.011 & .2152 & 0.365 & 0.005 & .2032 & 0.188 & 0.003\\
2454860 & .2400 & 0.530 & 0.012 & .2390 & 0.459 & 0.007 & .2381 & 0.319 & 0.004 & .2371 & 0.170 & 0.002\\
\hline
\end{tabular}
\begin{list}{}{}
\item[$^{\star}$ Observer:] L. Br\'at.
\end{list}
\end{table*}

\begin{table*} 
   \caption{$BV(RI)_{\rm C}$ photometric data obtained at Bia{\l}k\'ow
Observatory$^{\star}$ (Poland) during the 2008/9 eclipse ($E$~=~10).  The
0.6~m Cassegrain telescope with a~CCD camera was used.  Differential
magnitudes are given with respect to BD~+55$\degr$2690.  The columns
labelled $HJD+$ denote the fraction of the day.}
   \label{Phot.Bialkow.10.dat}
\centering 
\begin{tabular}{lrrrrrrrr}
\hline\hline
$HJD$ & $HJD+$ & $\Delta B$ & $HJD+$ & $\Delta V$ & $HJD+$ & $\Delta R_{\rm C}$ & $HJD+$ & $\Delta I_{\rm C}$\\
\hline
2454814 & .30228 & 0.521 & .31205 & 0.459 & .31825 & 0.348 & .29353 & 0.176\\
2454815 & .17723 & 0.518 & .18402 & 0.460 & .18773 & 0.346 & .17027 & 0.179\\
2454816 & .21367 & 0.505 & .20595 & 0.444 & ...    & ...   & .19731 & 0.170\\
2454831 & .16875 & 0.701 & .17475 & 0.617 & .17954 & 0.494 & .18382 & 0.318\\
2454834 & .22008 & 0.689 & .22197 & 0.607 & .22323 & 0.485 & .22447 & 0.315\\
2454837 & .20076 & 0.733 & .20672 & 0.660 & .21128 & 0.541 & .21535 & 0.368\\
2454838 & .18539 & 0.779 & .19092 & 0.708 & .19458 & 0.587 & .19824 & 0.410\\
2454840 & .29154 & 0.903 & .29343 & 0.823 & .29468 & 0.689 & .29593 & 0.501\\
2454843 & .18047 & 0.924 & .18543 & 0.841 & .18913 & 0.708 & .19283 & 0.517\\
2454844 & .17864 & 0.883 & .18479 & 0.812 & .18909 & 0.684 & .19280 & 0.494\\
2454845 & .18903 & 0.834 & .19402 & 0.765 & .19771 & 0.641 & .20141 & 0.463\\
\hline
\end{tabular}
\begin{list}{}{}
\item[$^{\star}$ Observers:] G. Kopacki, A. Majewska, A. Narwid, E. Niemczura, A. Pigulski, M. St\c{e}\'slicki
\end{list}
\end{table*}

\begin{table*} 
   \caption{$BV(RI)_{\rm C}$ photometric data obtained at Green Island
Observatory$^{\star}$ (North Cyprus) during the 2008/9 eclipse ($E$~=~10). 
The 0.35 m telescope (Meade 14$"$LX200R ) with a~Meade~DSI~II~Pro CCD camera
was used.  Differential magnitudes are given with respect to
BD~+55$\degr$2690 together with the corresponding standard deviations.  The
columns labelled $JD+$ denote the fraction of the day.}
   \label{Phot.GreenIsl.10.dat}
\centering 
\begin{tabular}{lrrrrrrrrrrrrrr}
\hline\hline
$JD$& $JD+$ & $\Delta$B & $\sigma_{\rm B}$ & $JD+$ & $\Delta$V & $\sigma_{\rm V}$ & $JD+$ & $\Delta$R & $\sigma_{{\rm R}_C}$ & $JD+$ & $\Delta$I & $\sigma_{{\rm I}_C}$\\
\hline
2454810 & .20398 & 0.500 & 0.005 & .19671 & 0.443 & 0.009 & .20988 & 0.325 & 0.012 & .21571 & 0.178 & 0.007\\
2454820 & .23238 & 0.488 & 0.005 & .23920 & 0.447 & 0.010 & .24587 & 0.326 & 0.006 & .25252 & 0.195 & 0.003\\
2454849 & .17309 & 0.702 & 0.015 & .18015 & 0.657 & 0.012 & .18716 & 0.526 & 0.008 & .19584 & 0.346 & 0.023\\
2454852 & .18026 & 0.599 & 0.004 & .18916 & 0.543 & 0.016 & .19609 & 0.428 & 0.006 & .20456 & 0.268 & 0.019\\
2454854 & .21422 & 0.561 & 0.005 & .22135 & 0.502 & 0.020 & .22853 & 0.369 & 0.003 & .23729 & 0.244 & 0.015\\
2454858 & .18768 & 0.509 & 0.010 & .19550 & 0.459 & 0.019 & .20248 & 0.343 & 0.008 & .21265 & 0.188 & 0.019\\
2454860 & .19138 & 0.503 & 0.001 & .18434 & 0.440 & 0.009 & .19978 & 0.323 & 0.008 & .17670 & 0.172 & 0.016\\
2454865 & .17912 & 0.496 & 0.005 & .18599 & 0.438 & 0.003 & .19291 & 0.316 & 0.005 & .20348 & 0.183 & 0.004\\
2454878 & ...    & ...   & ...   & .18347 & 0.455 & 0.009 & .18899 & 0.325 & 0.010 & .19432 & 0.169 & 0.009\\
\hline
\end{tabular}
\begin{list}{}{}
\item[$^{\star}$ Observer:] Y. \"O\u{g}men.
\end{list}
\end{table*}

\clearpage

\begin{table*} 
   \caption{$BV(RI)_{\rm C}$ photometric data obtained at Furzehill House
Observatory$^{\star}$ (Ilston, Swansea, United Kingdom) during the 2008/9
eclipse ($E$~=~10).  The 0.35 m Schmidt-Cassegrain telescope with
an~SXVF-H16 CCD camera was used.  Differential magnitudes are given with
respect to BD~+55$\degr$2690, together with the corresponding standard
deviations.  The columns labelled $HJD+$ denote the fraction of the day.}
   \label{Phot.Ilston.10.dat}
\centering 
\begin{tabular}{lrrrrrrrrrrrrrr}
\hline\hline
$HJD$& $HJD+$ & $\Delta$B & $\sigma_{\rm B}$ & $HJD+$ & $\Delta$V & $\sigma_{\rm V}$ & $HJD+$ & $\Delta$R & $\sigma_{{\rm R}_C}$ & $HJD+$ & $\Delta$I & $\sigma_{{\rm I}_C}$\\
\hline
2454815 & .39684 & 0.496 & 0.005 & .39406 & 0.472 & 0.004 & ...    & ...   & ...   & ...    & ...   & ...  \\
2454816 & .36785 & 0.478 & 0.009 & .35943 & 0.439 & 0.006 & ...    & ...   & ...   & ...    & ...   & ...  \\
2454827 & .26603 & 0.539 & 0.004 & .26977 & 0.512 & 0.003 & .27370 & 0.392 & 0.003 & .27740 & 0.221 & 0.009\\
2454828 & .26105 & 0.551 & 0.004 & .26495 & 0.506 & 0.003 & .26900 & 0.410 & 0.003 & .27280 & 0.236 & 0.007\\
2454829 & .27405 & 0.595 & 0.003 & .27958 & 0.532 & 0.003 & .28370 & 0.426 & 0.003 & .28984 & 0.256 & 0.005\\
2454831 & .25665 & 0.673 & 0.003 & .26239 & 0.622 & 0.004 & .26650 & 0.506 & 0.004 & .27228 & 0.333 & 0.006\\
2454834 & .28341 & 0.679 & 0.003 & .30354 & 0.604 & 0.003 & .29910 & 0.500 & 0.003 & .29261 & 0.329 & 0.005\\
2454835 & .25826 & 0.657 & 0.003 & .27667 & 0.614 & 0.003 & .27270 & 0.492 & 0.003 & .26649 & 0.331 & 0.005\\
2454837 & .25763 & 0.725 & 0.003 & .27688 & 0.678 & 0.003 & .27270 & 0.549 & 0.003 & .26606 & 0.383 & 0.005\\
2454838 & .26906 & 0.780 & 0.003 & .29043 & 0.754 & 0.004 & .28560 & 0.596 & 0.004 & .27763 & 0.439 & 0.005\\
2454839 & .33017 & 0.845 & 0.003 & .35110 & 0.773 & 0.004 & .34680 & 0.659 & 0.004 & .34021 & 0.481 & 0.005\\
2454852 & .32291 & 0.602 & 0.003 & .34376 & 0.555 & 0.004 & .33940 & 0.453 & 0.004 & .33353 & 0.277 & 0.006\\
2454855 & .30354 & 0.534 & 0.003 & .33404 & 0.515 & 0.006 & .32830 & 0.389 & 0.004 & .31159 & 0.222 & 0.005\\
2454858 & .30088 & 0.502 & 0.003 & .33170 & 0.459 & 0.004 & .32312 & 0.370 & 0.005 & .31639 & 0.202 & 0.006\\
2454866 & .29978 & 0.487 & 0.003 & .31625 & 0.455 & 0.004 & .31257 & 0.333 & 0.004 & .30715 & 0.182 & 0.005\\
2454869 & .30427 & 0.489 & 0.003 & .33609 & 0.437 & 0.004 & .32837 & 0.332 & 0.004 & .32114 & 0.157 & 0.006\\
2454871 & .32916 & 0.474 & 0.006 & .30698 & 0.466 & 0.005 & .30204 & 0.342 & 0.005 & .29362 & 0.168 & 0.005\\
2454878 & .29032 & 0.503 & 0.004 & .30275 & 0.461 & 0.006 & .29972 & 0.352 & 0.006 & .29484 & 0.197 & 0.008\\
\hline
\end{tabular}
\begin{list}{}{}
\item[$^{\star}$ Observer:] I. Miller.
\end{list}
\end{table*}

\begin{table*} 
   \caption{Photometry in the $V$ filter obtained at Observatorio
Astron\'omico "Las Pegueras"$^{\star}$ (Navas de Oro, Segovia, Spain) during
the 2008/9 eclipse ($E$~=~10).  The 0.35~m Reflector with a~CCD camera was
used.  Differential magnitudes are given with respect to BD~+55$\degr$2690.}
   \label{Phot.NavasDeOro.10.dat}
\centering 
\begin{tabular}{lr}
\hline\hline
$JD$ & $\Delta V$ \\
\hline
2454815.2834 & -0.43\\
2454817.2900 & -0.39\\
2454820.2953 & -0.42\\
2454823.2650 & -0.40\\
2454824.2668 & -0.39\\
2454825.2556 & -0.37\\
2454826.2524 & -0.35\\
2454836.3460 & -0.21\\
2454839.2727 & -0.07\\
2454842.2511 &  0.01\\
2454843.2532 &  0.00\\
2454846.2701 & -0.15\\
2454848.3420 & -0.20\\
2454849.2724 & -0.22\\
2454856.3162 & -0.37\\
2454863.3053 & -0.43\\
\hline
\end{tabular}
\begin{list}{}{}
\item[$^{\star}$ Observer:] T. A. Heras.
\end{list}
\end{table*}

\begin{table*}
   \caption{Photometry obtained at Ostrava Observatory$^{\star}$ (Czech
Republic) with standard $BV(RI)_{\rm C}$ filters during the 2008/9 eclipse
($E$~=~10).  The 0.3~m Schmidt-Cassegrain telescope with a~CCD camera has
been used.  Differential magnitudes are given with respect to
BD~+55$\degr$2690.  The columns labelled $JD+$ denote the fraction of the
day.}
   \label{Phot.Kocian.10.dat}
\centering 
\begin{tabular}{lrrrrrrrr}
\hline\hline
$JD$ & $JD+$ & $\Delta B$ & $JD+$ & $\Delta V$ & $JD+$ & $\Delta R_{\rm C}$ & $JD+$ & $\Delta I_{\rm C}$\\
\hline
2454830 & .3938 & 0.652 & .3913 & 0.581 & .3912 & 0.440 & .3928 & 0.248\\
\hline
\end{tabular}
\begin{list}{}{}
\item[$^{\star}$ Observer:] R. Koci\'an.
\end{list}
\end{table*}

\begin{table*}
   \caption{$BV(RI)_{\rm C}$ photometry obtained at Rozhen
Observatory$^{\star}$ (Bulgaria) during the 2008/9 eclipse ($E$~=~10).  The
2~m Ritchey-Chr\'etien telescope with a~CCD camera was used.  Differential
magnitudes are given with respect to BD~+55$\degr$2690.  The columns
labelled $JD+$ denote the fraction of the day.}
   \label{Phot.Rozhen2m.10.dat}
\centering 
\begin{tabular}{lrrrrrrrr}
\hline\hline
$JD$ & $JD+$ & $\Delta B$ & $JD+$ & $\Delta V$ & $JD+$ & $\Delta R_{\rm C}$ & $JD+$ & $\Delta I_{\rm C}$\\
\hline
2454827 & .370 & 0.52 & .365 & 0.47 & .363 & 0.36 & .361 & 0.21\\
\hline
\end{tabular}
\begin{list}{}{}
\item[$^{\star}$ Observers:] S. Peneva, E. Semkov.
\end{list}
\end{table*}

\begin{table*} 
   \caption{$UBV(RI)_{\rm C}$ photometry obtained at Rozhen
Observatory$^{\star}$ (Bulgaria) during and near the 2008/9 eclipse
($E$~=~10).  The 0.5/0.7~m Schmidt telescope with a~CCD camera was used. 
The differential magnitudes are given with respect to BD~+55$\degr$2690. 
The columns labelled $JD+$ denote the fraction of the day.}
   \label{Phot.Rozhen0507m.10.dat}
\centering 
\begin{tabular}{lrrrrrrrrrr}
\hline\hline
$JD$ & $JD+$ & $\Delta U$ & $JD+$ & $\Delta B$ & $JD+$ & $\Delta V$ & $JD+$ & $\Delta R_{\rm C}$ & $JD+$ & $\Delta I_{\rm C}$\\
\hline
2454762 & ...  & 0.12 & ...  & 0.46 & ...  & 0.42 & ...  & 0.32 & ...  & 0.16\\
2454764 & ...  & 0.12 & ...  & 0.46 & ...  & 0.41 & ...  & 0.31 & ...  & 0.15\\
2454830 & ...  & ...  & ...  & ...  & ...  & ...  & .275 & 0.48 & .270 & 0.29\\
2454842 & ...  & ...  & .191 & 0.95 & .186 & 0.86 & .182 & 0.73 & .176 & 0.52\\
2454843 & .191 & 0.62 & .186 & 0.94 & .183 & 0.85 & .180 & 0.73 & .175 & 0.54\\
2454844 & .190 & 0.57 & .186 & 0.90 & .183 & 0.82 & .180 & 0.69 & .176 & 0.49\\
2454845 & ...  & ...  & .294 & 0.81 & .269 & 0.75 & .265 & 0.64 & .250 & 0.45\\
\hline
\end{tabular}
\begin{list}{}{}
\item[$^{\star}$ Observers:] S. Peneva, E. Semkov.
\end{list}
\end{table*}

\begin{table*} 
   \caption{Photometry obtained at Rozhen Observatory$^{\star}$ (Bulgaria)
during the 2008/9 eclipse ($E$~=~10) with standard $UBV(RI)_{\rm C}$
(Bessell) filters.  The 0.6~m Cassegrain telescope with a~FLI~PL09000 CCD
camera was used.  Differential magnitudes are given with respect to
BD~+55$\degr$2690.  Each point is the mean value obtained from several
frames.  The columns labelled $HJD+$ denote the fraction of the day.}
   \label{Phot.Rozhen06m.10.dat}
\centering 
\begin{tabular}{lrrrrrrrrrr}
\hline\hline
$HJD$ & $HJD+$ & $\Delta U$ & $HJD+$ & $\Delta B$ & $HJD+$ & $\Delta V$ & $HJD+$ & $\Delta R_{\rm C}$ & $HJD+$ & $\Delta I_{\rm C}$\\
\hline
2454808 & .34867 & 0.190 & .35088 & 0.502 & .35202 & 0.451 & .35264 & 0.348 & .35308 & 0.146\\
2454808 & .41329 & 0.178 & .41551 & 0.494 & .41054 & 0.424 & .41115 & 0.332 & .41159 & 0.133\\
2454810 & .24350 & 0.197 & .24571 & 0.511 & .24685 & 0.454 & .24746 & 0.355 & .24791 & 0.159\\
2454832 & .17438 & 0.420 & .17467 & 0.698 & .17611 & 0.615 & .17721 & 0.501 & .17795 & 0.296\\
2454858 & ...    & ...   & .23000 & 0.533 & .23000 & 0.453 & .23000 & 0.362 & .23000 & 0.163\\
2454869 & .21210 & 0.185 & .21241 & 0.497 & .21034 & 0.445 & .21155 & 0.341 & .21494 & 0.155\\
2454886 & .21908 & 0.178 & .22383 & 0.483 & .22219 & 0.440 & .22340 & 0.326 & .22425 & 0.136\\
\hline
\end{tabular}
\begin{list}{}{}
\item[$^{\star}$ Observers:] D. Dimitrov, V. Popov.
\end{list}
\end{table*}

\begin{table*}
   \caption{$BV(RI)_{\rm C}$ photometry obtained at Kryoneri
Observatory$^{\star}$ (Greece) at the beginning of the 2008/9 eclipse
($E$~=~10).  The 1.2~m Cassegrain telescope with a~CCD camera was used. 
Differential magnitudes are given with respect to BD~+55$\degr$2691.  The
columns labelled $HJD+$ denote the fraction of the day.}
   \label{Phot.Kryoneri.10.dat}
\centering 
\begin{tabular}{lrrrrrrrr}
\hline\hline
$HJD$ & $HJD+$ & $\Delta B$ & $HJD+$ & $\Delta V$ & $HJD+$ & $\Delta R_{\rm C}$ & $HJD+$ & $\Delta I_{\rm C}$\\
\hline
2454774 & .2517 & -0.413 & .2792 & -0.458 & .2863 & -0.576 & .2911 & -0.761\\
2454775 & .3071 & -0.372 & .3043 & -0.456 & .2932 & -0.560 & .3011 & -0.748\\
\hline
\end{tabular}
\begin{list}{}{}
\item[$^{\star}$ Observers:] I. Bellas-Velidis, A. Dapergolas.
\end{list}
\end{table*}

\begin{table*}
   \caption{$BV(RI)_{\rm C}$ photometry obtained at Sonoita Research
Observatory$^{\star}$ (Arizona, USA) during and near the 2008/9 eclipse
($E$~=~10).  The 0.5~m Cassegrain telescope with an SBIG~STL~6303 CCD camera
was used.  Comparisons were made against two stars designated with the AAVSO
Unique Identifiers as: 000-BCQ-040, 000-BJJ-300.  The former is a~$d$ object
from Meinunger's comparison star sequence (Meinunger 1975).  The apparent
magnitudes are also given with respect to BD~+55$\degr$2690 (calculated with
the adoption of $BV(RI)_{\rm C}$ magnitudes of BD~+55$\degr$2690 given in
\citet{Mik2003}).  The columns labelled $HJD+$ denote the fraction of the
day.}
   \label{Phot.SRO.Staels.10.dat}
\centering 
\begin{tabular}{lllllllllllll}
\hline\hline
$HJD$ & $HJD+$ & $B$ & $\Delta B$ & $HJD+$ & $V$ & $\Delta V$ & $HJD+$ & $R_{\rm C}$ & $\Delta R_{\rm C}$ & $HJD+$ & $I_{\rm C}$ & $\Delta I_{\rm C}$\\
\hline
2454790 & .6019 & 11.140 & 0.460 & .6022 & 10.800 & 0.420 & .6003 & 10.510 & 0.420 & .6020  & 10.155 & 0.285 \\
2454791 & .7615 & 11.155 & 0.475 & .7618 & 10.805 & 0.425 & .7618 & 10.515 & 0.425 & .7618  & 10.165 & 0.295 \\
2454792 & .7133 & 11.170 & 0.490 & .7135 & 10.810 & 0.430 & .7136 & 10.525 & 0.435 & .7134  & 10.165 & 0.295 \\
2454801 & .5847 & 11.155 & 0.475 & .5850 & 10.815 & 0.435 & .5851 & 10.525 & 0.435 & .5850  & 10.180 & 0.310 \\
2454804 & .7281 & 11.160 & 0.480 & .7282 & 10.825 & 0.445 & .7282 & 10.585 & 0.445 & .7282  & 10.200 & 0.330 \\
2454806 & .6826 & 11.170 & 0.490 & .6827 & 10.815 & 0.435 & .6828 & 10.560 & 0.435 & .6827  & 10.195 & 0.325 \\
2454807 & .7415 & 11.170 & 0.490 & .7417 & 10.820 & 0.440 & .7417 & 10.610 & 0.440 & .7416  & 10.195 & 0.325 \\
2454809 & .7298 & 11.175 & 0.495 & .7300 & 10.830 & 0.450 & .7301 & 10.545 & 0.455 & .7300  & 10.190 & 0.320 \\
2454810 & .6454 & 11.185 & 0.505 & .6457 & 10.825 & 0.445 & .6458 & 10.555 & 0.465 & .6456  & 10.215 & 0.345 \\
2454811 & .6918 & 11.180 & 0.500 & .6920 & 10.830 & 0.450 & .6920 & 10.545 & 0.455 & .6919  & 10.200 & 0.330 \\
2454812 & .6638 & 11.175 & 0.495 & .6640 & 10.830 & 0.450 & .6640 & 10.545 & 0.455 & .6639  & 10.205 & 0.335 \\
2454816 & .7199 & 11.195 & 0.515 & .7201 & 10.835 & 0.455 & .7202 & 10.550 & 0.460 & .7201  & 10.210 & 0.340 \\
2454821 & .6856 & 11.190 & 0.510 & .6859 & 10.835 & 0.455 & .6859 & 10.540 & 0.450 & .6859  & 10.215 & 0.345 \\
2454829 & .5868 & 11.290 & 0.610 & .6801 & 10.895 & 0.515 & .6802 & 10.610 & 0.520 & .6801  & 10.270 & 0.400 \\
2454828 & .6799 & 11.260 & 0.580 & .5875 & 10.930 & 0.550 & .5876 & 10.640 & 0.550 & .5875  & 10.285 & 0.415 \\
2454830 & .5858 & 11.325 & 0.645 & .5861 & 10.950 & 0.570 & .5861 & 10.660 & 0.570 & .5860  & 10.290 & 0.420 \\
2454837 & .57280& 11.440 & 0.760 & .57295& 11.060 & 0.680 & .57310& 10.765 & 0.675 & .57280 & 10.415 & 0.545 \\
2454838 & .57290& 11.480 & 0.800 & .57310& 11.115 & 0.735 & .57320& 10.810 & 0.720 & .57310 & 10.460 & 0.590 \\
2454839 & .57145& 11.560 & 0.880 & .57170& 11.180 & 0.800 & .57175& 10.870 & 0.780 & .57165 & 10.515 & 0.645 \\
2454840 & .58085& 11.605 & 0.925 & .58115& 11.230 & 0.850 & .58120& 10.920 & 0.830 & .58105 & 10.550 & 0.680 \\
2454841 & .59040& 11.625 & 0.945 & .59065& 11.230 & 0.850 & .59065& 10.930 & 0.840 & .59050 & 10.560 & 0.690 \\
2454842 & .61965& 11.625 & 0.945 & .61995& 11.240 & 0.860 & .61995& 10.925 & 0.835 & .61980 & 10.545 & 0.675 \\
2454843 & .57510& 11.600 & 0.920 & .57540& 11.215 & 0.835 & .57540& 10.910 & 0.820 & .57530 & 10.555 & 0.685 \\
2454844 & .61630& 11.565 & 0.885 & .61655& 11.175 & 0.795 & .61660& 10.885 & 0.795 & .61645 & 10.530 & 0.660 \\
2454846 & .61500& 11.485 & 0.805 & .61530& 11.110 & 0.730 & .61530& 10.825 & 0.735 & .61515 & 10.470 & 0.600 \\
2454847 & .61645& 11.440 & 0.760 & .61665& 11.080 & 0.700 & .61670& 10.785 & 0.695 & .61655 & 10.435 & 0.565 \\
2454848 & .57710& 11.410 & 0.730 & .57735& 11.055 & 0.675 & .57740& 10.760 & 0.670 & .57725 & 10.410 & 0.540 \\
2454849 & .60140& 11.380 & 0.700 & .60165& 11.010 & 0.630 & .60170& 10.725 & 0.635 & .60155 & 10.370 & 0.500 \\
2454850 & .60925& 11.340 & 0.660 & .60955& 10.985 & 0.605 & .60955& 10.690 & 0.600 & .60940 & 10.345 & 0.475 \\
2454857 & .58720& 11.200 & 0.520 & .58745& 10.850 & 0.470 & .58750& 10.575 & 0.485 & .58740 & 10.215 & 0.345 \\
2454858 & .59370& 11.195 & 0.515 & .59395& 10.830 & 0.450 & .59400& 10.555 & 0.465 & .59380 & 10.210 & 0.340 \\
2454860 & .58890& 11.180 & 0.500 & .58915& 10.820 & 0.440 & .58920& 10.550 & 0.460 & .58905 & 10.205 & 0.335 \\
2454861 & .58450& 11.160 & 0.480 & .58470& 10.815 & 0.435 & .58475& 10.530 & 0.440 & .58460 & 10.200 & 0.330 \\
2454862 & .58525& 11.175 & 0.495 & .58550& 10.825 & 0.445 & .58550& 10.535 & 0.445 & .58545 & 10.200 & 0.330 \\
2454863 & .58500& 11.180 & 0.500 & .58525& 10.845 & 0.465 & .58530& 10.535 & 0.445 & .58515 & 10.190 & 0.320 \\
2454864 & .58570& 11.180 & 0.500 & .58595& 10.830 & 0.450 & .58600& 10.550 & 0.460 & .58585 & 10.200 & 0.330 \\
2454865 & .58600& 11.180 & 0.500 & .58625& 10.820 & 0.440 & .58630& 10.545 & 0.455 & .58615 & 10.205 & 0.335 \\
2454867 & .58970& 11.170 & 0.490 & .58990& 10.815 & 0.435 & .58995& 10.530 & 0.440 & .58980 & 10.190 & 0.320 \\
2454869 & .58750& 11.150 & 0.470 & .58825& 10.795 & 0.415 & .58830& 10.495 & 0.405 & .58815 & 10.165 & 0.295 \\
2454870 & .58935& 11.160 & 0.480 & .58960& 10.815 & 0.435 & .58965& 10.540 & 0.450 & .58950 & 10.195 & 0.325 \\
2454879 & .59660& 11.195 & 0.515 & .59685& 10.820 & 0.440 & .59690& 10.515 & 0.425 & .59675 & 10.205 & 0.335 \\
2454881 & .59585& 11.200 & 0.520 & .59615& 10.830 & 0.450 & .59615& 10.545 & 0.455 & .59590 & 10.210 & 0.340 \\
2454882 & .59465& 11.190 & 0.510 & .59490& 10.830 & 0.450 & .59495& 10.525 & 0.435 & .59485 & 10.205 & 0.335 \\
2454885 & .02960& 11.175 & 0.495 & .02985& 10.825 & 0.445 & .02990& 10.540 & 0.450 & .02975 & 10.180 & 0.310 \\
2454890 & .02490& 11.165 & 0.485 & .02520& 10.815 & 0.435 & .02525& 10.525 & 0.435 & .02510 & 10.185 & 0.315 \\
2454891 & .01930& 11.165 & 0.485 & .01955& 10.815 & 0.435 & .01960& 10.525 & 0.435 & .01945 & 10.185 & 0.315 \\
2454893 & .01010& 11.145 & 0.465 & .01040& 10.810 & 0.430 & .01045& 10.525 & 0.435 & .01030 & 10.180 & 0.310 \\
2454894 & .01105& 11.175 & 0.495 & .01125& 10.815 & 0.435 & .01135& 10.525 & 0.435 & .01105 & 10.200 & 0.330 \\
2454895 & .01540& 11.125 & 0.445 & .01570& 10.770 & 0.390 & .01575& 10.485 & 0.395 & .01560 & 10.160 & 0.290 \\
\hline
\end{tabular}
\begin{list}{}{}
\item[$^{\star}$ Observer:] B. Staels.
\end{list}
\end{table*}

\begin{table*}
   \caption{$BV(RI)_{\rm C}$ photometry obtained at Sonoita Research
Observatory$^{\star}$ (Arizona, USA) during and near the 2008/9 eclipse
($E$~=~10).  The 0.5~m Cassegrain telescope with an SBIG~STL~6303 CCD camera
was used.  The apparent magnitudes are also given with respect to
BD~+55$\degr$2690 (calculated by adopting $BV(RI)_{\rm C}$ magnitudes of
BD~+55$\degr$2690 from \citet{Mik2003}).}
   \label{Phot.SRO.HPO.10.dat}
\centering 
\begin{tabular}{llrrlrrlrrlrr}
\hline\hline
$JD$ & $B$ & $\Delta B$ & $\sigma_{\rm B}$ & $V$ & $\Delta V$ & $\sigma_{\rm V}$ & $R_{\rm C}$ & $\Delta R_{\rm C}$ & $\sigma_{{\rm R}_C}$ & $I_{\rm C}$ & $\Delta I_{\rm C}$ & $\sigma_{{\rm I}_C}$  \\
\hline
2454790.60 & 11.14 & 0.46 & 0.02 & 10.79 & 0.41 & 0.01 & 10.38 & 0.29 & ...  & 10    & 0.13 & 0.02\\
2454791.76 & 11.15 & 0.47 & 0.03 & 10.79 & 0.41 & 0.01 & 10.38 & 0.29 & 0.02 & 10.01 & 0.14 & 0.02\\
2454801.58 & 11.16 & 0.48 & 0.03 & 10.79 & 0.41 & 0.01 & 10.39 & 0.30 & ...  & 10.01 & 0.14 & 0.02\\
2454804.72 & 11.17 & 0.49 & 0.03 & 10.78 & 0.40 & 0.04 & 10.4  & 0.31 & 0.07 & 10.02 & 0.15 & 0.01\\
2454806.68 & 11.17 & 0.49 & 0.03 & 10.8  & 0.42 & 0.01 & 10.4  & 0.31 & 0.04 & 10.02 & 0.15 & 0.02\\
2454807.74 & 11.18 & 0.50 & 0.03 & 10.79 & 0.41 & 0.03 & 10.42 & 0.33 & 0.09 & 10.03 & 0.16 & 0.01\\
2454809.73 & 11.18 & 0.50 & 0.02 & 10.81 & 0.43 & 0.01 & 10.41 & 0.32 & 0.01 & 10.05 & 0.18 & 0.06\\
2454810.64 & 11.19 & 0.51 & 0.02 & 10.82 & 0.44 & 0.01 & 10.42 & 0.33 & 0.02 & 10.04 & 0.17 & 0.01\\
2454811.69 & 11.19 & 0.51 & 0.03 & 10.81 & 0.43 & 0.01 & 10.4  & 0.31 & 0.01 & 10.03 & 0.16 & 0.01\\
2454812.66 & 11.19 & 0.51 & 0.03 & 10.81 & 0.43 & 0.01 & 10.41 & 0.32 & 0.01 & 10.03 & 0.16 & 0.01\\
2454816.72 & 11.19 & 0.51 & 0.03 & 10.82 & 0.44 & 0.01 & 10.4  & 0.31 & 0.01 & 10.04 & 0.17 & 0.01\\
2454821.68 & 11.19 & 0.51 & 0.02 & 10.81 & 0.43 & 0.01 & 10.4  & 0.31 & 0.01 & 10.04 & 0.17 & 0.02\\
2454828.68 & 11.26 & 0.58 & 0.02 & 10.87 & 0.49 & 0.01 & 10.46 & 0.37 & 0.02 & 10.09 & 0.22 & 0.01\\
2454829.58 & 11.18 & 0.50 & ...  & 10.94 & 0.56 & 0.07 & 10.49 & 0.40 & 0.01 & 10.11 & 0.24 & 0.02\\
2454830.58 & 11.36 & 0.68 & 0.02 & 10.96 & 0.58 & 0.01 & 10.54 & 0.45 & 0.01 & 10.14 & 0.27 & 0.02\\
2454838.57 & 11.49 & 0.81 & 0.02 & 11.09 & 0.71 & 0.01 & 10.67 & 0.58 & 0.01 & 10.29 & 0.42 & 0.02\\
2454839.57 & 11.56 & 0.88 & 0.03 & 11.15 & 0.77 & 0.01 & 10.73 & 0.64 & 0.01 & 10.34 & 0.47 & 0.02\\
2454840.58 & 11.61 & 0.93 & 0.02 & 11.21 & 0.83 & 0.01 & 10.78 & 0.69 & 0.01 & 10.37 & 0.50 & 0.01\\
2454841.59 & 11.63 & 0.95 & 0.02 & 11.21 & 0.83 & 0.02 & 10.78 & 0.69 & 0.01 & 10.38 & 0.51 & 0.02\\
2454842.62 & 11.62 & 0.94 & 0.03 & 11.21 & 0.83 & 0.01 & 10.78 & 0.69 & 0.01 & 10.37 & 0.50 & 0.02\\
2454843.57 & 11.6  & 0.92 & 0.02 & 11.2  & 0.82 & 0.01 & 10.77 & 0.68 & 0.01 & 10.37 & 0.50 & 0.02\\
2454844.61 & 11.56 & 0.88 & 0.03 & 11.16 & 0.78 & 0.01 & 10.74 & 0.65 & 0.01 & 10.35 & 0.48 & 0.01\\
2454846.61 & 11.49 & 0.81 & 0.02 & 11.1  & 0.72 & 0.02 & 10.69 & 0.60 & 0.01 & 10.29 & 0.42 & 0.02\\
2454847.61 & 11.44 & 0.76 & 0.03 & 11.07 & 0.69 & 0.01 & 10.65 & 0.56 & 0.01 & 10.27 & 0.40 & 0.02\\
2454848.57 & 11.41 & 0.73 & 0.03 & 11.04 & 0.66 & 0.01 & 10.62 & 0.53 & 0.01 & 10.23 & 0.36 & 0.02\\
2454849.60 & 11.38 & 0.70 & 0.03 & 11    & 0.62 & 0.01 & 10.58 & 0.49 & 0.01 & 10.19 & 0.32 & 0.01\\
2454850.61 & 11.34 & 0.66 & 0.03 & 10.97 & 0.59 & 0.02 & 10.54 & 0.45 & 0.01 & 10.17 & 0.30 & 0.02\\
2454857.58 & 11.2  & 0.52 & 0.03 & 10.83 & 0.45 & 0.01 & 10.43 & 0.34 & 0.01 & 10.04 & 0.17 & 0.02\\
2454858.60 & 11.2  & 0.52 & 0.03 & 10.82 & 0.44 & 0.02 & 10.42 & 0.33 & 0.01 & 10.04 & 0.17 & 0.02\\
2454860.59 & 11.18 & 0.50 & 0.03 & 10.81 & 0.43 & 0.01 & 10.4  & 0.31 & 0.01 & 10.03 & 0.16 & 0.01\\
2454861.58 & 11.18 & 0.50 & 0.03 & 10.8  & 0.42 & 0.01 & 10.39 & 0.30 & 0.01 & 10.02 & 0.15 & 0.02\\
2454861.58 & 11.18 & 0.50 & 0.03 & 10.8  & 0.42 & 0.01 & 10.39 & 0.30 & 0.01 & 10.02 & 0.15 & 0.02\\
2454862.58 & 11.18 & 0.50 & 0.02 & 10.81 & 0.43 & 0.01 & 10.39 & 0.30 & 0.01 & 10.02 & 0.15 & 0.02\\
2454863.58 & 11.17 & 0.49 & 0.03 & 10.8  & 0.42 & 0.01 & 10.39 & 0.30 & 0.01 & 10.02 & 0.15 & 0.02\\
2454865.59 & 11.18 & 0.50 & 0.03 & 10.8  & 0.42 & 0.01 & 10.4  & 0.31 & 0.01 & 10.02 & 0.15 & 0.02\\
2454867.59 & 11.18 & 0.50 & ...  & 10.8  & 0.42 & ...  & 10.41 & 0.32 & ...  & 10.02 & 0.15 & ... \\ 
2454869.58 & 11.17 & 0.49 & 0.03 & 10.8  & 0.42 & 0.01 & 10.38 & 0.29 & 0.01 & 10.02 & 0.15 & 0.01\\
2454870.59 & 11.18 & 0.50 & 0.03 & 10.81 & 0.43 & 0.01 & 10.39 & 0.30 & 0.02 & 10.01 & 0.14 & 0.02\\
2454879.59 & 11.19 & 0.51 & 0.02 & 10.81 & 0.43 & 0.02 & 10.37 & 0.28 & 0.02 & 10.03 & 0.16 & 0.02\\
\hline
\end{tabular}
\begin{list}{}{}
\item[$^{\star}$ Observers:] L. Elder, J. Hopkins, J. Pye.
\end{list}
\end{table*}

\begin{landscape}
\begin{table} 
   \caption{Photometry obtained at Athens Observatory$^{\star}$ (Greece)
with standard $BV(RI)_{\rm C}$ (Bessell) filters during and near the 2008/9
eclipse ($E$~=~10).  The 0.4~m Cassegrain telescope with an
SBIG~ST-8XMEI~CCD camera was used.  Differential magnitudes are given with
respect to three comparison stars: $a$~=~BD~+55$\degr$2690,
$b$~=~GSC-3973:2150, and $c$~=~BD~+55$\degr$2691.  Each point is the mean
value obtained from several frames.  The columns labelled $HJD+$ denote the
fraction of the day.}
   \label{Phot.Athens.10.dat}
\tiny{
\begin{tabular}{|l|rrrr|rrrr|rrrr|rrrr|}
\hline\hline
      & \multicolumn{4}{c|}{$\Delta B$} & \multicolumn{4}{c|}{$\Delta V$} & \multicolumn{4}{c|}{$\Delta R_{\rm C}$} & \multicolumn{4}{c|}{$\Delta I_{\rm C}$} \\
\hline
$HJD$&$HJD+$&$v-a$&$v-b$&$v-c$&$HJD+$&$v-a$&$v-b$&$v-c$&$HJD+$&$v-a$&$v-b$&$v-c$&$HJD+$&$v-a$&$v-b$&$v-c$\\
\hline
2454791& .3793& 0.470& -0.293& -0.343& .3805& 0.428& -0.419& -0.431& .3812& 0.305& -0.602& -0.564& .3818& 0.152& -0.799& -0.721\\
2454792& .4465& 0.465& -0.317& -0.348& .4477& 0.423& -0.444& -0.440& .4484& 0.305& -0.603& -0.569& .4490& 0.153& -0.814& -0.723\\
2454796& .1741& 0.461& -0.295& -0.350& .1753& 0.421& -0.426& -0.435& .1759& 0.302& -0.599& -0.573& .1766& 0.138& -0.825& -0.736\\
2454798& .1742& 0.462& -0.300& -0.348& .1754& 0.425& -0.424& -0.436& .1761& 0.314& -0.596& -0.564& .1767& 0.145& -0.813& -0.728\\
2454804& .1734& 0.475& -0.292& -0.345& .1746& 0.432& -0.427& -0.433& .1752& 0.320& -0.596& -0.565& .1758& 0.148& -0.813& -0.712\\
2454809& .1717& 0.495& -0.268& -0.319& .1729& 0.455& -0.396& -0.408& .1736& 0.336& -0.577& -0.538& .1742& 0.166& -0.796& -0.708\\
2454824& .3406& 0.525& -0.239& -0.301& .3417& 0.479& -0.366& -0.389& .3424& 0.354& -0.546& -0.519& .3431& 0.184& -0.771& -0.688\\
2454825& .1721& 0.530& -0.234& -0.284& .1733& 0.488& -0.356& -0.378& .1740& 0.374& -0.525& -0.503& .1746& 0.194& -0.753& -0.676\\
2454851& .1886& 0.630& -0.137& -0.189& .1895& 0.583& -0.272& -0.289& .1901& 0.465& -0.442& -0.411& .1906& 0.286& -0.667& -0.589\\
2454882& .1990& 0.521& -0.255& -0.305& .2000& 0.457& -0.375& -0.388& .2005& 0.340& -0.559& -0.537& .2011& 0.168& -0.761& -0.670\\
2454883& .2032& 0.506& -0.258& -0.310& .2042& 0.437& -0.408& -0.420& .2047& 0.320& -0.586& -0.541& .2053& 0.142& -0.814& -0.739\\
\hline
\end{tabular}
}
\begin{list}{}{}
\item[$^{\star}$ Observers:] K. Gazeas, A. Liakos, P. Niarchos.
\end{list}
\end{table}

\begin{table} 
   \caption{Photometry obtained at Hankasalmi Observatory$^{\star}$
(Finland) with standard $BV(RI)_{\rm C}$ (Bessell) filters during the 2008/9
eclipse ($E$~=~10).  The 0.4~m RCOS telescope with an SBIG~STL-1001 CCD
camera was used.  Differential magnitudes are given with respect to three
comparison stars: $a$~=~BD~+55$\degr$2690, $b$~=~GSC-3973:2150, and
$c$~=~BD~+55$\degr$2691.  Each point is the mean value obtained from several
frames.  The columns labelled $HJD+$ denote the fraction of the day.}
   \label{Phot.Hankasalmi.10.dat}
\tiny{
\begin{tabular}{|l|rrrr|rrr|rrr|rrrr|}
\hline\hline
      & \multicolumn{4}{c|}{$\Delta B$} & \multicolumn{3}{c|}{$\Delta V$} & \multicolumn{3}{c|}{$\Delta R_{\rm C}$} & \multicolumn{4}{c|}{$\Delta I_{\rm C}$} \\
\hline
$HJD$&$HJD+$&$v-a$&$v-b$&$v-c$&$HJD+$&$v-b$&$v-c$&$HJD+$&$v-b$&$v-c$&$HJD+$&$v-a$&$v-b$&$v-c$\\
\hline
2454828 & .3851 & 0.579 & -0.195 & -0.245 & .3822 & -0.340 & -0.351 & .3872 & -0.506 & -0.470 & .3862 & 0.195 & -0.755 & -0.662\\
2454833 & .1808 & 0.705 & -0.057 & -0.112 & .1779 & -0.228 & -0.242 & .1827 & -0.400 & -0.370 & .1848 & 0.292 & -0.668 & -0.574\\
2454836 & .1388 & 0.693 & -0.061 & -0.118 & .1358 & -0.217 & -0.235 & .1406 & -0.381 & -0.359 & .1430 & 0.315 & -0.640 & -0.547\\
2454839 & .2096 & 0.854 &  0.093 &  0.034 & .2070 & -0.068 & -0.086 & .2116 & -0.250 & -0.221 & .2135 & 0.430 & -0.522 & -0.431\\
2454848 & .3753 & 0.768 & -0.020 & -0.080 & .3722 & -0.171 & -0.192 & .3772 & -0.351 & -0.316 & .3790 & 0.370 & -0.595 & -0.512\\
2454862 & .3108 & 0.502 & -0.265 & -0.313 & .3082 & -0.403 & -0.419 & .3129 & -0.569 & -0.542 & .3147 & 0.140 & -0.812 & -0.723\\
2454865 & .1540 & 0.496 & -0.262 & -0.317 & .1511 & -0.409 & -0.422 & .1559 & -0.567 & -0.539 & .1579 & 0.131 & -0.819 & -0.731\\
\hline
\end{tabular}
}
\begin{list}{}{}
\item[$^{\star}$ Observer:] A. Oksanen.
\end{list}
\end{table}

\begin{table} 
   \caption{Photometry obtained at Ostrava Observatory$^{\star}$ (Czech
Republic) with standard $BV(RI)_{\rm C}$ filters during the 2008/9 eclipse
($E$~=~10).  The 0.2~m Newton telescope with an SBIG~ST8-XME CCD camera has
been used.  Differential magnitudes are given with respect to three
comparison stars: $a$~=~BD~+55$\degr$2690, $b$~=~GSC-3973:2150, and
$c$~=~BD~+55$\degr$2691.  Each point is the mean value obtained from several
frames.  The columns labelled $HJD+$ denote the fraction of the day.}
   \label{Phot.Kucakova.10.dat}
\tiny{
\begin{tabular}{|l|rrrr|rrrr|rrrr|rrrr|}
\hline\hline
      & \multicolumn{4}{c|}{$\Delta B$} & \multicolumn{4}{c|}{$\Delta V$} & \multicolumn{4}{c|}{$\Delta R_{\rm C}$} & \multicolumn{4}{c|}{$\Delta I_{\rm C}$} \\
\hline
$HJD$&$HJD+$&$v-a$&$v-b$&$v-c$&$HJD+$&$v-a$&$v-b$&$v-c$&$HJD+$&$v-a$&$v-b$&$v-c$&$HJD+$&$v-a$&$v-b$&$v-c$\\
\hline
2454810 & .3006 & 0.475 & -0.245 & -0.317 & .2972 & 0.438 & -0.412 & -0.421 & .2983 & 0.327 & -0.564 & -0.540 & .3017 & 0.152 & -0.784 & -0.700\\
2454815 & .3856 & 0.504 & -0.223 & -0.295 & .3822 & 0.447 & -0.398 & -0.406 & .3833 & 0.346 & -0.548 & -0.520 & .3844 & 0.162 & -0.781 & -0.701\\
2454841 & .3133 & 0.929 &  0.209 &  0.139 & .3099 & 0.848 & -0.009 & -0.009 & .3087 & 0.709 & -0.184 & -0.153 & .3121 & 0.493 & -0.454 & -0.358\\
2454843 & .2011 & 0.925 &  0.200 &  0.137 & .1970 & 0.846 &  0.003 & -0.016 & .1982 & 0.724 & -0.167 & -0.152 & .1993 & 0.510 & -0.431 & -0.359\\
2454850 & .2206 & 0.662 & -0.070 & -0.125 & .2194 & 0.611 & -0.248 & -0.249 & .2206 & 0.494 & -0.404 & -0.379 & .2217 & 0.303 & -0.647 & -0.559\\
2454857 & .2009 & 0.504 & -0.221 & -0.282 & .1975 & 0.461 & -0.391 & -0.398 & .1986 & 0.352 & -0.543 & -0.509 & .1998 & 0.168 & -0.774 & -0.684\\
\hline
\end{tabular}
}
\begin{list}{}{}
\item[$^{\star}$ Observer:] H. Ku\v{c}\'akov\'a.
\end{list}
\end{table}
\end{landscape}

\begin{landscape}
\begin{table} 
   \caption{$UBV(RI)_{\rm C}$ photometry obtained at Krak\'ow
Observatory$^{\star}$ (Poland) during and near the 2008/9 eclipse
($E$~=~10).  The 0.5~m Cassegrain telescope with a~CCD camera was used.  The
differential magnitudes are given with respect to three comparison stars:
$a$~=~BD~+55$\degr$2690, $b$~=~GSC-3973:2150, and $c$~=~BD~+55$\degr$2691. 
Each point is the mean value obtained from several to tens of frames.  The
columns labelled $HJD+$ denote the fraction of the day.}
   \label{Phot.Krakow.10.dat}
\tiny{
\begin{tabular}{|l|rrrr|rrrr|rrrr|rrrr|rrrr|}
\hline\hline
      & \multicolumn{4}{c|}{$\Delta U$} & \multicolumn{4}{c|}{$\Delta B$} & \multicolumn{4}{c|}{$\Delta V$} & \multicolumn{4}{c|}{$\Delta R_{\rm C}$} & \multicolumn{4}{c|}{$\Delta I_{\rm C}$} \\
\hline
$HJD$&$HJD+$&$v-a$&$v-b$&$v-c$&$HJD+$&$v-a$&$v-b$&$v-c$&$HJD+$&$v-a$&$v-b$&$v-c$&$HJD+$&$v-a$&$v-b$&$v-c$&$HJD+$&$v-a$&$v-b$&$v-c$\\
\hline
2454751 & ...   & ...   & ...    & ...    & .5043 & 0.473 & -0.305 & -0.348 & .5045 & 0.420 & -0.424 & -0.432 & .5046 & 0.310 & -0.582 & -0.555 & ...   & ...   & ...    & ...   \\
2454753 & ...   & ...   & ...    & ...    & .2875 & 0.466 & -0.305 & -0.354 & .2899 & 0.415 & -0.424 & -0.430 & .2954 & 0.308 & -0.581 & -0.556 & ...   & ...   & ...    & ...   \\
2454761 & .3144 & 0.091 & -0.381 & -0.483 & .3174 & 0.455 & -0.293 & -0.341 & .3185 & 0.393 & -0.458 & -0.454 & .3195 & 0.295 & -0.599 & -0.565 & .3203 & 0.124 & -0.821 & -0.732\\
2454762 & ...   & ...   & ...    & ...    & .2463 & 0.465 & -0.290 & -0.340 & .2505 & 0.396 & -0.459 & -0.457 & .2533 & 0.297 & -0.603 & -0.569 & ...   & ...   & ...    & ...   \\
2454766 & ...   & ...   & ...    & ...    & .3773 & 0.449 & -0.301 & -0.352 & .3795 & 0.385 & -0.472 & -0.468 & .3820 & 0.287 & -0.623 & -0.582 & .3788 & 0.122 & -0.826 & -0.739\\
2454773 & ...   & ...   & ...    & ...    & .4528 & 0.479 & -0.292 & -0.345 & .4516 & 0.432 & -0.413 & -0.426 & .4478 & 0.311 & -0.587 & -0.558 & ...   & ...   & ...    & ...   \\
2454780 & ...   & ...   & ...    & ...    & .2080 & 0.478 & -0.295 & -0.345 & .2081 & 0.428 & -0.417 & -0.431 & .2089 & 0.316 & -0.576 & -0.551 & ...   & ...   & ...    & ...   \\
2454781 & ...   & ...   & ...    & ...    & .2612 & 0.470 & -0.294 & -0.349 & .2611 & 0.425 & -0.416 & -0.429 & .2614 & 0.316 & -0.575 & -0.552 & ...   & ...   & ...    & ...   \\
2454782 & ...   & ...   & ...    & ...    & .4971 & 0.477 & -0.301 & -0.345 & .4969 & 0.419 & -0.421 & -0.433 & .4973 & 0.307 & -0.590 & -0.565 & ...   & ...   & ...    & ...   \\
2454786 & ...   & ...   & ...    & ...    & .2532 & 0.471 & -0.293 & -0.358 & .2571 & 0.417 & -0.422 & -0.433 & .2590 & 0.304 & -0.587 & -0.559 & ...   & ...   & ...    & ...   \\
2454795 & ...   & ...   & ...    & ...    & .2542 & 0.477 & -0.292 & -0.344 & .2585 & 0.434 & -0.402 & -0.419 & .2504 & 0.323 & -0.570 & -0.546 & ...   & ...   & ...    & ...   \\
2454799 & .1689 & 0.104 & -0.360 & -0.474 & .1725 & 0.472 & -0.272 & -0.332 & .1731 & 0.403 & -0.445 & -0.446 & .1731 & 0.305 & -0.586 & -0.554 & .1729 & 0.132 & -0.811 & -0.721\\
2454807 & ...   & ...   & ...    & ...    & .1906 & 0.486 & -0.267 & -0.320 & .1909 & 0.420 & -0.437 & -0.438 & .1910 & 0.323 & -0.578 & -0.548 & .1911 & 0.148 & -0.804 & -0.709\\
2454814 & ...   & ...   & ...    & ...    & .2001 & 0.516 & -0.248 & -0.304 & .2004 & 0.458 & -0.385 & -0.389 & .2006 & 0.348 & -0.547 & -0.516 & ...   & ...   & ...    & ...   \\
2454816 & .2096 & 0.151 & -0.311 & -0.425 & .2100 & 0.502 & -0.245 & -0.304 & .2103 & 0.437 & -0.419 & -0.420 & .2105 & 0.341 & -0.558 & -0.529 & .2103 & 0.162 & -0.789 & -0.702\\
2454824 & .2160 & 0.165 & -0.300 & -0.410 & .1902 & 0.514 & -0.237 & -0.291 & .1912 & 0.444 & -0.417 & -0.409 & .1898 & 0.348 & -0.553 & -0.516 & .1903 & 0.170 & -0.778 & -0.685\\
2454830 & .3385 & 0.327 & -0.130 & -0.230 & .3572 & 0.667 & -0.084 & -0.137 & .3491 & 0.566 & -0.289 & -0.286 & .3494 & 0.456 & -0.441 & -0.404 & .3589 & 0.265 & -0.680 & -0.590\\
2454831 & .3318 & 0.389 & -0.089 & -0.194 & .3328 & 0.703 & -0.095 & -0.045 & .3329 & 0.600 & -0.253 & -0.249 & .3337 & 0.488 & -0.411 & -0.379 & .3330 & 0.295 & -0.655 & -0.559\\
2454832 & .2844 & 0.381 & -0.091 & -0.206 & .2854 & 0.697 & -0.055 & -0.108 & .2844 & 0.599 & -0.260 & -0.259 & .2861 & 0.483 & -0.414 & -0.382 & .2848 & 0.294 & -0.653 & -0.563\\
2454834 & .3050 & 0.360 & -0.102 & -0.210 & .3056 & 0.686 & -0.064 & -0.118 & .3059 & 0.594 & -0.264 & -0.265 & .3064 & 0.484 & -0.417 & -0.387 & .3069 & 0.294 & -0.655 & -0.563\\
2454835 & .2493 & 0.323 & -0.132 & -0.235 & .2497 & 0.670 & -0.083 & -0.134 & .2507 & 0.585 & -0.272 & -0.271 & .2506 & 0.478 & -0.416 & -0.383 & .2508 & 0.297 & -0.655 & -0.564\\
2454840 & .2018 & 0.565 &  0.099 &  0.000 & .2026 & 0.904 &  0.155 &  0.099 & .2028 & 0.800 & -0.054 & -0.051 & .2029 & 0.682 & -0.214 & -0.184 & .2031 & 0.470 & -0.475 & -0.385\\
2454841 & .2219 & 0.596 &  0.132 &  0.023 & .2224 & 0.941 &  0.192 &  0.130 & .2226 & 0.838 & -0.026 & -0.028 & .2228 & 0.712 & -0.198 & -0.165 & .2230 & 0.498 & -0.457 & -0.361\\
2454843 & .2471 & 0.604 &  0.137 &  0.020 & .2484 & 0.924 &  0.174 &  0.119 & .2486 & 0.827 & -0.033 & -0.033 & .2478 & 0.702 & -0.191 & -0.165 & .2489 & 0.495 & -0.445 & -0.365\\
2454844 & .2223 & 0.550 &  0.080 & -0.016 & .2260 & 0.882 &  0.137 &  0.082 & .2284 & 0.787 & -0.066 & -0.067 & .2298 & 0.673 & -0.222 & -0.191 & .2308 & 0.471 & -0.478 & -0.387\\
2454845 & .2407 & 0.495 &  0.041 & -0.073 & .2413 & 0.830 &  0.083 &  0.026 & .2414 & 0.742 & -0.091 & -0.105 & .2416 & 0.634 & -0.259 & -0.230 & .2418 & 0.430 & -0.508 & -0.416\\
2454850 & .2349 & 0.319 & -0.176 & -0.261 & .2357 & 0.667 & -0.081 & -0.134 & .2360 & 0.593 & -0.263 & -0.259 & .2368 & 0.480 & -0.420 & -0.368 & .2363 & 0.302 & -0.647 & -0.565\\
2454851 & .2346 & 0.302 & -0.169 & -0.268 & .3316 & 0.624 & -0.129 & -0.184 & .2316 & 0.552 & -0.301 & -0.299 & .2302 & 0.455 & -0.443 & -0.414 & .2292 & 0.267 & -0.675 & -0.589\\
2454857 & .2723 & 0.145 & -0.325 & -0.401 & .2757 & 0.517 & -0.236 & -0.294 & .2761 & 0.459 & -0.405 & -0.399 & .2756 & 0.352 & -0.554 & -0.520 & .2763 & 0.169 & -0.779 & -0.688\\
2454865 & .2306 & 0.130 & -0.332 & -0.446 & .2383 & 0.490 & -0.262 & -0.318 & .2387 & 0.420 & -0.439 & -0.435 & .2398 & 0.316 & -0.583 & -0.552 & .2383 & 0.144 & -0.804 & -0.712\\
2454882 & .2211 & 0.147 & -0.336 & -0.438 & .2219 & 0.494 & -0.262 & -0.311 & .2222 & 0.430 & -0.422 & -0.416 & .2224 & 0.304 & -0.570 & -0.539 & .2212 & 0.153 & -0.794 & -0.698\\
2454884 & .2302 & 0.128 & -0.341 & -0.456 & .2301 & 0.492 & -0.260 & -0.313 & .2317 & 0.426 & -0.433 & -0.430 & .2333 & 0.318 & -0.582 & -0.542 & .2323 & 0.141 & -0.806 & -0.716\\
2454891 & .6501 & 0.109 & -0.351 & -0.453 & .6512 & 0.475 & -0.277 & -0.327 & .6521 & 0.411 & -0.444 & -0.443 & .6519 & 0.311 & -0.586 & -0.550 & .6519 & 0.133 & -0.814 & -0.721\\
\hline
\end{tabular}
}
\begin{list}{}{}
\item[$^{\star}$ Observers:] E. Kuligowska, T. Kundera, M. Kurpi\'nska-Winiarska, A. Ku\'zmicz, T. Szyma\'nski, M. Winiarski, S. Zo{\l}a.
\end{list}
\end{table}

\begin{table} 
   \caption{CCD photometry obtained in $UBV(RI)_{\rm C}$ (Bessel) filters
with 0.81~m Tenagra-II telescope in Southern Arizona$^{\star}$ (USA) during
and near the 2008/9 eclipse ($E$~=~10).  Differential magnitudes are given
with respect to three comparison stars: $a$~=~BD~+55$\degr$2690,
$b$~=~GSC-3973:2150, and $c$~=~BD~+55$\degr$2691.  Each point is the mean value
obtained from several frames.  The columns labelled $HJD+$ denote the fraction
of the day.}
   \label{Phot.Lister.10.dat}
\tiny{
\begin{tabular}{|l|rrrr|rrrr|rrrr|rrrr|rrrr|}
\hline\hline
      & \multicolumn{4}{c|}{$\Delta U$} & \multicolumn{4}{c|}{$\Delta B$} & \multicolumn{4}{c|}{$\Delta V$} & \multicolumn{4}{c|}{$\Delta R_{\rm C}$} & \multicolumn{4}{c|}{$\Delta I_{\rm C}$} \\
\hline
$HJD$&$HJD+$&$v-a$&$v-b$&$v-c$&$HJD+$&$v-a$&$v-b$&$v-c$&$HJD+$&$v-a$&$v-b$&$v-c$&$HJD+$&$v-a$&$v-b$&$v-c$&$HJD+$&$v-a$&$v-b$&$v-c$\\
\hline
2454811 & .6217 & 0.202 & -0.299 & -0.465 & .6258 & 0.522 & -0.248 & -0.320 & .6303 & 0.474 & -0.385 & -0.405 & .6351 & 0.345 & -0.559 & -0.536 & .6399 & 0.182 & -0.776 & -0.700\\
2454820 & .6005 & 0.214 & -0.295 & -0.447 & .6035 & 0.515 & -0.259 & -0.333 & .6060 & 0.468 & -0.396 & -0.411 & .6081 & 0.340 & -0.562 & -0.539 & .6102 & 0.177 & -0.780 & -0.705\\
2454821 & .6524 & 0.214 & -0.298 & -0.450 & .6554 & 0.528 & -0.250 & -0.322 & .6579 & 0.480 & -0.385 & -0.399 & .6600 & 0.347 & -0.563 & -0.537 & .6621 & 0.179 & -0.779 & -0.698\\
2454822 & .5941 & 0.224 & -0.285 & -0.446 & .5970 & 0.536 & -0.238 & -0.315 & .5992 & 0.478 & -0.381 & -0.401 & .6012 & 0.351 & -0.550 & -0.528 & .6030 & 0.189 & -0.761 & -0.692\\
\hline
\end{tabular}
}
\begin{list}{}{}
\item[$^{\star}$ Observer:] T. Lister.
\end{list}
\end{table}
\end{landscape}

\addtocounter{table}{29}
\longtabL{29}{
\tiny{
\begin{landscape}
\begin{longtable}{|l|rrrr|rrrr|rrrr|rrrr|rrrr|}
   \caption{\label{Phot.Piwnice.10.dat} {\ $UBV(RI)_{\rm C}$ photometry}
obtained at Piwnice Observatory$^{\star}$ (Poland) during and near the
2008/9 eclipse ($E$~=~10).  The 0.6~m Cassegrain telescope with
SBIG~STL-1001 CCD camera was used.  Differential magnitudes are given with
respect to three comparison stars: $a$~=~BD~+55$\degr$2690,
$b$~=~GSC-3973:2150, and $c$~=~BD~+55$\degr$2691.  Each point is the mean value
obtained from several to tens of frames.  The columns labelled $HJD+$ denote
the fraction of the day.}\\
\hline\hline
      & \multicolumn{4}{c|}{$\Delta U$} & \multicolumn{4}{c|}{$\Delta B$} & \multicolumn{4}{c|}{$\Delta V$} & \multicolumn{4}{c|}{$\Delta R_{\rm C}$} & \multicolumn{4}{c|}{$\Delta I_{\rm C}$} \\
\hline
$HJD$&$HJD+$&$v-a$&$v-b$&$v-c$&$HJD+$&$v-a$&$v-b$&$v-c$&$HJD+$&$v-a$&$v-b$&$v-c$&$HJD+$&$v-a$&$v-b$&$v-c$&$HJD+$&$v-a$&$v-b$&$v-c$\\
\hline
\endfirsthead
\caption{continued.}\\
\hline\hline
      & \multicolumn{4}{c|}{$\Delta U$} & \multicolumn{4}{c|}{$\Delta B$} & \multicolumn{4}{c|}{$\Delta V$} & \multicolumn{4}{c|}{$\Delta R_{\rm C}$} & \multicolumn{4}{c|}{$\Delta I_{\rm C}$} \\
\hline
$HJD$&$HJD+$&$v-a$&$v-b$&$v-c$&$HJD+$&$v-a$&$v-b$&$v-c$&$HJD+$&$v-a$&$v-b$&$v-c$&$HJD+$&$v-a$&$v-b$&$v-c$&$HJD+$&$v-a$&$v-b$&$v-c$\\
\hline
\endhead
\hline
\endfoot
2453545 & .4537 & 0.131 & -0.343 & -0.429 & .4460 & 0.459 & -0.307 & -0.367 & .4407 & 0.411 & -0.437 & -0.452 & .4615 & 0.288 & -0.608 & -0.579 & .4647 & 0.109 & -0.840 & -0.746\\
2453580 & .5803 & 0.163 & -0.353 & -0.419 & .5708 & 0.439 & -0.313 & -0.365 & .5681 & 0.394 & -0.453 & -0.460 & .5735 & 0.294 & -0.625 & -0.583 & .5759 & 0.106 & -0.848 & -0.755\\
2453602 & .4483 & 0.158 & -0.349 & -0.427 & .4395 & 0.447 & -0.306 & -0.372 & .4365 & 0.405 & -0.442 & -0.457 & .4531 & 0.300 & -0.605 & -0.586 & .4545 & 0.123 & -0.830 & -0.755\\
2453613 & .5375 & 0.175 & -0.347 & -0.437 & .5421 & 0.459 & -0.306 & -0.370 & .5439 & 0.411 & -0.442 & -0.459 & .5450 & 0.304 & -0.598 & -0.578 & .5459 & 0.124 & -0.834 & -0.751\\
2453648 & .4016 & 0.163 & -0.337 & -0.424 & .4056 & 0.456 & -0.299 & -0.363 & .4074 & 0.419 & -0.434 & -0.447 & .4086 & 0.303 & -0.595 & -0.583 & .4093 & 0.121 & -0.829 & -0.746\\
2453663 & .3733 & 0.154 & -0.341 & -0.430 & .3709 & 0.451 & -0.299 & -0.355 & .3699 & 0.407 & -0.426 & -0.439 & .3716 & 0.298 & -0.585 & -0.562 & .3721 & 0.113 & -0.816 & -0.741\\
2453745 & .2238 & 0.148 & -0.365 & -0.449 & .2201 & 0.453 & -0.307 & -0.363 & .2177 & 0.405 & -0.443 & -0.456 & .2265 & 0.296 & -0.612 & -0.578 & .2275 & 0.106 & -0.844 & -0.755\\
2453760 & .2837 & 0.226 & -0.270 & -0.348 & .2778 & 0.468 & -0.304 & -0.364 & .2722 & 0.415 & -0.443 & -0.452 & .2907 & 0.295 & -0.598 & -0.584 & .2946 & 0.110 & -0.842 & -0.774\\
2453863 & .5217 & 0.165 & -0.338 & -0.428 & .5134 & 0.457 & -0.308 & -0.365 & .5071 & 0.402 & -0.439 & -0.452 & .5045 & 0.291 & -0.603 & -0.574 & .5031 & 0.104 & -0.833 & -0.750\\
2453868 & ...   & ...   & ...    & ...    & ...   & ...   & ...    & ...    & .5313 & 0.347 & -0.514 & -0.473 & .5305 & 0.275 & -0.616 & -0.575 & .5293 & 0.096 & -0.833 & -0.750\\
2453899 & .4656 & 0.161 & -0.360 & -0.437 & .4608 & 0.471 & -0.295 & -0.362 & .4569 & 0.433 & -0.421 & -0.452 & .4702 & 0.290 & -0.609 & -0.577 & .4719 & 0.112 & -0.835 & -0.751\\
2453940 & .4096 & 0.164 & -0.357 & -0.432 & .4029 & 0.468 & -0.297 & -0.364 & .3978 & 0.429 & -0.425 & -0.454 & .4163 & 0.296 & -0.602 & -0.580 & .4179 & 0.112 & -0.842 & -0.758\\
2453983 & .4526 & 0.186 & -0.330 & -0.411 & .4478 & 0.456 & -0.302 & -0.358 & .4435 & 0.396 & -0.452 & -0.456 & .4417 & 0.313 & -0.594 & -0.574 & .4399 & 0.120 & -0.832 & -0.753\\
2454024 & .4313 & 0.188 & -0.324 & -0.413 & .4273 & 0.450 & -0.315 & -0.357 & .4249 & 0.394 & -0.452 & -0.450 & .4355 & 0.305 & -0.594 & -0.568 & .4375 & 0.115 & -0.829 & -0.743\\
2454128 & .3870 & 0.169 & -0.333 & -0.401 & .3918 & 0.436 & -0.303 & -0.354 & .3795 & 0.393 & -0.443 & -0.446 & .3991 & 0.291 & -0.592 & -0.568 & .4012 & 0.114 & -0.832 & -0.748\\
2454249 & .4816 & 0.165 & -0.377 & -0.400 & .4730 & 0.492 & -0.287 & -0.354 & .4709 & 0.442 & -0.412 & -0.438 & .4755 & 0.308 & -0.594 & -0.574 & .4772 & 0.113 & -0.836 & -0.741\\
2454290 & .5118 & 0.177 & -0.353 & -0.417 & .5058 & 0.483 & -0.286 & -0.355 & .4974 & 0.437 & -0.414 & -0.431 & .5007 & 0.300 & -0.600 & -0.564 & .5032 & 0.110 & -0.842 & -0.740\\
2454321 & .4035 & 0.155 & -0.363 & -0.431 & .3989 & 0.497 & -0.270 & -0.345 & .3952 & 0.431 & -0.420 & -0.437 & .4075 & 0.298 & -0.599 & -0.559 & .4092 & 0.116 & -0.840 & -0.734\\
2454382 & .4395 & 0.172 & -0.346 & -0.392 & .4301 & 0.456 & -0.295 & -0.359 & .4271 & 0.417 & -0.426 & -0.453 & .4329 & 0.305 & -0.585 & -0.568 & .4352 & 0.133 & -0.815 & -0.733\\
2454407 & .3093 & 0.159 & -0.358 & -0.436 & .2999 & 0.453 & -0.310 & -0.357 & .2968 & 0.404 & -0.440 & -0.442 & .3032 & 0.299 & -0.601 & -0.564 & .3044 & 0.113 & -0.839 & -0.740\\
2454530 & ...   & ...   & ...    & ...    & .2293 & 0.460 & -0.306 & -0.355 & .2323 & 0.413 & -0.436 & -0.437 & .2351 & 0.299 & -0.608 & -0.576 & .2365 & 0.132 & -0.827 & -0.754\\
2454536 & ...   & ...   & ...    & ...    & .2699 & 0.461 & -0.316 & -0.356 & .2668 & 0.404 & -0.451 & -0.453 & .2727 & 0.292 & -0.604 & -0.570 & .2743 & 0.124 & -0.837 & -0.760\\
2454571 & .5846 & 0.186 & -0.336 & -0.415 & .5795 & 0.467 & -0.302 & -0.352 & .5777 & 0.418 & -0.429 & -0.433 & .5890 & 0.304 & -0.600 & -0.567 & .5907 & 0.126 & -0.831 & -0.734\\
2454606 & .4481 & 0.173 & -0.339 & -0.419 & .4439 & 0.459 & -0.308 & -0.361 & .4387 & 0.403 & -0.439 & -0.448 & .4405 & 0.291 & -0.608 & -0.575 & .4418 & 0.113 & -0.838 & -0.740\\
2454648 & .4318 & 0.171 & -0.340 & -0.428 & .4281 & 0.460 & -0.297 & -0.353 & .4228 & 0.408 & -0.434 & -0.442 & .4249 & 0.300 & -0.593 & -0.561 & .4261 & 0.120 & -0.819 & -0.731\\
2454621 & ...   & ...   & ...    & ...    & .3847 & 0.466 & -0.303 & -0.358 & .3805 & 0.399 & -0.437 & -0.449 & .3752 & 0.289 & -0.600 & -0.568 & .3665 & 0.122 & -0.822 & -0.741\\
2454662 & .4231 & 0.178 & -0.347 & -0.435 & .4125 & 0.463 & -0.299 & -0.354 & .4096 & 0.409 & -0.430 & -0.438 & .4157 & 0.311 & -0.582 & -0.572 & .4183 & 0.129 & -0.822 & -0.736\\
2454677 & .4145 & 0.177 & -0.341 & -0.424 & .4064 & 0.455 & -0.308 & -0.357 & .4045 & 0.411 & -0.440 & -0.450 & .4096 & 0.301 & -0.598 & -0.574 & .4110 & 0.131 & -0.827 & -0.744\\
2454700 & ...   & ...   & ...    & ...    & .3964 & 0.465 & -0.284 & -0.342 & .3939 & 0.425 & -0.417 & -0.432 & .4001 & 0.313 & -0.580 & -0.557 & .4041 & 0.139 & -0.809 & -0.727\\
2454720 & .4889 & 0.179 & -0.330 & -0.412 & .4840 & 0.467 & -0.292 & -0.344 & .4827 & 0.415 & -0.424 & -0.439 & .4853 & 0.312 & -0.594 & -0.581 & .4862 & 0.148 & -0.824 & -0.731\\
2454734 & .5773 & ...   & -0.362 & -0.423 & .5725 & 0.451 & -0.312 & -0.355 & .5707 & 0.415 & -0.441 & -0.444 & .5747 & 0.290 & -0.610 & -0.570 & .5759 & 0.109 & -0.840 & -0.737\\
2454735 & .2914 & 0.178 & -0.325 & -0.414 & .2875 & 0.481 & -0.274 & -0.338 & .2830 & 0.434 & -0.430 & -0.441 & .2845 & 0.320 & -0.578 & -0.571 & .2856 & 0.139 & -0.809 & -0.736\\
2454742 & .2763 & 0.189 & -0.333 & -0.411 & .2725 & 0.483 & -0.278 & -0.336 & .2700 & 0.437 & -0.415 & -0.443 & .2801 & 0.321 & -0.577 & -0.561 & .2812 & 0.135 & -0.810 & -0.732\\
2454750 & .2490 & 0.198 & -0.322 & -0.422 & .2430 & 0.470 & -0.287 & -0.347 & .2395 & 0.427 & -0.430 & -0.435 & .2569 & 0.329 & -0.580 & -0.574 & .2591 & 0.155 & -0.806 & -0.743\\
2454757 & .2543 & 0.178 & -0.335 & -0.427 & .2435 & 0.475 & -0.282 & -0.343 & .2441 & 0.420 & -0.430 & -0.433 & .2492 & 0.319 & -0.584 & -0.564 & .2514 & 0.141 & -0.804 & -0.738\\
2454763 & .2014 & 0.176 & -0.337 & -0.414 & .2047 & 0.466 & -0.294 & -0.348 & .2077 & 0.419 & -0.418 & -0.442 & .2125 & 0.311 & -0.586 & -0.562 & .2116 & 0.138 & -0.810 & -0.733\\
2454771 & .3691 & 0.187 & -0.335 & -0.411 & .3723 & 0.461 & -0.294 & -0.350 & .3741 & 0.432 & -0.435 & -0.441 & .3751 & 0.316 & -0.587 & -0.561 & .3759 & 0.137 & -0.815 & -0.731\\
2454780 & .2116 & ...   & -0.365 & -0.383 & .2206 & 0.473 & -0.268 & -0.327 & .2178 & 0.407 & -0.421 & -0.427 & .2260 & 0.335 & -0.560 & -0.561 & .2338 & 0.132 & -0.781 & -0.725\\
2454781 & ...   & ...   & ...    & ...    & .1675 & ...   &  ...   & -0.351 & .1729 & 0.426 & -0.431 & -0.449 & ...   & ...   & ...    & ...    & ...   & ...   & ...    & ...   \\
2454782 & ...   & ...   & ...    & ...    & .2186 & 0.470 & -0.284 & -0.343 & .1968 & 0.412 & -0.425 & -0.436 & ...   & ...   & ...    & ...    & ...   & ...   & ...    & ...   \\
2454784 & .2763 & 0.190 & -0.331 & -0.400 & .2687 & 0.477 & -0.284 & -0.334 & .2657 & 0.432 & -0.417 & -0.430 & .2726 & 0.320 & -0.588 & -0.555 & .2713 & 0.136 & -0.826 & -0.728\\
2454793 & .3080 & 0.170 & -0.333 & -0.431 & .3014 & 0.457 & -0.296 & -0.343 & .2986 & 0.416 & -0.427 & -0.432 & .3033 & 0.307 & -0.588 & -0.560 & .3045 & 0.134 & -0.816 & -0.731\\
2454801 & .2991 & 0.196 & -0.311 & -0.426 & .2899 & 0.487 & -0.266 & -0.322 & .2742 & 0.433 & -0.427 & -0.426 & .2927 & 0.328 & -0.574 & -0.547 & .2943 & 0.149 & -0.810 & -0.721\\
2454804 & .1929 & 0.189 & -0.322 & -0.403 & .1806 & 0.494 & -0.266 & -0.335 & .1776 & 0.445 & -0.416 & -0.436 & .1867 & 0.329 & -0.567 & -0.551 & .1843 & 0.142 & -0.805 & -0.729\\
2454810 & .3225 & 0.211 & -0.287 & -0.361 & .3182 & 0.482 & -0.269 & -0.319 & .3104 & 0.439 & -0.411 & -0.416 & .3128 & 0.329 & -0.568 & -0.535 & .3155 & 0.144 & -0.804 & -0.709\\
2454811 & ...   & ...   & ...    & ...    & .1511 & 0.499 & -0.261 & -0.319 & ...   & ...   & ...    & ...    & .1453 & 0.345 & -0.561 & -0.533 & .1476 & 0.156 & -0.800 & -0.722\\
2454814 & .3147 & 0.246 & -0.271 & -0.365 & .3011 & 0.520 & -0.231 & -0.304 & .2966 & 0.476 & -0.364 & -0.392 & .3061 & 0.365 & -0.533 & -0.511 & .3088 & 0.178 & -0.773 & -0.688\\
2454815 & .1837 & 0.244 & -0.263 & -0.372 & .1818 & 0.527 & -0.245 & -0.287 & .1778 & 0.456 & -0.381 & -0.404 & .1792 & 0.355 & -0.551 & -0.521 & .1802 & 0.153 & -0.803 & -0.719\\
2454822 & .1843 & 0.211 & -0.298 & -0.372 & .1736 & 0.505 & -0.250 & -0.311 & .1610 & 0.467 & -0.387 & -0.397 & .1681 & 0.348 & -0.558 & -0.529 & .1649 & 0.162 & -0.796 & -0.707\\
2454824 & .1751 & 0.222 & -0.272 & -0.368 & .1676 & 0.515 & -0.247 & -0.294 & .1612 & 0.457 & -0.395 & -0.390 & .1658 & 0.358 & -0.544 & -0.512 & .1639 & 0.178 & -0.775 & -0.689\\
2454830 & .1785 & 0.403 & -0.140 & -0.215 & .1723 & 0.635 & -0.075 & -0.131 & .1709 & 0.582 & -0.245 & -0.269 & .1754 & 0.461 & -0.431 & -0.406 & .1766 & 0.273 & -0.661 & -0.591\\
2454831 & .1749 & 0.455 & -0.075 & -0.111 & .1694 & 0.697 & -0.045 & -0.097 & .1671 & 0.630 & -0.221 & -0.223 & .1715 & 0.497 & -0.398 & -0.370 & .1723 & 0.272 & -0.667 & -0.564\\
2454834 & .4419 & ...   &  ...   & -0.112 & .4379 & 0.683 & -0.076 & -0.136 & .4365 & 0.617 & -0.233 & -0.250 & .4389 & 0.497 & -0.403 & -0.374 & .4395 & 0.305 & -0.657 & -0.548\\
2454837 & .3498 & 0.543 & -0.016 & -0.042 & .3485 & 0.743 & -0.015 & -0.068 & .3471 & 0.674 & -0.194 & -0.187 & .3435 & 0.554 & -0.353 & -0.319 & .3579 & 0.330 & -0.605 & -0.528\\
2454838 & .3027 & 0.568 &  0.011 & -0.013 & .2993 & 0.782 &  0.033 & -0.026 & .3053 & 0.718 & -0.138 & -0.146 & .3052 & 0.590 & -0.306 & -0.286 & .3047 & 0.390 & -0.565 & -0.474\\
2454839 & .3540 & 0.621 &  0.098 &  0.004 & .3465 & 0.847 &  0.090 &  0.039 & .3399 & 0.800 & -0.056 & -0.078 & .3422 & 0.655 & -0.244 & -0.214 & .3433 & 0.463 & -0.499 & -0.413\\
2454840 & ...   & ...   & ...    & ...    & .4393 & 0.928 &  0.184 &  0.081 & .4408 & 0.829 &  0.006 & -0.045 & .4412 & 0.673 & -0.208 & -0.182 & .4417 & 0.482 & -0.466 & -0.380\\
2454843 & .2597 & 0.676 &  0.128 &  0.070 & .2403 & 0.954 &  0.166 &  0.144 & .2333 & 0.875 &  0.020 & -0.002 & .2473 & 0.726 & -0.195 & -0.149 & .2510 & 0.505 & -0.481 & -0.372\\
2454844 & .1798 & 0.660 &  0.112 &  0.057 & .1846 & 0.884 &  0.130 &  0.060 & .1874 & 0.829 & -0.024 & -0.046 & .1898 & 0.694 & -0.203 & -0.180 & .1920 & 0.481 & -0.472 & -0.387\\
2454845 & .2835 & 0.570 &  0.050 & -0.023 & .2697 & 0.852 &  0.095 &  0.038 & .2724 & 0.771 & -0.075 & -0.092 & .2749 & 0.647 & -0.251 & -0.228 & .2772 & 0.452 & -0.502 & -0.443\\
2454849 & .2648 & 0.499 & -0.093 & -0.122 & .2600 & 0.715 & -0.059 & -0.108 & .2510 & 0.644 & -0.212 & -0.226 & .2539 & 0.531 & -0.389 & -0.344 & .2559 & 0.326 & -0.656 & -0.543\\
2454851 & .1954 & 0.372 & -0.148 & -0.227 & .1879 & 0.630 & -0.121 & -0.176 & .1926 & 0.582 & -0.267 & -0.273 & .1913 & 0.467 & -0.430 & -0.400 & .1903 & 0.262 & -0.675 & -0.586\\
2454860 & .2508 & 0.218 & -0.307 & -0.399 & .2458 & 0.504 & -0.259 & -0.317 & .2455 & 0.454 & -0.402 & -0.411 & ...   & ...   & ...    & ...    & .2456 & 0.167 & -0.792 & -0.714\\
2454865 & .2111 & 0.218 & -0.280 & -0.367 & .2170 & 0.484 & -0.263 & -0.318 & .2200 & 0.448 & -0.401 & -0.410 & .2224 & 0.334 & -0.563 & -0.538 & .2248 & 0.144 & -0.802 & -0.714\\
2454872 & .2273 & 0.202 & -0.313 & -0.380 & .2169 & 0.492 & -0.270 & -0.325 & .2227 & 0.445 & -0.405 & -0.414 & .2210 & 0.331 & -0.578 & -0.541 & .2192 & 0.148 & -0.814 & -0.718\\
2454878 & .2197 & 0.222 & -0.283 & -0.363 & .2331 & 0.519 & -0.240 & -0.296 & .2308 & 0.461 & -0.409 & -0.400 & .2287 & 0.344 & -0.555 & -0.532 & .2260 & 0.164 & -0.796 & -0.695\\
2454881 & ...   & ...   & ...    & ...    & .2281 & 0.500 & -0.254 & -0.305 & ...   & ...   & ...    & ...    & ...   & ...   & ...    & ...    & .2287 & 0.139 & -0.804 & -0.736\\
2454883 & .2266 & 0.225 & -0.316 & -0.374 & .2262 & 0.501 & -0.267 & -0.316 & .2264 & 0.453 & -0.405 & -0.408 & .2276 & 0.332 & -0.579 & -0.541 & .2278 & 0.144 & -0.812 & -0.726\\
2454884 & .2390 & 0.198 & -0.309 & -0.383 & .2391 & 0.502 & -0.259 & -0.313 & .2390 & 0.449 & -0.402 & -0.411 & .2383 & 0.319 & -0.554 & -0.523 & .2390 & 0.152 & -0.800 & -0.710\\
2454891 & .2865 & 0.204 & -0.324 & -0.394 & .2717 & 0.481 & -0.282 & -0.341 & .2807 & 0.439 & -0.416 & -0.434 & .2794 & 0.321 & -0.581 & -0.552 & .2741 & 0.142 & -0.818 & -0.729\\
2454903 & .2480 & 0.197 & -0.326 & -0.414 & .2412 & 0.476 & -0.292 & -0.334 & .2390 & 0.422 & -0.437 & -0.440 & .2431 &	0.310 & -0.596 & -0.568 & .2442 & 0.128 & -0.817 & -0.721\\
2454909 & .2744 & 0.179 & -0.349 & -0.418 & .2697 & 0.469 & -0.290 & -0.341 & .2651 & 0.417 & -0.426 & -0.438 & .2670 &	0.316 & -0.579 & -0.546 & .2682 & 0.121 & -0.809 & -0.723\\
2454922 & .5776 & 0.171 & -0.319 & -0.431 & .5723 & 0.467 & -0.302 & -0.357 & .5705 & 0.435 & -0.424 & -0.432 & .5739 &	0.315 & -0.582 & -0.559 & .5751 & 0.129 & -0.834 & -0.743\\
2454938 & .5187 & 0.170 & -0.365 & -0.409 & .5131 & 0.459 & -0.291 & -0.351 & .5110 & 0.402 & -0.431 & -0.442 & .5151 &	0.294 & -0.585 & -0.559 & .5162 & 0.112 & -0.827 & -0.729\\
2454950 & .5827 & 0.182 & -0.320 & -0.406 & .5794 & 0.465 & -0.299 & -0.356 & .5746 & 0.436 & -0.410 & -0.425 & .5763 &	0.303 & -0.591 & -0.569 & .5777 & 0.129 & -0.827 & -0.731\\
2455017 & .4503 & 0.145 & -0.357 & -0.439 & .4374 & 0.457 & -0.323 & -0.364 & .4330 & 0.399 & -0.450 & -0.446 & .4415 &	0.279 & -0.617 & -0.581 & .4397 & 0.112 & -0.835 & -0.742\\
2455027 & .4991 & 0.151 & -0.358 & -0.420 & .4951 & 0.461 & -0.312 & -0.357 & .4938 & 0.404 & -0.437 & -0.443 & .4962 &	0.288 & -0.602 & -0.556 & .4967 & 0.112 & -0.828 & -0.746\\
2455040 & .3724 & 0.184 & -0.363 & -0.437 & ...   & ...   & ...    & ...    & .3480 & 0.420 & -0.430 & -0.435 & ...   & ...   & ...    & ...    & ...   & ...   & ...    & ...   \\
2455063 & .3655 & 0.186 & -0.351 & -0.417 & .3580 & 0.427 & -0.319 & -0.348 & .3561 & 0.426 & -0.419 & -0.456 & .3602 &	0.282 & -0.612 & -0.579 & .3613 & 0.129 & -0.815 & -0.739\\
2455083 & .3257 & 0.168 & -0.342 & -0.402 & ...   & ...   & ...    & ...    & .3030 & 0.405 & -0.434 & -0.452 & .3040 &	0.287 & -0.606 & -0.575 & .3051 & 0.109 & -0.830 & -0.738\\
2455099 & .4112 & 0.158 & -0.351 & -0.444 & .4065 & 0.425 & -0.305 & -0.353 & .4048 & 0.412 & -0.433 & -0.442 & .4081 &	0.291 & -0.598 & -0.574 & .4087 & 0.110 & -0.833 & -0.733\\
2455111 & .2620 & 0.156 & -0.335 & -0.422 & .2521 & 0.445 & -0.291 & -0.342 & .2492 & 0.427 & -0.428 & -0.459 & .2567 &	0.301 & -0.588 & -0.565 & .2548 & 0.146 & -0.811 & -0.719\\
2455120 & .2429 & 0.161 & -0.356 & -0.431 & .2319 & 0.455 & -0.306 & -0.355 & .2289 & 0.414 & -0.441 & -0.440 & .2347 &	0.298 & -0.601 & -0.565 & .2368 & 0.117 & -0.835 & -0.744\\
2455138 & .2370 & 0.169 & -0.362 & -0.426 & .2298 & 0.452 & -0.318 & -0.360 & .2262 & 0.399 & -0.443 & -0.451 & .2239 &	0.287 & -0.606 & -0.574 & .2217 & 0.105 & -0.836 & -0.745\\
2455150 & .2016 & 0.168 & -0.345 & -0.423 & .1954 & 0.441 & -0.314 & -0.367 & .1922 & 0.399 & -0.454 & -0.459 & .1893 &	0.284 & -0.618 & -0.584 & .1871 & 0.102 & -0.854 & -0.754\\
2455156 & .2812 & 0.142 & -0.411 & -0.446 & .2775 & 0.458 & -0.308 & -0.357 & .2762 & 0.411 & -0.444 & -0.453 & .2784 &	0.298 & -0.604 & -0.569 & .2789 & 0.101 & -0.850 & -0.736\\
2455162 & .3936 & 0.171 & -0.344 & -0.396 & .3884 & 0.458 & -0.305 & -0.358 & .3854 & 0.416 & -0.438 & -0.445 & .3825 &	0.299 & -0.600 & -0.573 & .3806 & 0.115 & -0.834 & -0.744\\
2455164 & .1918 & 0.166 & -0.354 & -0.427 & .1857 & 0.465 & -0.303 & -0.353 & .1838 & 0.418 & -0.441 & -0.448 & .1879 &	0.300 & -0.603 & -0.571 & .1890 & 0.111 & -0.843 & -0.751\\
2455210 & .2149 & 0.142 & -0.390 & -0.391 & .2067 & 0.443 & -0.309 & -0.362 & .2045 & 0.397 & -0.447 & -0.454 & .2085 &	0.281 & -0.618 & -0.585 & .2124 & 0.091 & -0.872 & -0.753\\
2455233 & ...   & ...   & ...    & ...    & .2330 & 0.444 & -0.315 & -0.361 & .2305 & 0.401 & -0.448 & -0.451 & .2285 &	0.272 & -0.622 & -0.585 & .2271 & 0.086 & -0.864 & -0.759\\
2455253 & .2667 & 0.172 & -0.346 & -0.468 & .2605 & 0.453 & -0.328 & -0.377 & .2547 & 0.402 & -0.470 & -0.465 & .2512 &	0.269 & -0.645 & -0.606 & .2456 & 0.078 & -0.883 & -0.785\\
2455264 & ...   & ...   & ...    & ...    & .2437 & 0.452 & -0.317 & -0.369 & .2462 & 0.395 & -0.450 & -0.454 & .2479 &	0.284 & -0.614 & -0.590 & .2506 & 0.081 & -0.872 & -0.774\\
2455267 & ...   & ...   & ...    & ...    & .2491 & 0.438 & -0.323 & -0.377 & .2467 & 0.397 & -0.446 & -0.461 & .2515 &	0.275 & -0.618 & -0.589 & .2529 & 0.090 & -0.865 & -0.773\\
2455270 & ...   & ...   & ...    & ...    & .6493 & 0.438 & -0.315 & -0.361 & .6471 & 0.389 & -0.450 & -0.456 & .6513 &	0.266 & -0.616 & -0.585 & .6526 & 0.075 & -0.866 & -0.767\\
2455271 & ...   & ...   & ...    & ...    & .2712 & 0.446 & -0.318 & -0.367 & .2684 & 0.396 & -0.455 & -0.462 & .2734 &	0.261 & -0.623 & -0.596 & .2749 & 0.086 & -0.862 & -0.771\\
2455272 & ...   & ...   & ...    & ...    & .2729 & 0.454 & -0.309 & -0.361 & .2700 & 0.401 & -0.442 & -0.447 & .2757 &	0.275 & -0.613 & -0.584 & .2781 & 0.083 & -0.860 & -0.770\\
2455274 & .2602 & 0.163 & -0.348 & -0.415 & .2520 & 0.447 & -0.332 & -0.370 & .2545 & 0.404 & -0.443 & -0.457 & .2560 &	0.272 & -0.632 & -0.592 & .2487 & 0.088 & -0.871 & -0.777\\
2455294 & ...   & ...   & ...    & ...    & .5304 & 0.449 & -0.310 & -0.365 & .5282 & 0.401 & -0.447 & -0.457 & .5328 &	0.283 & -0.620 & -0.587 & ...   & ...   & ...    & ...   \\
2455303 & .5519 & 0.157 & -0.366 & -0.430 & .5448 & 0.461 & -0.317 & -0.371 & .5420 & 0.412 & -0.448 & -0.452 & .5473 &	0.284 & -0.623 & -0.586 & .5490 & 0.093 & -0.867 & -0.765\\
2455311 & .5037 & 0.162 & -0.356 & -0.436 & .4953 & 0.458 & -0.313 & -0.368 & .4929 & 0.404 & -0.449 & -0.459 & .4977 &	0.283 & -0.620 & -0.589 & .4993 & 0.088 & -0.876 & -0.771\\
\end{longtable}
\begin{list}{}{}
\item[$^{\star}$ Observers:] P. Dobierski, S. Fr\c{a}ckowiak, C. Ga{\l}an, G. Maciejewski, P. R\'o\.za\'nski, E. \'Swierczy\'nski, M. Wi\c{e}cek, P. Wychudzki.
\end{list}
\end{landscape}
}
}

\begin{landscape}
\begin{table} 
   \caption{$UBV(RI)_{\rm C}$ photometry obtained at Suhora
Observatory$^{\star}$ (Poland) during and near the 2008/9 eclipse
($E$~=~10).  The 0.6~m Cassegrain telescope with a~CCD camera was used.  The
differential magnitudes are given with respect to three comparison stars:
$a$~=~BD~+55$\degr$2690, $b$~=~GSC-3973:2150, and $c$~=~BD~+55$\degr$2691. 
Each point is the mean value obtained from several to tens of frames.  The
columns labelled $HJD+$ denote the fraction of the day.}
   \label{Phot.Suhora.10.dat}
\tiny{
\begin{tabular}{|l|rrrr|rrrr|rrrr|rrrr|rrrr|}
\hline\hline
      & \multicolumn{4}{c|}{$\Delta U$} & \multicolumn{4}{c|}{$\Delta B$} & \multicolumn{4}{c|}{$\Delta V$} & \multicolumn{4}{c|}{$\Delta R_{\rm C}$} & \multicolumn{4}{c|}{$\Delta I_{\rm C}$} \\
\hline
$HJD$&$HJD+$&$v-a$&$v-b$&$v-c$&$HJD+$&$v-a$&$v-b$&$v-c$&$HJD+$&$v-a$&$v-b$&$v-c$&$HJD+$&$v-a$&$v-b$&$v-c$&$HJD+$&$v-a$&$v-b$&$v-c$\\
\hline
2454758 & ...   & ...   & ...    & ...    & .6504 & 0.435 & -0.304 & -0.363 & .6507 & 0.403 & -0.450 & -0.451 & .6509 & 0.298 & -0.621 & -0.576 & .6504 & 0.122 & -0.854 & -0.746\\
2454759 & ...   & ...   & ...    & ...    & .5453 & 0.408 & -0.334 & -0.376 & .5448 & 0.366 & -0.513 & -0.469 & .5450 & 0.282 & -0.632 & -0.563 & .5450 & 0.095 & -0.869 & -0.744\\
2454770 & ...   & ...   & ...    & ...    & ...   & ...   & ...    & ...    & .5888 & 0.422 & -0.446 & -0.461 & .5886 & 0.304 & -0.602 & -0.571 & .5898 & 0.136 & -0.825 & -0.736\\
2454774 & .5766 & 0.150 & -0.301 & -0.424 & .5770 & 0.475 & -0.265 & -0.340 & .5772 & 0.407 & -0.451 & -0.458 & .5779 & 0.305 & -0.597 & -0.575 & .5775 & 0.124 & -0.820 & -0.729\\
2454789 & .1813 & 0.086 & -0.372 & -0.496 & .1930 & 0.451 & -0.287 & -0.349 & .1932 & 0.401 & -0.456 & -0.457 & .1933 & 0.300 & -0.599 & -0.564 & .1933 & 0.117 & -0.828 & -0.733\\
2454798 & ...   & ...   & ...    & ...    & .1804 & 0.447 & -0.289 & -0.344 & .1807 & 0.403 & -0.457 & -0.459 & .1804 & 0.300 & -0.600 & -0.565 & .1811 & 0.118 & -0.834 & -0.737\\ 
2454799 & ...   & ...   & ...    & ...    & .1689 & 0.460 & -0.260 & -0.335 & .1699 & 0.408 & -0.442 & -0.450 & .1701 & 0.308 & -0.583 & -0.556 & .1705 & 0.125 & -0.813 & -0.727\\
2454800 & .3391 & 0.174 & -0.328 & -0.428 & .3390 & 0.475 & -0.265 & -0.332 & .3399 & 0.408 & -0.452 & -0.452 & .3405 & 0.303 & -0.596 & -0.565 & .3401 & 0.121 & -0.827 & -0.736\\
2454801 & .3963 & 0.088 & -0.388 & -0.504 & .3960 & 0.471 & -0.270 & -0.335 & .3963 & 0.413 & -0.448 & -0.452 & .3965 & 0.313 & -0.587 & -0.554 & .3967 & 0.135 & -0.814 & -0.729\\
2454802 & .3645 & 0.092 & -0.380 & -0.487 & .3648 & 0.454 & -0.285 & -0.338 & .3651 & 0.410 & -0.456 & -0.451 & .3635 & 0.302 & -0.600 & -0.559 & .3654 & 0.124 & -0.830 & -0.730\\
2454803 & .1750 & 0.091 & -0.359 & -0.478 & .1758 & 0.462 & -0.273 & -0.330 & .1772 & 0.414 & -0.447 & -0.447 & .1769 & 0.313 & -0.589 & -0.553 & .1767 & 0.132 & -0.816 & -0.722\\
2454804 & .1658 & 0.085 & -0.363 & -0.484 & .1662 & 0.457 & -0.277 & -0.336 & .1664 & 0.409 & -0.450 & -0.453 & .1676 & 0.308 & -0.591 & -0.559 & .1667 & 0.125 & -0.824 & -0.730\\
2454807 & .1758 & 0.101 & -0.354 & -0.470 & .1769 & 0.464 & -0.267 & -0.324 & .1771 & 0.418 & -0.440 & -0.438 & .1772 & 0.314 & -0.584 & -0.547 & .1774 & 0.133 & -0.815 & -0.720\\
2454810 & .1602 & 0.108 & -0.336 & -0.462 & .1606 & 0.469 & -0.259 & -0.314 & .1610 & 0.419 & -0.433 & -0.435 & .1617 & 0.319 & -0.577 & -0.539 & .1612 & 0.135 & -0.809 & -0.717\\
2454814 & .1777 & 0.145 & -0.309 & -0.430 & .1789 & 0.503 & -0.228 & -0.294 & .1787 & 0.446 & -0.414 & -0.417 & .1788 & 0.343 & -0.553 & -0.523 & .1789 & 0.154 & -0.790 & -0.702\\
2454815 & .2924 & 0.146 & -0.316 & -0.431 & .2932 & 0.501 & -0.235 & -0.296 & .2934 & 0.453 & -0.412 & -0.414 & .2923 & 0.342 & -0.559 & -0.529 & .2924 & 0.152 & -0.800 & -0.701\\
2454816 & .1743 & 0.135 & -0.320 & -0.441 & .1743 & 0.493 & -0.243 & -0.307 & .1742 & 0.444 & -0.418 & -0.426 & .1752 & 0.334 & -0.563 & -0.534 & .1753 & 0.158 & -0.790 & -0.698\\
2454817 & ...   & ...   & ...    & ...    & .2193 & 0.511 & -0.218 & -0.284 & .2203 & 0.453 & -0.410 & -0.410 & .2250 & 0.343 & -0.557 & -0.528 & .2251 & 0.158 & -0.790 & -0.700\\
2454828 & .4590 & 0.297 & -0.196 & -0.296 & .4599 & 0.569 & -0.182 & -0.242 & .4602 & 0.484 & -0.377 & -0.373 & .4604 & 0.376 & -0.527 & -0.482 & .4606 & 0.199 & -0.756 & -0.658\\
2454829 & .2105 & 0.264 & -0.174 & -0.289 & .2107 & 0.603 & -0.124 & -0.195 & .2118 & 0.520 & -0.330 & -0.336 & .2114 & 0.416 & -0.471 & -0.449 & .2121 & 0.222 & -0.706 & -0.632\\
2454830 & .3338 & 0.338 & -0.123 & -0.241 & .3361 & 0.660 & -0.075 & -0.134 & .3365 & 0.573 & -0.285 & -0.285 & .3366 & 0.457 & -0.450 & -0.408 & .3368 & 0.247 & -0.704 & -0.595\\
2454831 & .2614 & 0.370 & -0.084 & -0.204 & .2667 & 0.698 & -0.040 & -0.100 & .2682 & 0.605 & -0.258 & -0.258 & .2692 & 0.488 & -0.417 & -0.383 & .2696 & 0.282 & -0.667 & -0.571\\
2454832 & .1930 & 0.355 & -0.077 & -0.208 & .1939 & 0.680 & -0.044 & -0.108 & .1942 & 0.600 & -0.250 & -0.259 & .1956 & 0.485 & -0.404 & -0.379 & .1946 & 0.287 & -0.652 & -0.567\\
2454835 & .1906 & 0.328 & -0.121 & -0.248 & .1912 & 0.654 & -0.083 & -0.132 & .1914 & 0.581 & -0.279 & -0.269 & .1915 & 0.463 & -0.439 & -0.383 & .1916 & 0.270 & -0.681 & -0.562\\
2454838 & .2140 & 0.431 & -0.008 & -0.123 & .2138 & 0.767 &  0.041 & -0.017 & .2140 & 0.688 & -0.159 & -0.160 & .2141 & 0.577 & -0.317 & -0.281 & .2139 & 0.380 & -0.564 & -0.470\\
2454840 & .2584 & 0.569 &  0.104 & -0.013 & .2583 & 0.897 &  0.165 &  0.099 & .2588 & 0.809 & -0.054 & -0.054 & .2585 & 0.684 & -0.215 & -0.185 & .2590 & 0.471 & -0.474 & -0.384\\
2454840 & .3564 & 0.683 &  0.335 &  0.122 & .3573 & 0.935 &  0.201 &  0.121 & .3575 & 0.817 & -0.046 & -0.055 & .3576 & 0.692 & -0.217 & -0.182 & .3577 & 0.483 & -0.470 & -0.378\\
2454841 & .2628 & 0.588 &  0.136 &  0.018 & .2637 & 0.928 &  0.192 &  0.128 & .2560 & 0.837 & -0.025 & -0.031 & .2641 & 0.712 & -0.189 & -0.159 & .2642 & 0.492 & -0.457 & -0.363\\
2454842 & .3841 & 0.728 &  0.259 &  0.070 & .3849 & 0.956 &  0.209 &  0.138 & .3851 & 0.848 & -0.019 & -0.025 & .3852 & 0.729 & -0.189 & -0.163 & .3854 & 0.490 & -0.464 & -0.364\\
2454843 & .1863 & 0.580 &  0.127 &  0.006 & .1880 & 0.914 &  0.187 &  0.120 & .1874 & 0.828 & -0.027 & -0.033 & .1824 & 0.705 & -0.192 & -0.161 & .1876 & 0.488 & -0.457 & -0.365\\
2454844 & .3335 & 0.544 &  0.079 & -0.041 & .3330 & 0.879 &  0.141 &  0.074 & .3332 & 0.759 & -0.095 & -0.075 & .3371 & 0.685 & -0.229 & -0.192 & .3335 & 0.470 & -0.479 & -0.393\\
2454845 & .1773 & 0.497 &  0.045 & -0.084 & .1927 & 0.818 &  0.095 &  0.031 & .1927 & 0.754 & -0.105 & -0.116 & .1929 & 0.625 & -0.262 & -0.233 & .1931 & 0.432 & -0.509 & -0.418\\
2454850 & .2687 & 0.319 & -0.143 & -0.251 & .2696 & 0.660 & -0.073 & -0.133 & .2699 & 0.593 & -0.261 & -0.259 & .2713 & 0.488 & -0.411 & -0.372 & .2702 & 0.290 & -0.652 & -0.560\\
2454851 & .3549 & 0.336 & -0.126 & -0.233 & .3558 & 0.633 & -0.125 & -0.186 & .3566 & 0.555 & -0.311 & -0.313 & .3561 & 0.454 & -0.454 & -0.422 & .3562 & 0.263 & -0.699 & -0.605\\
2454855 & .2659 & 0.197 & -0.242 & -0.386 & .2668 & 0.524 & -0.209 & -0.268 & .2658 & 0.475 & -0.389 & -0.394 & .2659 & 0.360 & -0.546 & -0.505 & .2661 & 0.183 & -0.776 & -0.678\\
2454858 & .3938 & 0.149 & -0.309 & -0.431 & .3945 & 0.512 & -0.238 & -0.308 & .3967 & 0.471 & -0.397 & -0.425 & ...   & ...   & ...    & ...    & ...   & ...   & ...    & ...   \\
2454858 & .6764 & 0.152 & -0.304 & -0.430 & .6780 & 0.498 & -0.244 & -0.309 & .6776 & 0.440 & -0.421 & -0.430 & .6777 & 0.332 & -0.563 & -0.538 & .6778 & 0.157 & -0.794 & -0.701\\
2454865 & .2305 & 0.137 & -0.302 & -0.421 & .2311 & 0.490 & -0.246 & -0.314 & .2313 & 0.421 & -0.435 & -0.435 & .2315 & 0.319 & -0.578 & -0.549 & .2316 & 0.140 & -0.804 & -0.718\\
2454868 & .2307 & 0.130 & -0.336 & -0.454 & .2321 & 0.481 & -0.256 & -0.317 & .2314 & 0.431 & -0.432 & -0.435 & .2316 & 0.329 & -0.574 & -0.544 & .2317 & 0.143 & -0.808 & -0.715\\
2454869 & .2276 & 0.111 & -0.344 & -0.457 & .2285 & 0.466 & -0.268 & -0.327 & .2288 & 0.410 & -0.453 & -0.447 & .2290 & 0.306 & -0.597 & -0.551 & .2291 & 0.097 & -0.846 & -0.732\\
2454872 & .6515 & 0.117 & -0.344 & -0.471 & .6524 & 0.485 & -0.263 & -0.316 & .6532 & 0.419 & -0.448 & -0.449 & .6540 & 0.312 & -0.592 & -0.556 & .6528 & 0.133 & -0.819 & -0.724\\
\hline
\end{tabular}
}
\begin{list}{}{}
\item[Observers:] M. Dr\'o\.zd\.z, J. Krzesi\'nski, W. Og{\l}oza, M. Siwak, M. Winiarski, S. Zo{\l}a.
\end{list}
\end{table}
\end{landscape}

\begin{table*} 
   \caption{$BVI_{\rm C}$ photometry obtained at GRAS Observatory$^{\star}$
(Mayhill, New Mexico, USA) during the 2008/9 eclipse ($E$~=~10).  The 0.3~m
GRAS-001 Telescope with an FLI~IMG~1024~DM CCD camera was used. 
Differential magnitudes are given with respect to three comparison stars:
$a$~=~BD~+55$\degr$2690, $b$~=~GSC-3973:2150, and $c$~=~BD~+55$\degr$2691. 
The columns labelled $HJD+$ denote the fraction of the day.}
   \label{Phot.GRAS.10.dat}
\centering 
\begin{tabular}{|l|rrrr|rrrr|rrrr|}
\hline\hline
      & \multicolumn{4}{c|}{$\Delta B$} & \multicolumn{4}{c|}{$\Delta V$} & \multicolumn{4}{c|}{$\Delta I_{\rm C}$} \\
\hline
$HJD$&$HJD+$&$v-a$&$v-b$&$v-c$&$HJD+$&$v-a$&$v-b$&$v-c$&$HJD+$&$v-a$&$v-b$&$v-c$\\
\hline
2454816 & .63926 & 0.502 & -0.249 & -0.314 & .63520 & 0.456 & -0.402 & -0.393 & ...    & ...   & ...    & ...   \\
2454821 & .58581 & 0.518 & -0.246 & -0.302 & .58205 & 0.450 & -0.417 & -0.423 & .58963 & 0.173 & -0.784 & -0.681\\
2454822 & .59271 & 0.526 & -0.214 & -0.281 & .58891 & 0.438 & -0.412 & -0.420 & .59650 & 0.170 & -0.783 & -0.696\\
2454825 & .63850 & 0.512 & -0.235 & -0.308 & .63147 & 0.446 & -0.408 & -0.422 & .64524 & 0.188 & -0.756 & -0.689\\
2454828 & .60494 & 0.514 & -0.223 & -0.310 & .59898 & 0.497 & -0.362 & -0.378 & ...    & ...   & ...    & ...   \\
2454829 & .70329 & 0.629 & -0.114 & -0.187 & .69616 & 0.528 & -0.325 & -0.350 & .65538 & 0.239 & -0.707 & -0.634\\
2454830 & .60893 & 0.651 & -0.100 & -0.162 & .60199 & 0.567 & -0.290 & -0.309 & .61564 & 0.276 & -0.669 & -0.605\\
2454831 & .60554 & 0.675 & -0.076 & -0.142 & .59850 & 0.577 & -0.279 & -0.301 & .61225 & 0.301 & -0.657 & -0.589\\
2454832 & .59797 & 0.708 & -0.039 & -0.119 & .59102 & 0.592 & -0.264 & -0.281 & .60470 & 0.305 & -0.653 & -0.574\\
2454833 & .59704 & 0.679 & -0.059 & -0.135 & .59008 & 0.591 & -0.266 & -0.289 & .60378 & 0.290 & -0.661 & -0.569\\
2454834 & .57529 & 0.655 & -0.087 & -0.150 & .56834 & 0.587 & -0.272 & -0.288 & .58201 & 0.297 & -0.653 & -0.576\\
2454835 & .61749 & 0.680 & -0.073 & -0.128 & .61434 & 0.611 & -0.232 & -0.245 & .62063 & 0.308 & -0.637 & -0.550\\
2454839 & .59620 & 0.848 &  0.101 &  0.037 & .58923 & 0.778 & -0.076 & -0.080 & .60181 & 0.491 & -0.457 & -0.423\\
2454840 & .57656 & 0.935 &  0.193 &  0.125 & .56960 & 0.807 & -0.045 & -0.067 & .58327 & 0.520 & -0.435 & -0.371\\
2454841 & .56866 & 0.929 &  0.178 &  0.116 & .56562 & 0.847 &  0.004 & -0.009 & .57172 & 0.494 & -0.452 & -0.376\\
2454843 & .57533 & 0.905 &  0.155 &  0.092 & .56830 & 0.810 & -0.048 & -0.062 & .58206 & 0.488 & -0.460 & -0.385\\
2454844 & .57236 & 0.897 &  0.141 &  0.080 & .56526 & 0.786 & -0.075 & -0.088 & .57802 & 0.492 & -0.485 & -0.399\\
2454845 & .57694 & 0.822 &  0.071 &  0.013 & .56986 & 0.739 & -0.112 & -0.138 & .58367 & 0.427 & -0.531 & -0.440\\
2454846 & .57689 & 0.823 &  0.069 & -0.004 & .56980 & 0.736 & -0.116 & -0.142 & .58362 & 0.429 & -0.516 & -0.448\\
2454847 & .58281 & 0.754 &  0.006 & -0.060 & .58032 & 0.695 & -0.147 & -0.158 & .58587 & 0.390 & -0.541 & -0.470\\
2454849 & .57644 & 0.683 & -0.058 & -0.124 & .57338 & 0.629 & -0.207 & -0.224 & .57956 & 0.310 & -0.617 & -0.538\\
2454850 & .59034 & 0.673 & -0.082 & -0.127 & .58326 & 0.597 & -0.254 & -0.274 & .59704 & 0.303 & -0.632 & -0.569\\
2454851 & .58991 & 0.627 & -0.120 & -0.183 & .58276 & 0.560 & -0.287 & -0.306 & .59657 & 0.272 & -0.675 & -0.605\\
2454852 & .58999 & 0.605 & -0.154 & -0.208 & .58291 & 0.548 & -0.312 & -0.336 & .59667 & 0.253 & -0.700 & -0.625\\
2454853 & .62551 & 0.574 & -0.187 & -0.253 & .66065 & 0.520 & -0.343 & -0.354 & .62100 & 0.218 & -0.728 & -0.657\\
2454855 & .57829 & 0.552 & -0.192 & -0.263 & .58145 & 0.478 & -0.379 & -0.400 & .57515 & 0.203 & -0.750 & -0.678\\
2454856 & .58972 & 0.558 & -0.206 & -0.286 & .58252 & 0.487 & -0.376 & -0.395 & .59638 & 0.194 & -0.759 & -0.674\\
2454857 & .58990 & 0.531 & -0.217 & -0.282 & .58281 & 0.464 & -0.406 & -0.416 & .59656 & 0.183 & -0.777 & -0.694\\
2454859 & .59015 & 0.507 & -0.250 & -0.314 & .58307 & 0.455 & -0.420 & -0.430 & .59682 & 0.170 & -0.798 & -0.704\\
2454860 & .59062 & 0.514 & -0.233 & -0.307 & .58356 & 0.451 & -0.413 & -0.432 & .59729 & 0.149 & -0.794 & -0.715\\
2454861 & .59023 & 0.509 & -0.245 & -0.309 & .58312 & 0.446 & -0.420 & -0.436 & .59691 & 0.163 & -0.788 & -0.726\\
2454862 & .58953 & 0.522 & -0.235 & -0.304 & .58245 & 0.454 & -0.409 & -0.431 & .59621 & 0.164 & -0.774 & -0.712\\
2454863 & .58992 & 0.512 & -0.247 & -0.307 & .58284 & 0.447 & -0.416 & -0.438 & .59658 & 0.157 & -0.811 & -0.731\\
2454865 & .57222 & 0.517 & -0.246 & -0.318 & .56513 & 0.462 & -0.404 & -0.418 & .57888 & 0.163 & -0.795 & -0.732\\
2454866 & .58951 & 0.518 & -0.241 & -0.306 & .58243 & 0.465 & -0.399 & -0.420 & .59618 & 0.166 & -0.764 & -0.708\\
2454867 & .57230 & 0.516 & -0.252 & -0.315 & .56520 & 0.465 & -0.405 & -0.416 & .57896 & 0.154 & -0.800 & -0.728\\
2454868 & .58957 & 0.509 & -0.256 & -0.318 & .58249 & 0.457 & -0.408 & -0.428 & .59623 & 0.166 & -0.801 & -0.726\\
2454869 & .57300 & 0.494 & -0.253 & -0.301 & .56591 & 0.452 & -0.419 & -0.427 & .57970 & 0.148 & -0.788 & -0.716\\
2454872 & .58669 & 0.495 & -0.243 & -0.317 & .57960 & 0.457 & -0.411 & -0.427 & .59336 & 0.153 & -0.803 & -0.723\\
2454874 & .58378 & 0.509 & -0.252 & -0.320 & .57669 & 0.463 & -0.406 & -0.431 & .59044 & 0.156 & -0.794 & -0.728\\
2454875 & .58655 & 0.524 & -0.252 & -0.266 & .57946 & 0.443 & -0.425 & -0.431 & .59433 & 0.145 & -0.818 & -0.722\\
2454876 & .57465 & 0.528 & -0.234 & -0.305 & .60310 & 0.451 & -0.401 & -0.412 & .58132 & 0.173 & -0.772 & -0.703\\
2454877 & .58635 & 0.534 & -0.227 & -0.260 & .57927 & 0.454 & -0.403 & -0.417 & .59302 & 0.160 & -0.791 & -0.696\\
\hline
\end{tabular}
\begin{list}{}{}
\item[$^{\star}$ Observer:] G. Myers.
\end{list}
\end{table*}

\begin{table*}
   \caption{$BV$ photometry obtained at Rolling Hills Observatory$^{\star}$
(Florida, USA) during the 2008/9 eclipse ($E$~=~10).  The 0.25~m (10")
LX~200 telescope with an SBIG~ST-9XE CCD camera was used.  Differential
magnitudes are given with respect to three comparison stars:
$a$~=~BD~+55$\degr$2690, $b$~=~GSC-3973:2150, and $c$~=~BD~+55$\degr$2691. 
The columns labelled $JD+$ denote the fraction of the day.}
   \label{Phot.Dvorak.10.dat}
\centering 
\begin{tabular}{|l|rrrr|rrrr|}
\hline\hline
      & \multicolumn{4}{c|}{$\Delta B$} & \multicolumn{4}{c|}{$\Delta V$}\\
\hline
$JD$&$JD+$&$v-a$&$v-b$&$v-c$&$JD+$&$v-a$&$v-b$&$v-c$\\
\hline
2454800 & .6208 & 0.493 & -0.263 & -0.314 & .6176 & 0.418 & -0.441 & -0.443\\
2454804 & .5321 & 0.488 & -0.277 & -0.329 & .5290 & 0.417 & -0.442 & -0.445\\
2454808 & .4982 & 0.505 & -0.247 & -0.315 & .4955 & 0.430 & -0.429 & -0.431\\
2454810 & .6058 & 0.506 & -0.230 & -0.299 & .6034 & 0.438 & -0.430 & -0.426\\
2454812 & .4833 & 0.486 & -0.251 & -0.317 & .4803 & 0.426 & -0.431 & -0.436\\
2454813 & .4903 & 0.518 & -0.230 & -0.285 & .4861 & 0.447 & -0.411 & -0.416\\
\hline
\end{tabular}
\begin{list}{}{}
\item[$^{\star}$ Observer:] S. Dvorak.
\end{list}
\end{table*}

\begin{table*}
\caption{Shifts onto our reference system (Krak\'ow Observatory) for the
photometric data obtained during and near the 2003 eclipse ($E$~=~9).  The
shifts were applied to the differential magnitudes shown in
Tables~\ref{Phot.Athens.9.dat}--\ref{Phot.Skinakas.9.dat}.}
\label{shifts.9}
\centering			
\begin{tabular}{llllrrrrr}
\hline\hline
Observatory		& country		& Tel. type		& Diameter	&$shift_{\rm U}$&$shift_{\rm B}$&$shift_{\rm V}$&$shift_{{\rm R}_C}$&$shift_{{\rm I}_C}$\\
\hline
Athens			& Greece		& Cassegrain		& 0.4m          &...    &-0.034 &-0.005 &~0.000 &-0.032 \\
Bia\l{}k\'ow		& Poland		& Cassegrain		& 0.6m          &...    &...    &+0.038 &...    &-0.005 \\
Krak\'ow		& Poland		& Cassegrain		& 0.5m 	        &~0.0   &~0.0   &~0.0   &~0.0   &~0.0   \\
Kryoneri		& Greece		& Cassegrain		& 1.2m 	        &-0.009 &+0.029 &+0.037 &+0.007 &-0.010 \\
Piszk\'estet\"o		& Hungary		& Schmidt		& 0.6/0.9m      &...    &+0.013 &+0.038 &+0.040 &-0.002 \\
Piwnice			& Poland		& Cassegrain		& 0.6m          &+0.009 &-0.018 &+0.021 &+0.053 &+0.016 \\
Rozhen			& Bulgaria		& Ritchey-Chr\'etien	& 2m            &~0.0   &~0.0   &+0.068 &~0.0   &~0.0   \\
Rozhen			& Bulgaria		& Schmidt		& 0.5/0.7m      &...    &...    &+0.062 &...    &...    \\
Rozhen			& Bulgaria		& Cassegrain		& 0.6m          &+0.091 &+0.061 &+0.032 &...    &...    \\
Skinakas		& Greece		& Ritchey-Chr\'etien	& 1.3m 	        &-0.036 &+0.005 &+0.020 &+0.007 &-0.008 \\
\hline                                                                                                                          
\end{tabular}
\end{table*}

\begin{table*}
\caption{Shifts onto our reference system (Krak\'ow Observatory) for the
photometric data obtained during and near the 2008/9 eclipse ($E$~=~10). 
The shifts were applied to the differential magnitudes from
Tables~\ref{Phot.Altan.10.dat}--\ref{Phot.SRO.HPO.10.dat} and reduced to
BD~+55$\degr$2690 average values shown in
Tables~\ref{Phot.Athens.10.dat}--\ref{Phot.Dvorak.10.dat}.}
\label{shifts.10}
\centering			
\begin{tabular}{llllrrrrr}
\hline\hline
Observatory		& country		& Tel. type		& Diameter   &$shift_{\rm U}$&$shift_{\rm B}$&$shift_{\rm V}$&$shift_{{\rm R}_C}$&$shift_{{\rm I}_C}$\\
\hline
Altan, Mt Giant		& Czech Republic	& Reflector		& 0.2m       &...    &-0.025 &-0.030 &-0.008 &-0.020 \\
Athens			& Greece		& Cassegrain		& 0.4m       &...    &-0.004 &-0.025 &-0.009 &-0.022 \\
Bia\l{}k\'ow		& Poland		& Cassegrain		& 0.6m       &...    &+0.004 &-0.014 &-0.006 &-0.022 \\
Green Island		& North Cyprus		& Ritchey-Chr\'etien 	& 0.35m      &...    &-0.001 &-0.018 &-0.001 &-0.031 \\
Hankasalmi		& Finland		& RCOS			& 0.4m       &...    &-0.009 &-0.009 &-0.012 &+0.009 \\
Furzehill, Swansea	& United Kingdom	& Schmidt-Cassegrain	& 0.35m      &...    &+0.013 &-0.020 &-0.020 &-0.036 \\
Krak\'ow		& Poland		& Cassegrain		& 0.5m 	     &~0.0   &~0.0   &~0.0   &~0.0   &~0.0   \\
\hline
\end{tabular}
\end{table*}


\addtocounter{table}{35}
\longtabL{35}{
\begin{landscape}
\begin{longtable}{lrrrrrrrrrrrrrrrrrrrr}
   \caption{\label{Mean.E.9.10.dat} The mean points of the average
$UBV(RI)_{\rm C}$ light curves obtained during and near the 2003 ($E$~=~9)
and 2008/9 ($E$~=~10) eclipses.  Magnitudes and the corresponding standard
deviations are given.  Mean magnitudes were obtained by averaging the data
($var - a$): in the case of the 2003 eclipse from
Tables~\ref{Phot.Athens.9.dat}--\ref{Phot.Piszkesteto.9.dat} and in the case
of the 2008/9 eclipse from
Tables~\ref{Phot.Altan.10.dat}--\ref{Phot.Dvorak.10.dat}.  Finally, to
obtain magnitudes in the standard Johnson-Cousins photometric system, the
comparison star ($a$=BD+55$^{\circ}2690$) magnitudes (according to
\citet{Mik2003}) have been added.  The columns labelled $JD+$ denote the
fraction of the day.}\\
\hline\hline
$JD$&$JD+$&$U$&$\sigma_{\rm U}$&$N_{\rm U}$&$JD+$&$B$&$\sigma_{\rm B}$&$N_{\rm B}$&$JD+$&$V$&$\sigma_{\rm V}$&$N_{\rm V}$&$JD+$&$R_{\rm C}$&$\sigma_{{\rm R}_C}$&$N_{{\rm R}_C}$&$JD+$&$I_{\rm C}$&$\sigma_{{\rm I}_C}$&$N_{{\rm I}_C}$\\
\hline
\endfirsthead
\caption{continued.}\\
\hline\hline
$JD$&$JD+$&$U$&$\sigma_{\rm U}$&$N_{\rm U}$&$JD+$&$B$&$\sigma_{\rm B}$&$N_{\rm B}$&$JD+$&$V$&$\sigma_{\rm V}$&$N_{\rm V}$&$JD+$&$R_{\rm C}$&$\sigma_{{\rm R}_C}$&$N_{{\rm R}_C}$&$JD+$&$I_{\rm C}$&$\sigma_{{\rm I}_C}$&$N_{{\rm I}_C}$\\
\hline
\endhead
\hline
\endfoot
2452520 &.44 &10.925 &...   &1  &.44 &11.141	&...	&1  &.44	&10.824	&...	&1  &.44	&10.396	&...	&1  &.44	& 9.984	&...	&1\\
2452528 &.48 &10.905 &...   &1  &.48 &11.111	&...	&1  &.48	&10.789	&...	&1  &.48	&10.382	&...	&1  &.48	& 9.976	&...	&1\\
2452537 &.39 &10.884 &...   &1  &.39 &11.117	&...	&1  &.39	&10.787	&...	&1  &.39	&10.375	&...	&1  &.39	& 9.957	&...	&1\\
2452550 &.48 &10.910 &...   &1  &.48 &11.139	&...	&1  &.48	&10.805	&...	&1  &.48	&10.390	&...	&1  &.48	& 9.959	&...	&1\\
2452567 &.44 &10.877 &...   &1  &.44 &11.092	&...	&1  &.44	&10.804	&...	&1  &.44	&10.382	&...	&1  &.44	& 9.971	&...	&1\\
2452618 &.44 &10.918 &...   &1  &.44 &11.126	&...	&1  &.44	&10.800	&...	&1  &.44	&10.400	&...	&1  &.44	& 9.991	&...	&1\\
2452644 &.30 &10.912 &...   &1  &.30 &11.120	&...	&1  &.30	&10.799	&...	&1  &.30	&10.376	&...	&1  &.30	& 9.970	&...	&1\\
2452706 &... &...    &...   &...&... &...   	&...	&...&...	&...   	&...	&...&.32	&10.370	&...	&1  &.32	& 9.967	&...	&1\\
\end{longtable}
\end{landscape}
}

\addtocounter{table}{36}
\longtabL{36}{
\begin{landscape}
\begin{longtable}{lllrlllrl}
   \caption{\label{Spect.list.8910.dat} List of spectra obtained during and
near the last three eclipses (from $E$~=~8 to $E$~=~10) and names of the
corresponding FITS files.}\\
\hline\hline
Date [UT]           & HJD-2400000 & Phase   & Epoch & Observatory   & Instrument                       & Spectral Region       & Res. power & Name of FITS file                 \\
\hline
\endfirsthead
\caption{continued.}\\
\hline\hline
Date [UT]           & HJD-2400000 & Phase   & Epoch & Observatory   & Instrument                       & Spectral Region       & Res. power & Name of FITS file                 \\
\hline
\endhead
\hline
\endfoot
1998-07-03 ~~1:05:30 & 50997.5455  & 8.12387 &  8    & Rozhen        & Ritchey-Chr\'etien/2.0  Coude   & H$\alpha$             & 15000      & 980703.Rozhen.Ha.fits             \\
1998-09-30  20:46:37 & 51087.3657  & 8.16769 &  8    & Rozhen        & Ritchey-Chr\'etien/2.0  Coude   & H$\alpha$             & 15000      & 980930.Rozhen.Ha.fits             \\
1998-09-30  23:14:09 & 51087.4682  & 8.16774 &  8    & Rozhen        & Ritchey-Chr\'etien/2.0  Coude   & \ion{Na}{i}           & 15000      & 980930.Rozhen.NaI.fits            \\
2002-11-19  18:08:37 & 52598.2560  & 8.90473 &  9    & Rozhen        & Ritchey-Chr\'etien/2.0  Coude   & \ion{Na}{i}           & 15000      & 021119.Rozhen.NaI.fits            \\
2002-11-19  18:47:34 & 52598.2830  & 8.90474 &  9    & Rozhen        & Ritchey-Chr\'etien/2.0  Coude   & H$\alpha$             & 15000      & 021119.Rozhen.Ha.fits             \\
2002-11-20  16:03:32 & 52599.1691  & 8.90517 &  9    & Rozhen        & Ritchey-Chr\'etien/2.0  Coude   & \ion{Na}{i}           & 15000      & 021120.Rozhen.NaI.fits            \\
2002-11-20  17:11:14 & 52599.2161  & 8.90520 &  9    & Rozhen        & Ritchey-Chr\'etien/2.0  Coude   & H$\alpha$             & 15000      & 021120.Rozhen.Ha.fits             \\
2003-04-12 ~~2:54:15 & 52741.6210  & 8.97466 &  9    & Rozhen        & Ritchey-Chr\'etien/2.0  Coude   & \ion{Na}{i}           & 15000      & 030412.Rozhen.NaI.fits            \\
\end{longtable}
\end{landscape}
}

\end{appendix}

\end{document}